\documentclass[pdftex,twocolumn,epjc3]{svjour3}          

\usepackage{amsmath,amsfonts,amssymb}
\RequirePackage[T1]{fontenc}

\newcommand{\feynslash}[1]{/\hspace*{-2mm} #1}

\smartqed  

\RequirePackage{graphicx}
\RequirePackage{mathptmx}      
\RequirePackage{flushend}
\RequirePackage[numbers,sort&compress]{natbib}
\RequirePackage[colorlinks,citecolor=blue,urlcolor=blue,linkcolor=blue]{hyperref}

\journalname{Eur. Phys. J. C}

\begin{document}

\title{The Quark-Gluon Vertex and the QCD Infrared Dynamics}


\author{Orlando Oliveira\thanksref{e1,addr1,addr2}
        \and
        Wayne de Paula\thanksref{e2,addr2}
        \and
        Tobias  Frederico\thanksref{e3,addr2}
        \and
        J. P. B. C de Melo\thanksref{e4,addr3}
}

\thankstext{e1}{e-mail: orlando@uc.pt}
\thankstext{e2}{e-mail: wayne@ita.br}
\thankstext{e3}{e-mail: tobias@ita.br}
\thankstext{e4}{e-mail: joao.mello@cruzeirodosul.edu.br}

\institute{CFisUC, Departamento de F\'{\i}sica, Universidade de Coimbra, 3004-516 Coimbra, Portugal\label{addr1}
          \and
          Dep. de F\'\i sica, Instituto Tecnol\'ogico da Aeron\'autica, Centro T\'ecnico Aeroespacial, 12.228-900 S\~ao Jos\'e dos Campos, S\~ao Paulo, Brazil\label{addr2}
          \and
           Laboratorio de F\'\i sica Te\'orica e Computacional, Universidade Cruzeiro do Sul, S\~ao Paulo, 01506-000 SP, Brazil\label{addr3}
}

\date{Received: date / Accepted: date}

\maketitle

\begin{abstract}
The Dyson-Schwinger quark equation is solved for the quark-gluon vertex using the most recent lattice data available in the Landau gauge for the quark, gluon and ghost propagators, the full set of longitudinal tensor structures in the Ball-Chiu vertex, taking into account a recently derived normalisation for a quark-ghost kernel form factors and the gluon contribution for the tree level quark-gluon vertex identified on a recent study of the lattice soft gluon limit. A solution for the inverse problem is computed after the Tikhonov linear regularisation of the integral equation, that implies solving a modified Dyson-Schwinger equation. We get longitudinal form factors that are strongly enhanced at the infrared region, deviate significantly from the tree level results for quark and gluon momentum below 2 GeV and at higher momentum approach their perturbative values. The computed quark-gluon vertex favours kinematical configurations where the quark momentum $p$ and the gluon momentum $q$ are small and parallel. Further, the quark-gluon vertex is dominated by the form factors associated to the tree level vertex $\gamma_\mu$ and  to the operator  $2 \, p_\mu + q_\mu$. The higher rank tensor structures provide small contributions to the vertex.
\end{abstract}

\section{Introduction}

The interaction of quarks and gluons is described by Quantum Chromodynamics~\cite{Marciano:1977su,Alkofer:2000wg,Fischer:2006ub,Binosi:2009qm}, 
a renormalisable gauge theory associated to the color gauge group SU(3). 
Of its correlation functions, the quark-gluon vertex has a fundamental role in hadron phenomenology, in the understanding of chiral symmetry breaking 
mechanism and the realisation of confinement. 
Despite its relevance for strong interactions, our knowledge of the quark-gluon vertex  from first principles calculations is relatively poor.
At the perturbative level, only recently a full calculation of the twelve form factors associated to this vertex
was published~\cite{Bermudez:2017bpx} but only for some kinematical configurations,
namely the symmetric configuration (equal incoming, outgoing quark and gluon  squared momenta),
the on-shell configuration (quarks on-shell with vanishing gluon momentum) and what the authors called its asymptotic limit.
In particular, the vertex asymptotic limit was used to investigate  \textit{ans\"atze} that can be found in the 
literature~\cite{Ball:1980ay,Curtis:1990zs,Bashir:2011dp,Qin:2013mta,Aslam:2015nia,Binosi:2016wcx} with the aim to test their description of the ultraviolet regime.

At the non-perturbative level, the quark-gluon vertex has been studied within continuum approaches to 
QCD by several authors~\cite{Alkofer:2008tt,Aguilar:2010cn,Aguilar:2011xe,Windisch:2012de,Hopfer:2013via,Aguilar:2014lha,Sanchis-Alepuz:2015qra,Pelaez:2015tba,Binosi:2014aea,Williams:2015cvx,Binosi:2016wcx,Aguilar:2016lbe,Aguilar:2018epe}. 
Typically, the computation is performed after writing the vertex in terms of other QCD vertices and propagators and taking into account its perturbative tail. 
Most of the computations  include only a fraction of the twelve form factors required
to fully describe the quark-gluon vertex. 
In~\cite{Aguilar:2014lha,Aguilar:2016lbe,Aguilar:2018epe} the authors look for a first principle determination of the vertex by solving the theory at the non-perturbative level,
gathering information on the vertex from QCD symmetries  and relying on one-loop dressed perturbation theory.
The vertex has also been investigated perturbatively within massive QCD, i.e. using the Curci-Ferrari model~\cite{Pelaez:2015tba},
and all its (perturbative) tensor structures form factors accessed for some kinematical configurations.

Lattice simulations, both for quenched~\cite{Skullerud:2001aw,Skullerud:2003qu,Kizilersu:2006et} and full QCD~\cite{Oliveira:2016muq}, 
were also used to investigate the quark-gluon vertex. Again, only a limited set of kinematical configurations were accessed and, in particular, its the soft gluon limit, 
defined by a vanishing gluon momenta, was mostly explored. 
For full QCD so far only a single form factor, that associated with the tree level tensor structure, was measured on lattice simulation in the soft gluon limit.

One can also find in the literature attempts to combine continuum non-per\-tur\-ba\-ti\-ve QCD equations with  lattice simulations
to study the quark-gluon vertex. Indeed, in~\cite{Rojas:2013tza} a generalised Ball-Chiu vertex was used in the quark gap equation, together with lattice
results for the quark, gluon and ghost propagators to investigate the quark-gluon vertex. In \cite{Oliveira2018a}, the full QCD lattice data for
$\lambda_1$ was studied relying on continuum information about the vertex. 

The use of continuum equations with results coming from lattice simulations requires high quality lattice data to feed the continuum equations that should be 
solved for the vertex. In this approach, the computation of a solution of the continuum equations requires assuming some type of functional dependence for 
various propagator functions. 
In recent years, there has been an effort to improve the quality of the lattice data, in the sense of being closer to the continuum and producing simulations
with large statistical ensembles, both for propagators and for vertex functions. This approach that combines lattice information and continuum equations
relies strongly on the effort to access high precision lattice simulations.

For the practitioner oftentimes it is sufficient to have a good model of the vertex that should incorporate the perturbative tail to describe the ultraviolet regime,
some ``guessing'' for the infrared region and, hopefully, comply with perturbative 
renormalisation~\cite{Ball:1980ay,Curtis:1990zs,Bashir:1994az,Bashir:2011dp,Bermudez:2017bpx}.
A popular and quite successful model was set in~\cite{Maris99a}, named the Maris-Tandy model,
that assumes a bare quark-gluon vertex and introduces an effective gluon propagator that is strongly enhanced at infrared scales and recovers
the one-loop behaviour at higher momentum. This model simplifies considerably the momentum dependence of the combined 
effective gluon propagator and quark-gluon vertex and assumes that the dominant momentum dependence is associated only with the gluon momentum.
Such type of vertex that appears in the Dyson-Schwinger and the Bethe-Salpeter equations can be seen as a reinterpretation of the full vertex tensor structure, 
after rewriting its main components in a way that formally can be associated with the effective gluon propagator.
Although the Maris-Tandy model is quite successful for phenomenology, it is not able to describe the full set of hadronic properties 
and fails to explain the mass splittings of the $\rho$ and $a_1$ parity partners, underestimates weak decay constants of heavy-light mesons and cannot
reproduce simultaneously the mass spectrum and decay constants of radially excited vector mesons to point out some known limitations.
For a more complete description see, for example,~\cite{Bashir:2012fs,Mojica:2017tvh,Serna:2018dwk} and references therein. 
Several authors have tried to improve the Maris-Tandy model either by studying its dependence on the various parameters, see e.g.~\cite{Mojica:2017tvh}, or
by changing its functional dependence at low momenta, see e.g.~\cite{Qin:2011dd}, to achieve either a better description of Nature or good agreement with
the results from lattice simulations.

The goal of the present work is to explore further the quark-gluon vertex in the non-perturbative regime from first principles calculations 
combining continuum methods with results coming from lattice simulations.
Our approach follows the spirit of the calculation performed in~\cite{Rojas:2013tza} that solves the quark Dyson-Schwinger equation for the vertex.
In~\cite{Rojas:2013tza} the quark-gluon vertex was described as a generalised Ball-Chiu vertex and single unknown form factor,
function only of the gluon momentum, was considered. The current work goes beyond this approximation and includes the full set of longitudinal form factors
that appear in the Ball-Chiu vertex. 
In this work we disregard any contribution due to the transverse form factors and 
consider the Landau gauge quark-gluon vertex, to profit from the recent high quality lattice data for the quark, the gluon and the ghost propagators.
Moreover, our computation also incorporate the recent analysis of the full QCD lattice simulation for the quark-gluon vertex in the soft gluon limit
that identifies an important contribution, for the infrared vertex, associated with the gluon propagator~\cite{Oliveira2018a}.
As in other studies, we rely on a Slavnov-Taylor identity to write the vertex longitudinal form factors as a function of the quark wave function, 
the running quark mass, the quark-ghost kernel form factors and the ghost propagator. 
The normalisation of the quark-ghost kernel form factors $X_0$ (see below for definitions) derived in~\cite{Aguilar:2014lha} for the soft gluon limit is also
taken into account when solving the quark gap equation for the vertex.
The normalisation of $X_0$ played an important role in the analysis of the full QCD lattice data analysis for $\lambda_1$, 
the form factor associated with the tree level tensor structure $\gamma_\mu$, performed in~\cite{Oliveira2018a} that identified
an important contribution for $\lambda_1$ coming from the gluon propagator.

Our solution for the quark-gluon vertex returns a $X_0$ that deviates only slightly from the normalisation condition referred above.
However, the longitudinal form factors describing the quark-gluon vertex are strongly enhanced in the infrared region. 
The enhancement of the four longitudinal form factors occurs for quark and gluon momentum below 2 GeV and can be traced back to
ghost contribution introduced by the Slavnov-Taylor identity and the gluon dependence of the ansatz. 
At high momentum the form factors seem to approach their perturbative values. 
The matching with the perturbative tail is not perfect and this result can be understood partially due to the regularisation for the mathematical problem,
i.e. the Tikhonov regularisation, and partially to the parametrisation of the vertex.
Indeed, by calling the gluon propagator to describe the various form factors, the inversion of the Dyson-Schwinger equations is
quite sensitive to the low momentum scales, where the gluon propagator is much larger, and less sensitive to the ultraviolet regime. 
In order to overcome this problem, we considered a relatively large cutoff in the inversion and in this way add information on the perturbative tail.

The computed quark-gluon vertex is a function of the angle between the quark four momentum $p$ and the gluon four momentum $q$ that, clearly, 
favours kinematical configurations where $p$ and $q$ of the order of 1 GeV or below.
The enhancement occurs preferably at momenta of $\sim \, \Lambda_{QCD}$.
Furthermore, the vertex is enhanced when all momenta entering the vertex (see Fig. 1) tends to be parallel  in pairs,  
solving in this way the compromise  that the momenta are restricted to a region around $\Lambda_{QCD}$.
Within our solution for the quark-gluon vertex, the dominant form factors are associated with the tree level vertex $\gamma_\mu$ and the operator $2 \, p_\mu + q_\mu$.
The higher rank tensor structures give sub-leading contributions to the vertex.

In the current work, the vertex is written using the Ball-Chiu construction. It is known that the Ball-Chiu vertex has kinematical
singularities for the Landau gauge~\cite{Kizilersu:1995iz} that are associated to the transverse form factors (see definitions below).
These singularities can be avoided by considering a different tensor basis for the full vertex as described, for example, in~\cite{Kizilersu:1995iz}. 
However, the singularities are not associated to the longitudinal form factors and our calculation only takes into account this class of form factors.

The paper is organised as follows. In Sec.~\ref{Sec:notacao} we introduce the notation for the propagators, the Dyson-Schwinger equations and the quark-gluon 
vertex. Moreover, we use a Slavnov-Taylor identity to rewrite the vertex in terms of the quark propagators functions and the quark-ghost kernel. The parametrisation
of the quark-ghost kernel is also discussed. In Sec.~\ref{Sec:DSE_sca_vec} the scalar and vector components of the DSE in Minkowski space are given,
together with the corresponding kernels.
In Sec.~\ref{Sec:DSEeuclidean} the DSE are rewritten in Euclidean space, introduce the vertex  \textit{ansatz} and 
perform a scaling analysis of the integral equations. 
In Sec.~\ref{Sec:LattFunctions} we give the details of the lattice data used in the current work for the various propagators
and on the functions that parametrise the lattice data. The kernels for the Euclidean space DSE are discussed in Sec.~\ref{Sec:quarkghostkernel},
together with the solutions for the vertex of the gap equation. The quark-gluon vertex form factors are reported in Sec.~\ref{Sec:lambda1_lambda4} for several
kinematical configurations.
Finally, on Sec.~\ref{Sec:summary} we summarise and conclude.

\section{The Quark Gap Equation and the Quark-Gluon Vertex \label{Sec:notacao}}

\begin{figure}[t] 
   \centering
   \includegraphics[scale=1]{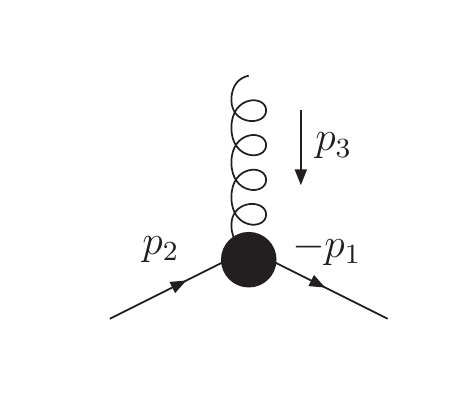} 
   \caption{The quark-gluon vertex.}
   \label{FIG:quark_gluon_vertex}
\end{figure}

In this section the notation used through out the article is defined. In this first part of this work, the equations discussed 
are written in Minkowski space with the diagonal metric $g = ( 1, \, -1, \, -1, \, -1)$.
Let us follow the notation of~\cite{Davydychev} for the quark-gluon vertex represented in  Fig.~\ref{FIG:quark_gluon_vertex}
that considers all momenta are incoming and, therefore, verify
\begin{equation}
p_1+p_2+p_3 = 0 \ .
\end{equation}
The one-particle irreducible Green's function associated to the vertex reads
\begin{equation}
  \Gamma^a_\mu (p_1, p_2, p_3) = g \, t^a \, \Gamma_\mu (p_1, p_2, p_3) \ ,
  \label{Eq:FullVertex}
\end{equation}
where $g$ is the strong coupling constant and $t^a$ are the color matrices in the fundamental representation. 

The quark propagator is diagonal in color and its spin-Lorentz structure is given by
\begin{eqnarray}
  S(p) & = &\frac{i}{A(p^2) \feynslash  p\ -  B(p^2) }
  = i \, \frac{A(p^2) \feynslash  p\ + B(p^2) }{ A^2(p^2) \,  p^2 - B^2(p^2)} \nonumber \\
  & = &
  i \, Z(p^2) \, \frac{ \feynslash  p\ + M(p^2) }{ p^2 - M^2(p^2)}
  \, ,
\end{eqnarray}
where $Z(p^2) = 1/A(p^2)$ stands for the quark wave function and $M(p^2) = B(p^2) / A(p^2)$ is the renormalisation group invariant running
quark mass. 

\begin{figure}[t] 
   \centering
   \includegraphics[scale=0.72]{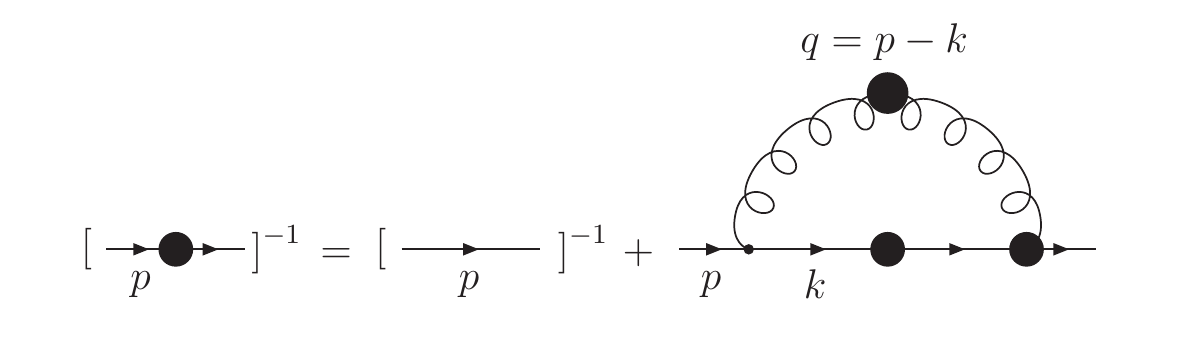} 
   \caption{The Dyson-Schwinger equation for the quark. The solid blobs denote dressed propagators and vertices.}
   \label{FIG:SDE}
\end{figure}

The Dyson-Schwinger equation for the quark propagator, also named the quark gap equation, is represented in Fig.~\ref{FIG:SDE}
and can be written as
\begin{equation}
S^{-1}(p)  =   -i \, Z_2 (\, \feynslash  p - m^{\mathrm{bm}}) + \Sigma (p^2) \ ,
\label{DSEquark}
\end{equation}
where $Z_2$ is the quark renormalisation constant, $m^{\mathrm{bm}}$ the bare current quark mass and the quark self-energy is given by
\begin{eqnarray}
 \Sigma (p^2) & = & Z_1 \, \int {d^4 q\over(2\pi)^4} ~ \times  \nonumber \\ 
 & & \times D_{\mu\nu}^{a b}(q)  ~ (\, i\, g\, t^b \gamma_\nu\,) ~ S(p-q) ~  \Gamma^a_\mu (-p, \, p-q, \, q) ,
 \label{DSEselfenergy}
\end{eqnarray}
where $Z_1$ is a combination of several renormalisation constants, $D_{\mu\nu}^{a b}(q)$ is the gluon propagator that, in the Landau gauge,
is given by
\begin{equation}
   D^{ab}_{\mu\nu} (q) =  -i\,  \delta^{ab} \left( g_{\mu\nu} - \frac{q_\mu q_\nu}{q^2} \right) D ( q^2 ) \ ;
\end{equation}   
below both $D^{ab}_{\mu\nu} (q)$ and 
$D (q^2)$ will be referred to as the gluon propagator.

A key ingredient in gap equation (\ref{DSEquark}) is the quark-gluon vertex. Indeed, it is only after knowing $\Gamma^a_\mu$ or, equivalently $\Gamma_\mu$, that 
$Z(p^2)$ and $M(p^2)$ can be computed.
The Lorentz structure of the quark-gluon vertex $\Gamma_\mu$ can be decomposed into 
longitudinal $\Gamma^{(L)}$ and transverse $\Gamma^{(T)}$ components relative to the gluon momenta, i.e. one writes
\begin{equation}
\Gamma_{\mu}(p_1, \, p_2, \, p_3)  = \Gamma^{(L)}_{\mu}(p_1, \, p_2, \, p_3) + \Gamma^{(T)}_{\mu}(p_1, \, p_2, \, p_3),
\label{Eq_vertex}
\end{equation}
where, by definition,  
\begin{equation}
p_{3}^{\mu} ~ \Gamma^{(T)}_{\mu}(p_1, \, p_2, \, p_3)= 0.
\end{equation}
By choosing a suitable tensor basis in the spinor-Lorentz space, $\Gamma_\mu$ can be written as a sum of scalar form factors that
multiply each of the elements of the basis. The full vertex $\Gamma_\mu$ requires twelve  form factors and
for the Ball and Chiu basis~\cite{Ball:1980ay} it reads
\begin{eqnarray}
   \Gamma^\mathrm{L}_\mu (p_1, p_2, p_3) & = & -i\, \sum^4_{i=1} \lambda_i (p_1, p_2, p_3) ~ L^{(i)}_\mu (p_1 , p_2) \, \\ 
   \label{EQ:Lvertex} 
   \Gamma^\mathrm{T}_\mu (p_1, p_2, p_3)  & = & -i\, \sum^8_{i=1} \tau_i (p_1, p_2, p_3) ~ T^{(i)}_\mu (p_1 , p_2) \ .
   \label{EQ:Tvertex} 
\end{eqnarray}
The operators associated to the  longitudinal vertex are
\begin{eqnarray}
 L^{(1)}_\mu (p_1 , p_2) & =  & \gamma_\mu \ , \nonumber \\
 L^{(2)}_\mu (p_1 , p_2) & =  & ( \feynslash p_1 - \feynslash p_2 ) \left( p_1 - p_2 \right)_\mu \ , \nonumber \\
 L^{(3)}_\mu (p_1 , p_2) & =  & ( p_1 - p_2  )_\mu \, \mathbb{I}_D \ ,  \nonumber \\
 L^{(4)}_\mu (p_1 , p_2) & = & \sigma_{\mu\nu} \left( p_1 - p_2 \right)^\nu \ , \label{Eq:LongOperators}
\end{eqnarray}
while those associated to the transverse part of the vertex read
\begin{eqnarray}
 T^{(1)}_\mu (p_1 , p_2)  & = &   \left [ p_{1\mu} \left(p_2 \cdot p_3 \right) -  p_{2\mu} \left(p_1 \cdot p_3 \right) \right ] \mathbb{I}_D  \ , \nonumber \\
 T^{(2)}_\mu (p_1 , p_2)  & =  &   - T^{(1)}_\mu (p_1 , p_2)~ \left( \feynslash p_1 - \feynslash p_2 \right) \ ,  \nonumber \\
 T^{(3)}_\mu (p_1 , p_2)  & = &  p^2_3 \,  \gamma_\mu - p_{3\mu} \  \feynslash p_3 \ , \nonumber
\end{eqnarray}
\begin{eqnarray}
 T^{(4)}_\mu (p_1 , p_2)  & = &     T^{(1)}_\mu  (p_1 , p_2) ~ \sigma_{\alpha\beta}\, p^\alpha_1 p^\beta_2 \ , \nonumber \\
 T^{(5)}_\mu (p_1 , p_2)  & = &   \sigma_{\mu\nu} \, p^\nu_3 \ ,  \nonumber \\
 T^{(6)}_\mu (p_1 , p_2)  & = &   \gamma_\mu \left( p^2_1 - p^2_2 \right) + \left( p_1 - p_2 \right)_\mu \,  \feynslash p_3 \ , \nonumber \\
 T^{(7)}_\mu (p_1 , p_2) & =  & -{1\over 2} \left( p^2_1 - p^2_2 \right) 
                      \left[ \gamma_\mu \left( \feynslash p_1 - \feynslash p_2 \right) - \left( p_1 - p_2 \right)_\mu \mathbb{I}_D \right]  \nonumber \\
                      & &
                      \hspace{1cm} -
                      \left( p_1 - p_2 \right)_\mu \sigma_{\alpha\beta} \, p^\alpha_1 \, p^\beta_2 \ , \hspace*{1cm} \nonumber \\
 T^{(8)}_\mu (p_1 , p_2) & =  &  - \gamma_\mu \, \sigma_{\alpha\beta} \, p^\alpha_1 \, p^\beta_2 + ( p_{1\mu} \, \feynslash  p_2 -  p_{2\mu} \, \feynslash p_1) \ , 
\end{eqnarray}
%
%
where $\sigma_{\mu\nu} = {1\over 2} [\gamma_{\mu},\gamma_{\nu}]$.

\subsection{QCD Symmetries and the Quark-Gluon Vertex}

The global and local symmetries of QCD constrain the full vertex $\Gamma_\mu$ and connect several of the Green's functions theory. 
For example, the global symmetries of QCD require that the form factors $\lambda_i$ and $\tau_i$ to be either symmetric or anti-symmetric 
under exchange of the two first momenta; see, e.g., ref.~\cite{Davydychev} and references therein. 
On the  other hand,  gauge symmetry implies that the Green functions also satisfy the Slavnov-Taylor identities (STI) ~\cite{Slavnov:1972fg,Taylor:1971ff}.
These identities play a major role in our understanding of QCD and, in particular, the longitudinal part of the quark-gluon vertex is constrained by the 
following identity
\begin{eqnarray}
   p^\mu_3 ~ \Gamma_\mu ( p_1, p_2, p_3 ) & = & F(p^2_3) ~ \Big [ S^{-1} ( -p_1) \, H( p_1, p_2, p_3 ) ~ \nonumber \\
     & & \hspace{1.2cm}
    -  ~  \overline H (p_2, p_1, p_3) \, S^{-1}(p_2) \Big ]  , \nonumber \\
    & & 
\label{STI}
\end{eqnarray}
where the ghost-dressing function $F(q^2)$ is related to the ghost two-point correlation function as
\begin{equation}
  D^{a b} (q^2) = -\, \delta^{ab} \, D_{gh} (q^2) = -\, \delta^{ab} \, \frac{F(q^2)}{q^2}  
\end{equation} 
and $H$ and $\overline H$ are associated to the quark-ghost kernel.
As discussed in~\cite{Davydychev}, these functions can be parametrised in terms of four form factors as
\begin{eqnarray}
  H( p_1, p_2, p_3 ) & = &  X_0\, \mathbb{I}_D +  X_1 \, \feynslash p_1 +  X_2 \,\feynslash p_2 + X_3 \, \sigma_{\alpha\beta} p^\alpha_1 p^\beta_2\, ,  \nonumber \\
 \overline H( p_2, p_1, p_3 ) & = & \overline X_0\, \mathbb{I}_D -  \overline X_2\, \feynslash p_1 - \overline X_1\, \feynslash p_2 +
     \overline X_3 \, \sigma_{\alpha\beta} p^\alpha_1 p^\beta_2\,  , \nonumber \\
     & & 
     \label{Eq:quarkghostkernel}
\end{eqnarray}
where $X_i \equiv X_i ( p_1, p_2, p_3 )$ and $ \overline X_i \equiv X_i( p_2, p_1, p_3 )$. 

The STI given in Eq.~(\ref{STI}) can be solved with respect to the vertex~\cite{Aguilar:2010cn} to write the longitudinal form factors $\lambda_i$  
in terms of the quark propagator functions $A(p^2)$, $B(p^2)$ and the quark-ghost kernel functions $X_i$ and $\overline X_i$ as
\begin{eqnarray}
& & \lambda_1 (p_1, p_2, p_3)  =  \frac{F(p^2_3)}{2} \Bigg\{  \nonumber \\
  & & \hspace{1cm} A(p^2_1) \left[ X_0 + \left( p^2_1 - p_1 \cdot p_2 \right) X_3 \right] \nonumber \\
               & & 
               \hspace{2cm}
              +  ~  A(p^2_2) \left[ \overline X_0 + \left( p^2_2 - p_1 \cdot p_2 \right) \overline X_3 \right]  \nonumber \\       
              & & \hspace{3cm}
              + ~ B(p^2_1) \left[X_1 + X_2 \right]     \nonumber \\
              & & \hspace{4cm} +  ~B(p^2_2) \left[ \overline X_1 + \overline X_2 \right]  \quad\Bigg\} \ ,
              \label{EQ:lambda_1} \\
& & \lambda_2 (p_1, p_2, p_3)  = 
               \frac{F(p^2_3)}{2\! \left( p^2_2 - p^2_1 \right) } \Bigg\{ \nonumber \\
               & & \hspace{1cm} A(p^2_1)\! \left[ \left( p^2_1 + p_1\! \cdot p_2 \right) X_3  - X_0 \right] 
               \nonumber \\
               & & \hspace{2cm}
               +  ~  A(p^2_2) \left[ \overline X_0 - \left( p^2_2 + p_1\! \cdot p_2 \right) \overline X_3 \right]  \nonumber \\  
               & & \hspace{3cm}
               +  ~ B(p^2_1) \left[X_2 - X_1 \right]  \nonumber \\
               & & \hspace{4cm} 
               +  ~ B(p^2_2) \left[ \overline X_1 - \overline X_2 \right]  \quad \Bigg\} \ ,
                \label{EQ:lambda_2} \\
& & \lambda_3 (p_1, p_2, p_3)  = 
                   \frac{F(p^2_3)}{p^2_1 - p^2_2} \Bigg\{ \nonumber \\
                   & & \hspace{1cm} A(p^2_1) \left[ p^2_1 \, X_1 + p_1 \cdot p_2 \ X_2\right] \nonumber \\
                   & & \hspace{2cm}
                   - ~   A(p^2_2) \left[ p^2_2\ \overline X_1 + p_1 \cdot p_2 \ \overline X_2 \right]   \nonumber \\ 
                     & &  \hspace{3cm}
                      +   ~ B(p^2_1) \, X_0 \nonumber \\
                      & & \hspace{4cm}
                      - ~  B(p^2_2) \, \overline X_0  \quad\Bigg\}  \ ,
                \label{EQ:lambda_3}  \\
& & \lambda_4 (p_1, p_2, p_3)  =  - \frac{F(p^2_3)}{2} \Bigg\{ A(p^2_1) \, X_2 -  A(p^2_2) \, \overline X_2 \nonumber \\
   & & \hspace{3.5cm}
                       +  ~ B(p^2_1) \, X_3  - B(p^2_2) \, \overline X_3 \quad\Bigg\}    \ .                                  
\label{EQ:lambda_4} 
\end{eqnarray}
A nice feature of the above solution for the various form factors $\lambda_i$, that can be checked by direct inspection, is that the symmetry requirements on the
$\lambda_i$ due to charge conjugation are automatically satisfied independently of the functions $A$, $B$, $X_i$ and $\overline X_i$. 
This is a particularly important point when modelling the vertex.

\section{Decomposing the Dyson-Schwinger Equation into its Scalar and Vector Components \label{Sec:DSE_sca_vec}}

The Dyson-Schwinger equation for the quark propagator is written in (\ref{DSEquark}), with the quark self-energy being given by (\ref{DSEselfenergy}).
This equation can be projected into its scalar and vector components by taking appropriate traces. 

The scalar part of the equation is given by the trace of (\ref{DSEquark}) which, after some algebra, reduces to
\begin{eqnarray}
 & &  i B(p^2)  ~ = ~ i Z_2\, m^\mathrm{bm} \nonumber \\
  & & + ~ C_F Z_1 g^2 \, 
         \int \frac{d^4 q}{(2 \, \pi)^4} ~ \frac{ \Delta(q^2) }{\left[ A(k^2) \right]^2  \, k^2 - \left[ B(k^2) \right]^2}  \Bigg\{  
         \nonumber \\
  & & 
          2 \, h(p,q) \, \bigg( 2 \, \left[ B(k^2) \right] ~ \lambda_2 -  \, \left[ A(k^2) \right] ~ \left(  \lambda_3 + \lambda_4 \right) \bigg) \nonumber \\
 & &   \hspace{0.5cm} +  3 \bigg(  \left[ B(k^2) \right]  \lambda_1 + \left[ A(k^2) \right]  \left( 2 p^2 + q^2 - 3 (p \cdot  q) \right) \lambda_4\bigg) 
  \Bigg\} , \nonumber \\
 \label{Eq:DSEMinkScalar}
\end{eqnarray}
after insertion of the  vertex decomposition (\ref{Eq_vertex}), taking into account only its longitudinal part,  where $k = p -q$,
\begin{equation}
   h(p,q) = \frac{ p^2  \, q^2 \, -  \, (p\cdot q)^2 }{ q^2 } \ ,
\end{equation}
 $\lambda_i \equiv \lambda_i (-p , \, p-q, \, q) $ and $C_F = 4/3$ is the Casimir invariant associated to the SU(3) fundamental representation.

The vector component of (\ref{DSEquark}) is obtained after multiplication by $\feynslash  p\ $ and then taking the trace of the resulting equation to
arrive on
\begin{eqnarray}
 & &  -i p^2 A(p^2)   =   -i Z_2\, p^2 \nonumber \\
  & & 
  + \, C_F Z_1 g^2  \int \frac{d^4 q}{(2 \, \pi)^4} ~ \frac{ \Delta(q^2) }{\left[ A(k^2) \right]^2  \, k^2 - \left[ B(k^2) \right]^2} \times  
  \nonumber \\
  & &
  \times \Bigg\{ 2 \, h(p,q)  \bigg( \left[ A(k^2) \right]  ~ \left[ \lambda_1 +  \lambda_2 \, \left(2p^2 + q^2 - 2 p\cdot q  \right) \right]  \nonumber \\
  & & \hspace{5cm}
                                                       + \left[ B(k^2) \right] \left(\lambda_4 - \lambda_3\right) \bigg) \nonumber \\
  & & \hspace{0.75cm}
 + \, 3 \, \bigg(  \left[ A(k^2) \right] ~ \lambda_1 ~ \left( ( p\cdot q ) - p^2  \right)  \nonumber \\
 & & \hspace{3.5cm}
 + \left[ B(k^2) \right] ~ \lambda_4 ~ ( (p \cdot q) - 2 p^2  ) \bigg) \Bigg\}
 \ .  \nonumber \\
 & & \label{Eq:DSEMinkVector} 
\end{eqnarray}

The two equations (\ref{Eq:DSEMinkScalar}) and (\ref{Eq:DSEMinkVector}) can be simplified further by modelling the quark-gluon vertex.
For example, in~\cite{Aguilar:2010cn,Rojas:2013tza} the vertex was parametrised using the solution of the Slavnov-Taylor identity 
(\ref{EQ:lambda_1})--(\ref{EQ:lambda_4}) and setting $X_1 = X_2 = X_3 = 0$. The rationale for such a choice comes from perturbation theory
which gives, at tree level, $X_0 = 1$ and $X_1 = X_2 = X_3 = 0$. This ansatz, that ignores all form factor associated to the quark-ghost kernel but $X_0$, 
assumes that at the non-perturbative level the hierarchy of the form factors follows its relative importance  observed in the high momentum regime. 
Furthermore, in order to compute a solution of the Dyson-Schwinger equations it was introduced a further restriction on $X_0$, that it depends 
only on the incoming gluon momenta, i.e. that $X_0 = X_0 (q^2)$. 

In order to solve the Dyson-Schwinger equations for the vertex, it will be assumed that the form factors associated to the quark-ghost kernel, see Eq. (\ref{Eq:quarkghostkernel}),
factorize as
\begin{equation}
   X_i (p^2_1, p^2_2, p^2_3) =  g_i(p^2_1, p^2_2) \, Y_i (p^2_3) \ ,
   \label{Eq:fact}
\end{equation}   
where $g_i(p^2_1, p^2_2) = g_i(p^2_2, p^2_1)$ are symmetric functions of its arguments. This type of factorisation is compatible, for example, with the 
Maris-Tandy quark-gluon description of the quark-gluon vertex~\cite{Maris99a} and simplifies considerably the analysis of the solutions of the equations to be solved.
In~\cite{Aguilar:2014lha} it was proved that, to all-orders,
 \begin{equation}
   X_0 (p,  \, -p, \, 0 ) = 1 ~\mbox{ and }~ X_1 (p,  \, -p, \, 0 ) = X_2(p,  \, -p, \, 0 )  \label{Eqxyasy}
\end{equation}
that, for the ansatz (\ref{Eq:fact}) implies that
\begin{eqnarray}
    g(p^2,p^2) \, Y_0(0)  & & =  1 \ , \nonumber \\
   g_1(p^2, p^2) \, Y_1(0)  & & =  g_2(p^2, p^2) ~ Y_2(0) \ .
   \label{Eq:all_orders}
\end{eqnarray}
A solution that complies with the second relation given in Eq. (\ref{Eqxyasy}) is to assume that $X_1 = X_2$ for any kinematical configuration. 
Note also that by choosing the $g_i$ to be symmetric functions of the arguments, the form factors $X_i$ and $\overline X_i$ become identical.
If one takes into account these relations into the ansatz for the quark-gluon vertex, then the solutions of the Slavnov-Taylor identities 
(\ref{EQ:lambda_1})--(\ref{EQ:lambda_4}) become
\begin{eqnarray}
& & \lambda_1 (-p, \, p-q, \, q )  =  \frac{F(q^2)}{2} \Bigg\{ \nonumber \\
& &  \bigg[ A(p^2)  + A( k^2 ) \bigg] \, g_0(p^2, k^2) \, Y_0(q^2) \nonumber \\
   & & \hspace{0.75cm} + \, 2 \, \bigg[ B(p^2)  + B( k^2 ) \bigg] \, g_1(p^2, k^2) \, Y_1(q^2)  \nonumber \\
   & & \hspace{1.5cm} + ~ \bigg[ A(p^2) \bigg( 2p^2 - (pq) \bigg) \nonumber \\
                          & & \hspace{3cm} + A(k^2) \bigg( 2 p^2 + q^2 - 3 (pq) \bigg)   \bigg]  
                          \times \nonumber \\
              & & \hspace{5cm} \times
 \, g_3(p^2, k^2) \, Y_3(q^2)  \Bigg\} \, , \nonumber \\ \label{Vertex:L1}
\end{eqnarray}
\begin{eqnarray}              
& & \lambda_2 (-p, \, p-q, \, q ) =  \frac{F(q^2)}{2( q^2 - 2 (p \cdot q))} \times \nonumber \\
& &  \times \Bigg\{  \Delta A ~ ~   g_0(p^2, k^2) \, Y_0(q^2)  \nonumber \\
 & & \hspace{0.75cm}
                                         + \bigg[ A(p^2) \, (pq)  - A(k^2) \bigg( q^2 -  (pq) \bigg) \bigg] \times \nonumber \\
 & & \hspace{3.5cm} \times \,  g_3 (p^2, k^2) \, Y_3(q^2) \Bigg\}  \, , \label{Vertex:L2}
\end{eqnarray}
\begin{eqnarray}                                            
& & \lambda_3 (-p, \, p-q, \, q ) =  \frac{F(q^2)}{q^2 - 2 (p \cdot q)}  \times \nonumber \\
& & \times \Bigg\{ \Delta B ~ ~  g_0 (p^2, k^2) \,  Y_0(q^2) \nonumber \\
   & & 
              \hspace{0.75cm} + \, \bigg[ A( k^2 ) \, \bigg( q^2 -   (p q) \bigg) - A(p^2) \, ( p q )   \bigg] \times \nonumber \\
              & & \hspace{4cm} \times \,  g_1 (p^2, k^2) \, Y_1(q^2) \Bigg\}  \, ,  \label{Vertex:L3}
\end{eqnarray}
\begin{eqnarray}
& & \lambda_4 (-p, \, p-q, \, q )  =  \frac{F(q^2)}{2} \times \nonumber \\
& & \hspace{0.4cm} \times \Bigg\{  \Delta A ~ ~ g_1 (p^2, k^2) \, Y_1(q^2)  + \Delta B \,\, g_3 (p^2, k^2) \, Y_3(q^2) \Bigg\}  \, , 
 \nonumber \\
 \label{Vertex:L4} 
\end{eqnarray}
where
\begin{equation}
\Delta A = A(k^2) - A(p^2) \, , \hspace{1cm}  \Delta B = B(k^2) - B(p^2) \, ,
\end{equation}
and $k = p - q$.
The scalar component of the Dyson-Schwinger equations is now
\begin{eqnarray}
  i B(p^2)  &  = & i ~ Z_2\, m^\mathrm{bm} \nonumber \\
  & & + ~ C_F Z_1 g^2 \, 
         \int \frac{d^4 q}{(2 \, \pi)^4} ~ \frac{ \Delta(q^2) ~ F(q^2)}{\left[ A(k^2) \right]^2  \, k^2 - \left[ B(k^2) \right]^2} \times 
         \nonumber \\
  & & 
         \times \Bigg\{ g_0(p^2, k^2) \, Y_0 (q^2) \, \mathcal{K}^{(0)}_B (p,q) \nonumber \\
   & & \hspace{1cm} + ~ g_1(p^2, k^2) \, Y_1(q^2) \, \mathcal{K}^{(1)}_B (p,q) \nonumber \\
   & & \hspace{1.5cm}
                               + ~  g_3(p^2, k^2) \, Y_3(q^2) \, \mathcal{K}^{(3)}_B (p,q) \, \Bigg\} 
 \label{Eq:DSEMinkScalarX}
\end{eqnarray}
where the kernels are defined as
\begin{eqnarray}
  \mathcal{K}^{(0)}_B (p,q)   & = &
   \frac{2 \, h(p,q)}{q^2 - 2(pq)} \, \bigg( B(k^2) \, \Delta A \, -  \, A(k^2)  \,\Delta B \bigg)  \nonumber \\
   & &  \hspace{1cm} +  \frac{3}{2}  \, B(k^2)  \, \bigg[ A(k^2) + A(p^2) \bigg]   , 
\end{eqnarray}   
%
\begin{eqnarray}   
 & & \mathcal{K}^{(1)}_B (p,q)    =  \nonumber \\
 & & = ~ h(p,q) \,  A(k^2) ~ 
  \Bigg[  2 \frac{ A(p^2) (p q) - A(k^2) \big(q^2 -   (p q) \big) }{q^2 - 2 (p q) } - \Delta A \Bigg]
  \nonumber \\
  & &  \hspace{1.2cm} + \, 3 \, B(k^2)  \bigg(  B(p^2) + B(k^2)  \bigg) \nonumber \\
  & & \hspace{2.4cm}  + \, \frac{3}{2} \, A(k^2) \, \Delta A \, \bigg( 2 p^2 + q^2 - 3(p q) \bigg)   , 
\end{eqnarray}  
%
\begin{eqnarray}
 & & \mathcal{K}^{(3)}_B (p,q)    =  \nonumber \\
 & &
 h(p,q) \,  \Bigg\{ 2 \, B(k^2)  \, \frac{ A(p^2) (p q) - A(k^2) \big(q^2 -   (p q) \big) }{q^2 - 2 (p q) } \nonumber \\
 & & \hspace{6cm} - A(k^2) \, \Delta B \Bigg\}
  \nonumber \\
  & &  \hspace{1cm} + \, \frac{3}{2} \, B(k^2)  \Bigg[ A(p^2) \bigg( 2p^2 - (pq) \bigg) \nonumber \\
  & & \hspace{3.5cm} + ~ A(k^2) \bigg( 2p^2 + q^2 - 3 (pq) \bigg) \Bigg] \nonumber \\
  & &  \hspace{1cm} + ~ \frac{3}{2} \, A(k^2) \, \Delta B \, \bigg( 2 p^2 + q^2 - 3(p q) \bigg)  \, . 
\end{eqnarray}
Similarly, the vector component of the Dyson-Schwinger equations reduces to
\begin{eqnarray}
  & &  -i p^2 A(p^2)   =   -i ~ Z_2\, p^2 \nonumber \\
  & & + \, C_F Z_1 g^2  \int \frac{d^4 q}{(2 \, \pi)^4} ~ \frac{ \Delta(q^2) ~F(q^2) }{\left[ A(k^2) \right]^2  \, k^2 - \left[ B(k^2) \right]^2} \times  
  \nonumber \\
  & & 
  \times \Bigg\{ g_0(p^2, k^2) \, Y_0(q^2) \, \mathcal{K}^{(0)}_A (p,q) \nonumber \\
  & &
  \hspace{1.5cm} + \, g_1(p^2, k^2) \, Y_1 (q^2) \, \mathcal{K}^{(1)}_A (p,q)  \nonumber \\
  & & \hspace{2.6cm}
       +  \, g_3(p^2, k^2) \, Y_3 (q^2) \, \mathcal{K}^{(3)}_A (p,q) \Bigg\}
         \label{Eq:DSEMinkVectorX}
\end{eqnarray}
with the kernels given by
\begin{eqnarray}
 & &  \mathcal{K}^{(0)}_A (p,q)  =  \nonumber \\
 & & 
   =  h(p,q)  \Bigg\{ A(k^2) \,  \bigg[ A(p^2) + A(k^2)  + \frac{2p^2 + q^2 - 2(p q)}{q^2 - 2(p q)} \, \Delta A \bigg] \nonumber \\
    & & \hspace{4cm} 
                                                - 2 B(k^2)  \, \frac{\Delta B}{q^2 - 2 (p  q)}  \Bigg\} \nonumber \\
    &   &                                        \hspace{0.75cm} - \frac{3}{2}  \,  A(k^2)   \,\bigg( A(p^2) + A(k^2) \bigg) \bigg(p^2 - ( p q ) \bigg)  \ ,
\end{eqnarray}
%
\begin{eqnarray} 
 & &  \mathcal{K}^{(1)}_A (p,q)  =  \nonumber \\
 & & =  h(p,q) \Bigg\{ 2 \,  A(k^2) \, \bigg( B(p^2) + B(k^2)  \bigg)  \nonumber \\
 & & 
  \hspace{1cm} + ~ B(k^2) \bigg[ \Delta A + 2 \frac{A(p^2) (p  q)  - A(k^2) (q^2 -  (p q)) }{q^2 - 2 (p q)}
                                                                                       \bigg] \Bigg\}
                                                                                       \nonumber \\
 & &   \hspace{0.4cm}  - \, 3 \, A(k^2) \, \bigg( B(p^2) + B(k^2)  \bigg)   \, \bigg(p^2 - (p  q) \bigg)    \nonumber \\
 & & \hspace{1cm}
               \, + \, \frac{3}{2}  \, B(k^2)  \, \Delta A \, \bigg( (p q) - 2 \, p^2 \bigg) \, , 
\end{eqnarray}
%
\begin{eqnarray} 
  & & \mathcal{K}^{(3)}_A (p,q)  =  \nonumber \\
  & &  h(p,q) \Bigg\{  A(k^2)  \Bigg[ A(p^2) \bigg(2p^2 - (pq) \bigg)  \nonumber \\
  & & \hspace{2.5cm} + A(k^2) \bigg( 2p^2 + q^2 - 3 (pq) \bigg) \nonumber \\
                  & & \hspace{2.5cm}   +\, \frac{2p^2 + q^2 - 2(pq)}{q^2 - 2(pq)} \times \nonumber \\
                  & & \hspace{3cm} \times  \bigg( A(p^2) (pq) - A(k^2) (q^2 - (pq)) \bigg) \Bigg] \nonumber \\
                  & & \hspace{2.8cm} + \, B(k^2)  ~ ~ \Delta B \Bigg\} \nonumber \\
   & & + \, \frac{3}{2} \bigg( (pq) - p^2 \bigg) \, A(k^2) \times  \nonumber \\
   & & \hspace{0.4cm} \times \Bigg( A(p^2) \bigg( 2p^2 - (pq) \bigg) + A(k^2) \bigg( 2 p^2 + q^2 - 3(pq) \bigg) \Bigg)  \nonumber \\
   & & + \, \frac{3}{2} \bigg( (pq) - 2 p^2 \bigg) \, B(k^2) \, \Delta B \, . 
\end{eqnarray}

\section{The Dyson-Schwinger Equations in Euclidean Space \label{Sec:DSEeuclidean}}

As already stated, our goal is to solve the Dyson-Schwinger equations for the quark-ghost kernel, said otherwise for the quark-gluon vertex, and 
this requires the knowledge of the quark, gluon and ghost propagators. For the propagators we will rely on lattice inputs that provide first principles 
non-perturbative results and also demand that the above expressions should be rewritten in Euclidean space. 
The Wick rotation to go from Minkowski to Euclidean space is achieved by making use of the following substitutions 
\begin{equation}
   \begin{tabular}{ll@{\hspace{0.8cm}}ll} 
     $p^2$ & $\rightarrow -p_E^2$               & $(p \cdot q)$ & $ \rightarrow - (p_E \cdot q_E)$ \\
     $A(p^2)$ & $\rightarrow A_E (-p_E^2)$   &  $B(p^2)$ & $ \rightarrow B_E (-p_E^2) $ \\
     $\int_{q}$ & $ \rightarrow i \int_{q_E} $     & $\Delta(p^2)$ & $ \rightarrow - \Delta_E(-p_E^2)$
    \end{tabular}
\end{equation}
on Eqs. (\ref{Eq:DSEMinkScalarX})  and (\ref{Eq:DSEMinkVectorX}). For completeness, we provide now all expressions in Euclidean space.

The scalar component of the Dyson-Schwinger equations reads
\begin{eqnarray}
  & & B(p^2)   =   Z_2\, m^\mathrm{bm} \nonumber \\
  & & + \, C_F Z_1 g^2 \, 
         \int \frac{d^4 q}{(2 \, \pi)^4} ~ \frac{ \Delta(q^2) ~ F(q^2)}{\left[ A(k^2) \right]^2  \, k^2 + \left[ B(k^2) \right]^2} \times 
         \nonumber \\
  & & \hspace{0.25cm}
 \times \Bigg\{ g_0(p^2, k^2) \, Y_0 (q^2) \, \mathcal{K}^{(0)}_B (p,q) \nonumber \\
  & & \hspace{1.25cm} + \, g_1(p^2, k^2) \, Y_1(q^2) \, \mathcal{K}^{(1)}_B (p,q) \nonumber \\
         & & \hspace{2.25cm}
                               + \, g_3(p^2, k^2) \, Y_3(q^2) \, \mathcal{K}^{(3)}_B (p,q) \, \Bigg\}    \label{Eq:DSEMinkScalarXEuclidean}
\end{eqnarray}
while its vector component is given by
\begin{eqnarray}
  & & p^2 A(p^2)   =   Z_2\, p^2 \nonumber \\
  & & + \, C_F Z_1 g^2  \int \frac{d^4 q}{(2 \, \pi)^4} ~ \frac{ \Delta(q^2) ~F(q^2) }{\left[ A(k^2) \right]^2  \, k^2 + \left[ B(k^2) \right]^2} \times  
  \nonumber \\
  & & \hspace{0.25cm}
 \times \Bigg\{ g_0(p^2, k^2) \, Y_0 (q^2) \, \mathcal{K}^{(0)}_A (p,q) \nonumber \\
 & & \hspace{1.25cm} + \, g_1(p^2, k^2) \, Y_1(q^2) \, \mathcal{K}^{(1)}_A (p,q) \nonumber \\
         & & \hspace{2.25cm}
                               + \, g_3(p^2, k^2) \, Y_3(q^2) \, \mathcal{K}^{(3)}_A (p,q) \, \Bigg\} 
 \ .  \label{Eq:DSEMinkVectorXEuclidean}
\end{eqnarray}
The kernels appearing in Eqs. (\ref{Eq:DSEMinkScalarXEuclidean}) and (\ref{Eq:DSEMinkVectorXEuclidean}) are
\begin{eqnarray}
  \mathcal{K}^{(0)}_B (p,q)   & = &
   \frac{2 \, h(p,q)}{q^2 - 2(pq)} \, \bigg[ B(k^2) \, \Delta A \, -  \, A(k^2)  \,\Delta B \bigg]  \nonumber \\
   & & \hspace{0.5cm} + \, \frac{3}{2}  \, B(k^2)  \, \bigg[ A(k^2) + A(p^2) \bigg]  \, , 
   \label{K0:B}
\end{eqnarray}  
%
\begin{eqnarray}
& &  \mathcal{K}^{(1)}_B (p,q)    =  \nonumber \\
& & = 
 h(p,q) \,  A(k^2)  ~ \Bigg[  \Delta A - 2 \frac{ A(p^2) (p q) - A(k^2) \big(q^2 -   (p q) \big) }{q^2 - 2 (p q) }  \Bigg]
  \nonumber \\
  & & \hspace{1cm} + \, 3 \, B(k^2)  \bigg[  B(p^2) + B(k^2)  \bigg] \nonumber \\
  & &  \hspace{2cm} + \, \frac{3}{2} \, A(k^2) \, \Delta A \, \bigg( 3(p q) - 2 p^2 - q^2  \bigg)  \, , 
   \label{K1:B}
\end{eqnarray} 
%
\begin{eqnarray}
 & & \mathcal{K}^{(3)}_B (p,q)    = \nonumber \\
 & & = 
 h(p,q) \,  \Bigg\{ A(k^2) ~ ~  \Delta B \nonumber \\
 & & \hspace{1.9cm} - 2 \, B(k^2)  \, \frac{ A(p^2) (p q) - A(k^2) \big(q^2 -   (p q) \big) }{q^2 - 2 (p q) }   \Bigg\}
  \nonumber \\
  & &  \hspace{0.5cm} + \, \frac{3}{2} \, B(k^2)  \Bigg[ A(p^2) \bigg(  (pq) - 2p^2  \bigg)  \nonumber \\
  & & \hspace{3.5cm} + A(k^2) \bigg(  3 (pq) - 2p^2 - q^2  \bigg) \Bigg] \nonumber \\
  & &  \hspace{1.2cm} + \, \frac{3}{2} \, A(k^2) \, \Delta B \, \bigg( 3(p q)  - 2 p^2 - q^2 \bigg)  \, . 
     \label{K3:B}
\end{eqnarray}
\begin{eqnarray}
 & &  \mathcal{K}^{(0)}_A (p,q)  = \nonumber \\
 & & 
    h(p,q)  \Bigg\{ -  A(k^2) \,  \bigg[ A(p^2) + A(k^2)  + \frac{2p^2 + q^2 - 2(p q)}{q^2 - 2(p q)} \, \Delta A \bigg] \nonumber\\
    & & \hspace{4cm}
                                                - 2 B(k^2)  \, \frac{\Delta B}{q^2 - 2 (p  q)}  \Bigg\} \nonumber \\
    &   &               \hspace{0.75cm}                      + \,  \frac{3}{2}  \,  A(k^2)   \,\bigg[ A(p^2) + A(k^2) \bigg] \bigg(p^2 - ( p q ) \bigg)  \ ,
       \label{K0:A}
\end{eqnarray}
%
\begin{eqnarray}
 & &  \mathcal{K}^{(1)}_A (p,q)  = \nonumber \\
 & &  h(p,q) \Bigg\{ - \, 2 \,  A(k^2) \, \bigg( B(p^2) + B(k^2)  \bigg) \nonumber \\
  & & \hspace{1cm}
  - \, B(k^2) \bigg[ \Delta A + 2 \frac{A(p^2) (p  q)  - A(k^2) (q^2 -  (p q)) }{q^2 - 2 (p q)}
                                                                                       \bigg] \Bigg\}
                                                                                       \nonumber \\
 & &  \hspace{0.75cm}    + \, 3 \, A(k^2) \, \bigg[ B(p^2) + B(k^2)  \bigg]   \, \bigg(p^2 - (p  q) \bigg)   \nonumber \\
 & &  \hspace{1.5cm}
               \, + \, \frac{3}{2}  \, B(k^2)  \, \Delta A \, \bigg( 2 \, p^2 - (p q)  \bigg) \, , 
   \label{K1:A}               
\end{eqnarray}
%
\begin{eqnarray} 
 & &  \mathcal{K}^{(3)}_A (p,q)  =  \nonumber \\
 & & =  h(p,q) \Bigg\{  A(k^2)  \Bigg[ A(p^2) \bigg(2p^2 - (pq) \bigg) \nonumber \\
 & &          \hspace{2.75cm} + ~ A(k^2) \bigg( 2p^2 + q^2 - 3 (pq) \bigg) \nonumber \\
                  & & \hspace{2.75cm}   + ~ \frac{2p^2 + q^2 - 2(pq)}{q^2 - 2(pq)} \times \nonumber \\
                  & & \hspace{3cm} \bigg( A(p^2) (pq) - A(k^2) (q^2 - (pq)) \bigg) \Bigg] \nonumber \\
                  & & \hspace{6cm} - \, B(k^2) \, \Delta B \Bigg\} \nonumber \\
   & & \hspace{0.5cm} + \, \frac{3}{2} \bigg( (pq) - p^2 \bigg) \, A(k^2)  \times \\
   & & \hspace{1.8cm} \times \Bigg[ A(p^2) \bigg( 2p^2 - (pq) \bigg)  \nonumber \\
   & & \hspace{3.5cm} + ~ A(k^2) \bigg( 2 p^2 + q^2 - 3(pq) \bigg) \Bigg]  \nonumber \\
   & & \hspace{1.2cm} + \, \frac{3}{2} \bigg(  2 p^2 - (pq)  \bigg) \, B(k^2) \, \Delta B \, . 
   \label{K3:A}
\end{eqnarray}

The computation of any solution of the above equations, using lattice inputs for the propagators, requires the use of the renormalised Dyson-Schwinger equations
and, therefore, all quantities appearing on these equations should be finite. This requirement constrains the integrand functions $g_i (p^2,$ $(p-q)^2) \, Y_i (q^2)$
and, in particular,  its possible behaviour in the limits where $q \rightarrow 0$ and $p \rightarrow + \infty$.

\begin{figure}[t] 
   \centering
   \includegraphics[width=3.5in]{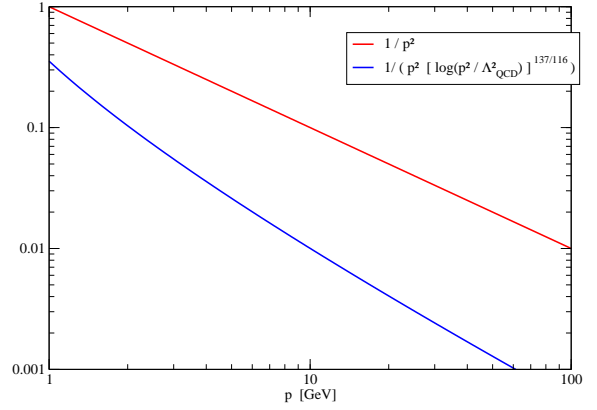} 
   \caption{Large $p$ behaviour and logarithmic corrections. The plotted function uses $\Lambda_{QCD} = 0.3$ GeV.}
   \label{fig:large_q_log}
\end{figure}

Let us start by investigating the ultraviolet limit of the integrand functions appearing 
in Eqs. (\ref{Eq:DSEMinkScalarXEuclidean}) and (\ref{Eq:DSEMinkVectorXEuclidean}). In the large $q$ limit it follows that
\begin{equation}
\frac{ q^3 \, D (q^2) \, F(q^2) }{A^2(k^2) \, k^2 + B^2(k^2)} \longrightarrow \frac{1}{q} \ ,
\end{equation}
up to logarithmic corrections associated to the various propagators. In this limit, the integrand function appearing on the scalar equation
(\ref{Eq:DSEMinkScalarXEuclidean}) read
\begin{eqnarray}
  & &  g_0 \left(p^2, (p-q)^2 \right) \, Y_0 (q^2) ~  \frac{1}{q}  ~ \Bigg\{ \frac{3}{2} B(q^2) \bigg[ A(q^2) + A(p^2) \bigg] \Bigg\} \nonumber  \\
  & &   + ~ g_1 \left(p^2, (p-q)^2 \right) \, Y_1 (q^2) ~ q ~ \Bigg\{ \frac{3}{2} A(q^2) \bigg[ A(p^2) - A(q^2) \bigg] \Bigg\}  \nonumber \\
   & &  + ~ g_3 \left(p^2, (p-q)^2 \right) \, Y_3 (q^2) ~  q ~ \times \nonumber \\
   & & \hspace{2cm} \times \Bigg\{ \frac{3}{2} \bigg[ A(q^2) B(p^2) - 2 \, A(q^2) B(q^2)  \bigg] \Bigg\} \ .
\end{eqnarray}
The requirement of having a finite integral demands that at large $q$
\begin{eqnarray}
   & & g_1 \left(p^2, (p-q)^2 \right) \, Y_1 (q^2)   \approx \frac{1}{q^2} \nonumber \\
   & &    g_3 \left(p^2, (p-q)^2 \right) \, Y_3 (q^2) \approx \frac{1}{q^2} 
   \label{Eq:lei_escala_1}
\end{eqnarray} 
or that these functions are proportional to a higher negative power of q.
The logarithmic corrections, not taken into account in this analysis, are sufficient to avoid the
UV logarithmic divergence suggested by the naive power counting. Indeed, these logarithmic corrections 
introduced by the renormalisation group analysis are, for large momenta, of type
$\left( \log (q^2/ \Lambda^2) \right)^\gamma$, with $\gamma$ standing for the anomalous dimensions.
Our large $q$ analysis should take into account the logarithmic corrections coming from  the gluon, the ghost and the quark propagators
that for $N_f = 2$ result in $\gamma = \gamma_{glue} + \gamma_{ghost} + \gamma_{quark} = -137/116$. Then, assuming a large 
$q$ behaviour as in (\ref{Eq:lei_escala_1}) times the log correction, the integration function at high momenta becomes
\begin{equation}
     \frac{1}{q^2} ~ \bigg[ \log \left( \frac{q^2}{ \Lambda^2_{QCD}} \right) \bigg]^{ - \, \frac{137}{116}} \ ,
\end{equation}
resulting in a finite value for the integral. The difference between the naive power counting and taking
into account the log corrections is illustrated on Fig.~\ref{fig:large_q_log}, where one can observe the effect due to the log corrections that
suppress further the integrand function at high momenta.
In what concerns the quark-ghost kernel form factor $Y_0(q^2)$ at high energies, the power counting analysis is compatible with 
having a $Y_0 (q^2) = 1$ at 
large momenta as required by perturbation theory and by the all-orders result summarised in Eq. (\ref{Eq:all_orders}).

The same analysis for the vector component (\ref{Eq:DSEMinkVectorXEuclidean}) gives, up to logarithmic corrections,
\begin{eqnarray}
  & &  g_0 \left(p^2, (p-q)^2 \right) \, Y_0 (q^2) ~  \frac{1}{q}  ~ \bigg\{ \cdots \bigg\} 
          \nonumber  \\
  & &
 \hspace{1cm} + ~ g_1 \left(p^2, (p-q)^2 \right) \, Y_1 (q^2) ~ \frac{1}{q} ~ \bigg\{ \cdots  \bigg\}  \nonumber \\
   & &  \hspace{1.5cm} + ~ g_3 \left(p^2, (p-q)^2 \right) \, Y_3 (q^2) ~  q ~ \bigg\{ \cdots \bigg\} \ ,
\end{eqnarray}
where $\{ \cdots \}$ stand for finite expressions involving $A(p^2)$, $A(q^2)$, $B(p^2)$ and $B(q^2)$. 
The conditions given in (\ref{Eq:lei_escala_1}) are sufficient to ensure a finite result associate to the UV integration over $q$ for the 
vector component of the Dyson-Schwinger equations. 

The dynamics of QCD generates infrared mass scales for the quark and gluon propagators, see Secs.~\ref{Sec:gluonghost} and~\ref{Sec:quarkpropfits},
that eliminate possible infinities associated to the low momentum  limit in the integral of the quark gap equation and the analysis of the infrared limit does not 
add any new constraints.

For full QCD, the $\lambda_1$ form factor was computed in the soft gluon limit, i.e. vanishing gluon momenta, using 
lattice simulations in~\cite{Oliveira:2016muq}. The analysis of the lattice data performed in~\cite{Oliveira2018a} shows that the lattice data
is well described by
\begin{equation}
   \lambda_1(p^2) = A(p^2) \bigg\{ a + b \, D(p^2) \bigg\}
\end{equation}   
where $a$ and $b$ are constants, that in terms of $Y_1$ and $Y_3$ translates into
\begin{equation}
 2 \, M(p^2) \, Y_1(p^2) - 2 \, p^2 \, Y_3 (p^2) \propto D (p^2) \ ,
 \label{Eq:seila44}
\end{equation}
where $M(p^2) = B(p^2) / A(p^2)$. This result suggests to write
\begin{equation}
   X_1 ( p^2, (p-q)^2, q^2) = D \left( \frac{ p^2  + (p-q)^2 }{2}  \right) ~ Y_1 (q^2)
   \label{Eq:X1ansatz}
\end{equation}
that in the high $q$ limit gives $X_1 \sim Y_1 (q^2)/q^2$ and regularises the ultraviolet behaviour in agreement with
the discussion summarised in (\ref{Eq:lei_escala_1}). 
Similarly, equation (\ref{Eq:seila44})  also suggests
\begin{equation}
   X_3 ( p^2, (p-q)^2, q^2) =  D \left( \frac{ p^2  + (p-q)^2 }{2}  \right) ~ Y_3 (q^2)
   \label{Eq:X3ansatz}
\end{equation}
giving at high $q$ momenta a $X_3 \sim \tilde{X}_3 (q^2) / q^2$ and, in this way, the ultraviolet problems referred in (\ref{Eq:lei_escala_1}) are solved.
Furthermore, for large quark momentum the ansatz (\ref{Eq:X1ansatz}) and (\ref{Eq:X3ansatz}) give
\begin{equation}
\frac{ q^3 \, D (q^2) \, F(q^2) }{A^2(k^2) \, k^2 + B^2(k^2)} \quad \underset{p \rightarrow + \infty}{\xrightarrow{\hspace*{2cm}}} \quad  
\frac{ q^3 \, D (q^2) \, F(q^2) }{A^2(p^2) \, p^2 }  
\end{equation}
implying the vanishing of the kernels (\ref{K0:B}) --  (\ref{K3:A}) for sufficiently large $p$.

In short, our ansatz for the quark-gluon vertex used to solve the Dyson-Schwinger equations reads
\begin{eqnarray}
   X_0 ( p^2, (p-q)^2, q^2 ) & = & X_0( q^2) \ ,  \label{Eq:AnsatzX0} \\
   X_1 ( p^2, (p-q)^2, q^2 ) & = & D \left(  \frac{p^2 + (p-q)^2}{2} \right)  ~ Y_1 (q^2)  \ , \label{Eq:AnsatzX1}  \\
   X_3 ( p^2, (p-q)^2, q^2 ) & = & D \left(  \frac{p^2 + (p-q)^2}{2} \right)  ~ Y_3 (q^2) \ . \label{Eq:AnsatzX3} 
\end{eqnarray}
The quark gap equation should be solved taking into account the constraint (\ref{Eq:all_orders}) that demands 
\begin{equation}
   X_0 (0) ~  = ~  X_0 ( q \rightarrow + \infty ) ~  =  ~ 1 \ .
   \label{Eq:bcX0}
\end{equation}
The Landau gauge lattice gluon propagator as given by lattice simulations~\cite{Duarte:2016iko,Dudal:2018cli} retuns a $D(q^2)$ that
is strongly enhanced at low momenta. It follows from Eqs. (\ref{Eq:AnsatzX1}) and (\ref{Eq:AnsatzX3}) that within the ansatz considered here, one expects
$X_1$ and $X_3$ to rise significantly for small $p^2 + (p-q)^2 = 2 \, p^2 + q^2 - 2 p \cdot q$. On the other hand, at high momenta, the form factors should approach 
its perturbative value. At tree level in perturbation theory the quark-ghost kernel form factors read $X_0 = 1$ and $X_1 = X_3 = 0$, suggesting that $X_1$ and $X_3$
give  marginal contributions to the full vertex at sufficient high energy. 
At the qualitative level, the guessed behaviour associated with the ansatz (\ref{Eq:AnsatzX0}) -- (\ref{Eq:AnsatzX3}) reproduce the computed
quark-ghost kernel form factors computed in~\cite{Aguilar:2018epe} using the Dyson-Schwinger equations and one-loop dressed perturbation theory
for the quark-ghost kernel.
For the inversion of the Dyson-Schwinger equations, i.e. from the numerical point of view,
given the strong enhancement of the gluon propagator at low momenta, this can mean a poorer resolution of $X_1$ and $X_3$ in the ultraviolet regime.

\section{Preparing to Solve the Euclidean Dyson-Schwinger Equations \label{Sec:LattFunctions}}

The computation of a solution of the Dyson-Schwinger equations requires parameterising either the quark-gluon vertex, if one aims to look at the quark propagator,
or the quark propagator functions to extract information on the quark-gluon vertex. In both these cases, a complete description of the
gluon and ghost propagators is assumed explicitly. 

In the current work, we aim to solve the gap equation for the quark-gluon vertex and, therefore, the knowledge of the various propagators over all range of
momenta appearing in the integral equation is required.
This is achieved fitting the Landau gauge lattice propagators with model functions that are compatible with the results of 1-loop renormalisation group improved 
perturbation theory. 
In this way, it is ensured that the perturbative tails are taken into account properly in the parameterisation of the propagators.
The parameterisations considered here are compared to those of~\cite{Rojas:2013tza} in \ref{Sec:prop_fits}.
As can be seen on Fig.~\ref{fig:comparefitsprops}, the differences between the two sets of curves are more quantitative than qualitative.

\subsection{Landau gauge lattice gluon and ghost propagators \label{Sec:gluonghost}}

\begin{figure}[t] 
   \centering
   \includegraphics[width=3.5in]{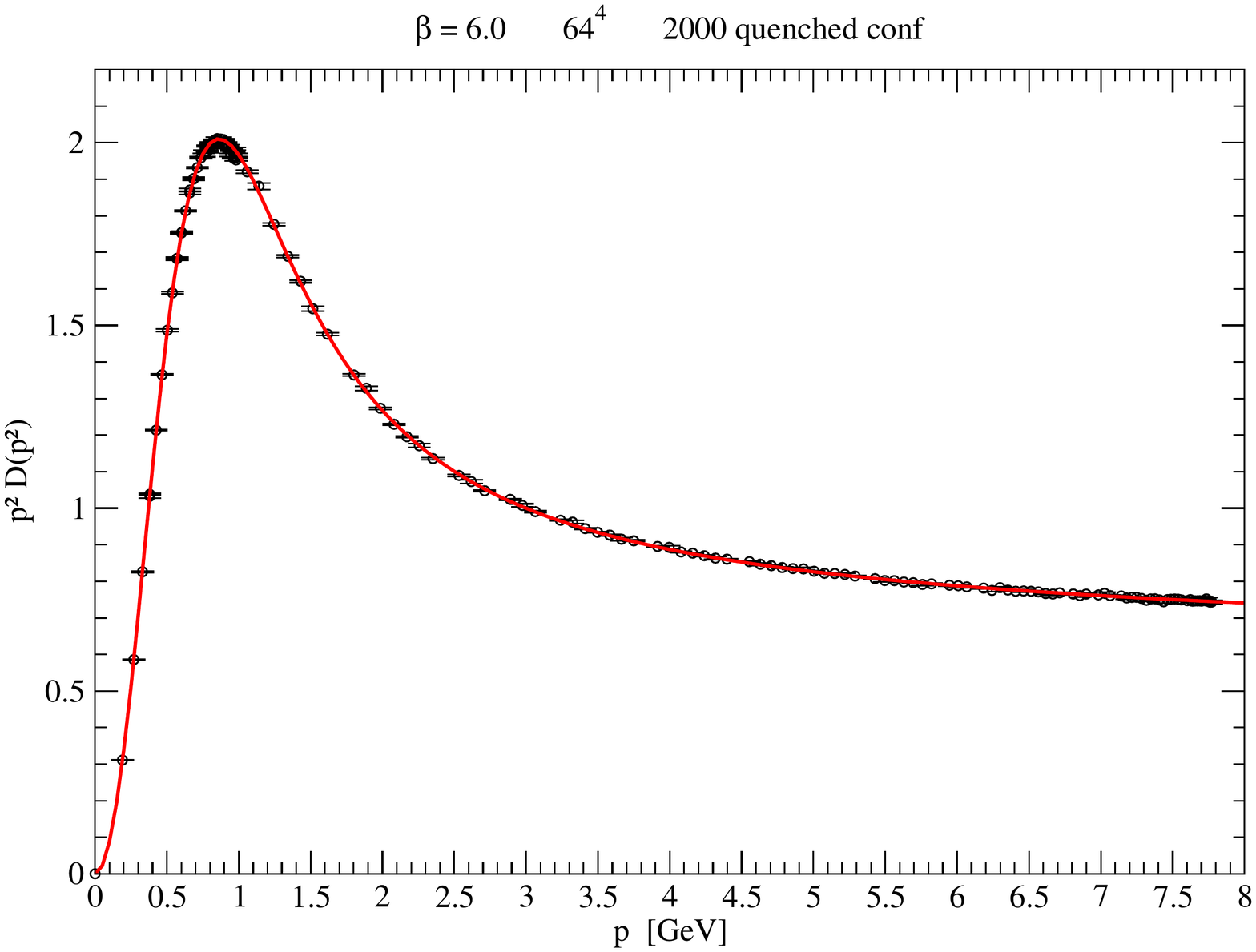} ~
   \includegraphics[width=3.5in]{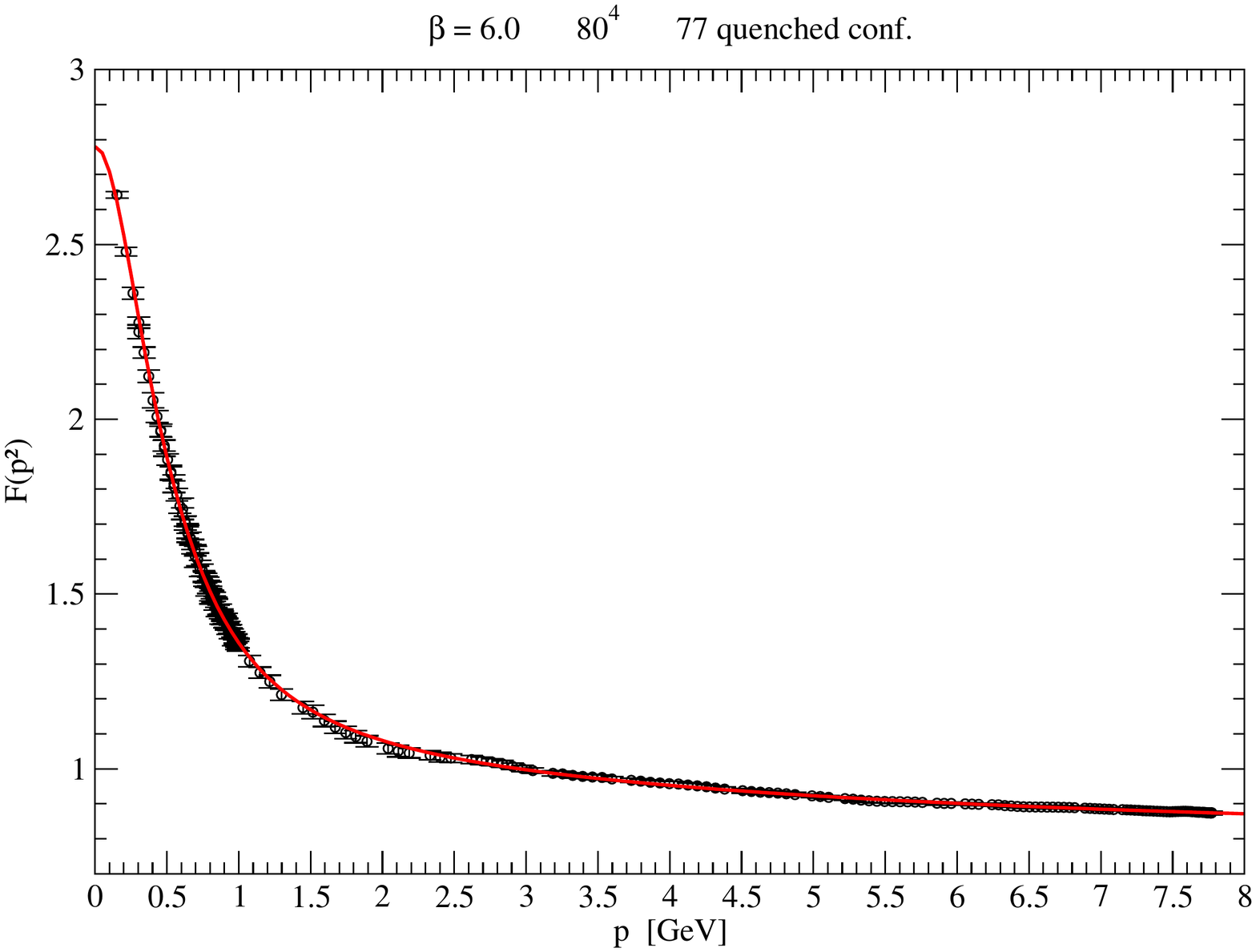} 
   \caption{Pure Yang-Mills gluon (top) and ghost (bottom) lattice dressing functions and the corresponding fit functions used herein. See text for details.}
   \label{fig:gluon_ghost_dress_func}
\end{figure}

The lattice gluon propagator has been computed in the Landau gauge both for full QCD and for the pure Yang-Mills. The gluon propagator
is well known for the pure Yang-Mills theory and it was calculated in~\cite{Dudal:2018cli} for large statistical ensembles and for large physical volumes
$\sim (6.6 \mbox{ fm})^4$ and $\sim (8.2 \mbox{ fm})^4$. Furthermore, in~\cite{Dudal:2018cli} the authors provide global fits to the lattice data that 
reproduce the 1-loop renormalisation group summation of the leading logarithmic behaviour. Of the various expressions given there, we will use to solve 
the integral Dyson-Schwinger equations the following fit to the $(6.6 \mbox{ fm})^4$ volume result
\begin{equation}
   D (p^2) = Z \, \frac{ p^2 + M^2_1}{p^4+ M^2_2 \, p^2 + M^4_3} \left[ \omega \,  \ln \left(\frac{p^2 + m^2_0}{\Lambda^2_{QCD}}\right) + 1 \right]^{\, \gamma} \ ,
   \label{Eq:global_gluon_Fit}
\end{equation}
with the gluon anomalous dimension being
$\gamma = -13/22$, $Z  = 1.36486 \pm 0.00097$, $M^2_1   =  2.510 \pm 0.030 \mbox{ GeV}^2$, $M^2_2   =  0.471 \pm 0.014 \mbox{ GeV}^2$, 
$M^4_3   =  0.3621 \pm 0.0038  \mbox{ GeV}^4$, $m^2_0   =  0.216 \pm 0.026  \mbox{ GeV}^2$ using $\Lambda_{QCD} = 0.425 \mbox{ GeV}$ and where
$\omega = 33 \,  \alpha_s ( \mu ) / 12 \pi$ with a strong coupling constant $\alpha_s ( \mu = 3 \mbox{ GeV} ) = 0.3837$;  see~\cite{Dudal:2018cli} for 
details. This  fit to the lattice data has an associated $\chi^2/\text{d.o.f.} = 3.15$. The authors provide fits with better values for the
$\chi^2/\text{d.o.f.}$ However, given that the level of precision achieved on lattice simulations for the quark propagator is considerably smaller
than for the gluon propagator, one should not distinguish between the various fitting functions provided in~\cite{Dudal:2018cli}. Our option
considers the simplest functional form given in that work.

The lattice data for the Landau gauge gluon dressing function $p^2 D (p^2)$, renormalised in the MOM-scheme
at the mass scale $\mu = 3$ GeV and the fit associated to Eq. (\ref{Eq:global_gluon_Fit})
can be seen on the top part of Fig.~\ref{fig:gluon_ghost_dress_func}.

For the ghost propagator we take the data reported in~\cite{Duarte:2016iko} for the $80^4$ lattice simulation and fit the lattice data to the functional
form
\begin{eqnarray}
   & & D_{gh} (p^2)  =  \frac{F(p^2)}{p^2} \nonumber \\
   & & =  \frac{Z}{p^2} ~ \frac{ p^4 + M^2_2 \, p^2 + M^4_1}{p^4+ M^2_4 \, p^2 + M^4_3}  ~
            \left[ \omega \,  \ln \left(\frac{p^2 + \frac{m^4_1}{p^2 + m^2_0}}{\Lambda^2_{QCD}}\right) + 1 \right]^{\, \gamma_{gh}} \ ,
   \label{Eq:global_ghost_Fit}
\end{eqnarray}
getting $Z  =  1.0429 \pm 0.0054$, $M^4_1 = 18.2 \pm 5.7 \mbox{ GeV}^4$, $M^2_2 = 33.4 \pm 6.4 \mbox{ GeV}^2$,
$M^4_3 = 6.0 \pm 2.7 \mbox{ GeV}^4$, $M^2_4 = 29.5 \pm 5.7$ GeV$^2$,
$m^4_1 = 0.237 \pm 0.049$, $m^2_0 = 0.09 \pm 0.42 \mbox{ GeV}^2$
with a $\chi^2/\text{d.o.f.} = 0.27$. In the above expression the ghost anomalous dimension reads $\gamma_{gh} = - 9/44$ with
$\omega$ and $\Lambda_{QCD}$ taking the same values as in the gluon fitting function (\ref{Eq:global_gluon_Fit}).
The lattice data, renormalised in the MOM-scheme at the mass scale $\mu = 3$ GeV,  and the fitting curve (\ref{Eq:global_ghost_Fit}) can be
seen on the bottom of Fig. \ref{fig:gluon_ghost_dress_func}.

 \subsection{Lattice Quark Propagator \label{Sec:quarkpropfits}}

For the quark propagator we consider the result of a $N_f = 2$ full QCD simulation in the Landau gauge~\cite{Oliveira:2016muq,Oliveira:2018lln} for
$\beta = 5.29$, $\kappa = 0.13632$ and for a $32^3 \times 64$ lattice. For this particular lattice setup,
the corresponding bare quark mass is 8 MeV and the pion mass reads $M_\pi = 295$ MeV. 

\begin{figure}[t] 
   \centering
   \includegraphics[width=3.5in]{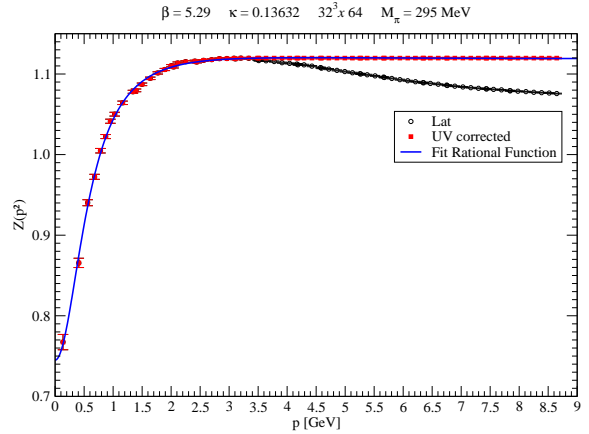} ~
   \includegraphics[width=3.5in]{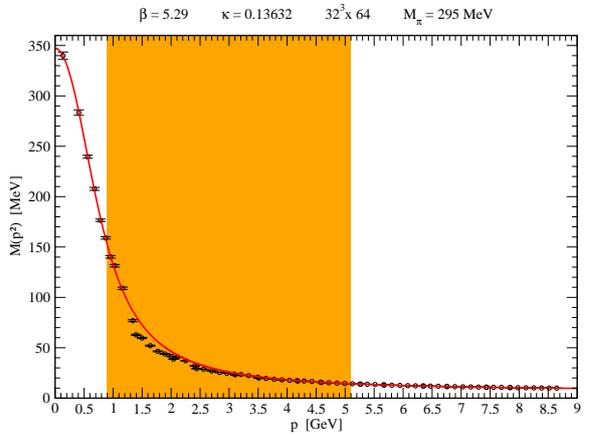} 
   \caption{Quark wave function (top) and running mass (bottom) lattice functions from full QCD simulations with $N_f = 2$.}
   \label{fig:Z_M_func_fullQCD}
\end{figure}

Our fittings to the lattice data, see below, take into account that the lattice data is not free of lattice artefacts; see ~\cite{Oliveira:2016muq} and ~\cite{Oliveira:2018lln}
for details. At high momenta the lattice quark wave function $Z(p^2)$ is a decreasing function of momenta, a behaviour that is not compatible with perturbation
theory that predicts a constant $Z(p^2)$ in the Landau gauge. As reported in~\cite{Oliveira:2016muq,Oliveira:2018lln}, the analysis of the lattice artefacts relying on the H4
method suggests that, indeed, $Z(p^2)$ is constant at high $p$. In order to be compatible with perturbation theory, we identify the region
of momenta where $Z(p^2)$ is constant and, for momenta above this plateaux, we replace the lattice estimates of $Z(p^2)$ by constant values, i.e.
the higher value of the quark wave function belonging to the plateaux.
The original lattice data and the ultraviolet corrected lattice data can be seen on the left of Fig.~\ref{fig:Z_M_func_fullQCD}. The UV corrected lattice data
is then fitted to the rational function
\begin{equation}
    Z(p^2) = Z_0 ~ \frac{ p^4 + M^2_2 \, p^2 + M^4_1 }{p^4 + M^2_4 \, p^2 + M^4_3}
\end{equation} 
giving $Z_0 = 1.11824 \pm 0.00036$, $M^4_1 = 1.41 \pm 0.18 \mbox{ GeV}^4$, 
$M^2_2 = 6.28 \pm1.00 \mbox{ GeV}^2$, $M^4_3 = 2.11 \pm 0.28 \mbox{ GeV}^4$,
$M^2_4 = 6.20$ $\pm 0.98$ GeV$^2$ for a $\chi^2/\text{d.o.f.} = 0.74$. The solid red line on Fig.~\ref{fig:Z_M_func_fullQCD} (top) 
refers to the fit just described.

The removal of the lattice artefacts for the running quark mass is more delicate when compared to the evaluation of the quark wave 
function lattice artefacts~\cite{Skullerud:2000un,Skullerud:2001aw,Oliveira:2018lln}. 
The lattice data published in~\cite{Oliveira:2016muq,Oliveira:2018lln} and reported on Fig.~\ref{fig:Z_M_func_fullQCD} (bottom) was obtained using the 
so called hybrid corrections to  reduce the lattice effects~\cite{Skullerud:2001aw} . The hybrid method results in a smoother mass function 
when compared to the one obtained by applying the multiplicative corrections. 
The differences on the corrected running mass between the two methods occur for momenta above 1 GeV, 
with the multiplicative corrected running mass being larger than the corresponding hybrid estimation; see Appendix on~\cite{Oliveira:2018lln}. 
The running mass provided by the two methods, corrected for the lattice artefacts, seems to converge to the same values at large momentum. 

The running mass reported on Fig.~\ref{fig:Z_M_func_fullQCD} (bottom) is not smooth enough to be fitted. To model the lattice running mass
in a way that reproduces the ultraviolet and the infrared lattice data and is compatible with the perturbative behaviour at high moment, we remove
some of the lattice data at intermediate momenta. On  Fig.~\ref{fig:Z_M_func_fullQCD} the data in the region with an orange background was
not taken into account in the global fit of the running quark mass. The remaining lattice data was fitted to
\begin{equation}
 M(p^2) = \frac{ m_q ( p^2) }{ \left[ A + \log( p^2 + \lambda \, m^2_q(p^2) ) \right]^{\gamma_m} }
 \label{Eq:RunningMass}
\end{equation}
where $\gamma_m = 12/29$ is the quark anomalous dimension for $N_f = 2$ and
\begin{equation}
        m_q(p^2) = M_q ~\frac{ p^2 + m^2_1}{ p^4 + m^2_2 \, p^2 + m^4_3 } + m_0 \ .
         \label{Eq:RunningQuarkMass}
\end{equation}
The fitted parameters are
 $M_q = 349 \pm 10 \mbox{ MeV GeV}^2$, $m^2_1  = 1.09 \pm 0.43 \mbox{ GeV}^2$, $m^2_2  = 0.92 \pm 0.28 \mbox{ GeV}^2$, 
$m^4_3 = 0.42 \pm 0.15 \mbox{ GeV}^4$,
$m_0 = 10.34 \pm 0.63$ MeV and $A = -2.98 \pm 0.25$ for a $\chi^2/\text{d.o.f.} = 1.97$ after setting $\lambda = 1 \mbox{ GeV}^2/\mbox{MeV}^2$.
 The fit function and the full lattice running
quark mass data can be seen on Fig.~\ref{fig:Z_M_func_fullQCD} (bottom). Note that in (\ref{Eq:RunningMass}) 
and in (\ref{Eq:RunningQuarkMass}) $p$ is given in GeV and $m_q(p^2)$ and $M(p^2)$ are given in MeV.

\clearpage
\section{Solving the Dyson-Schwinger Equations \label{Sec:quarkghostkernel}}

\begin{figure}[t] 
 \vspace{-2.6cm}
   \centering
   \vspace{0.5cm}
      \includegraphics[width=3.5in]{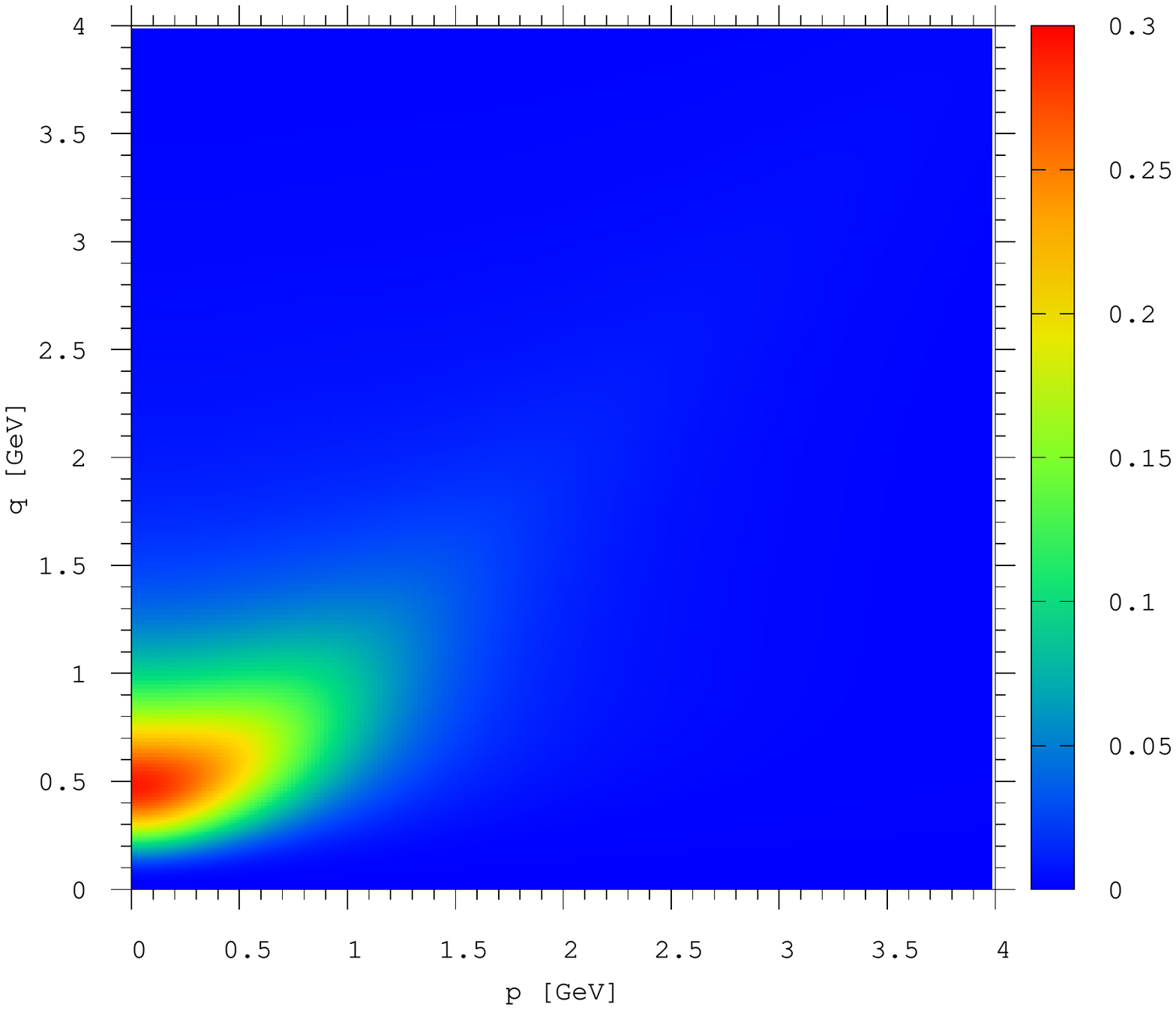}  \\
      \vspace{-4cm}
      \includegraphics[width=3.5in]{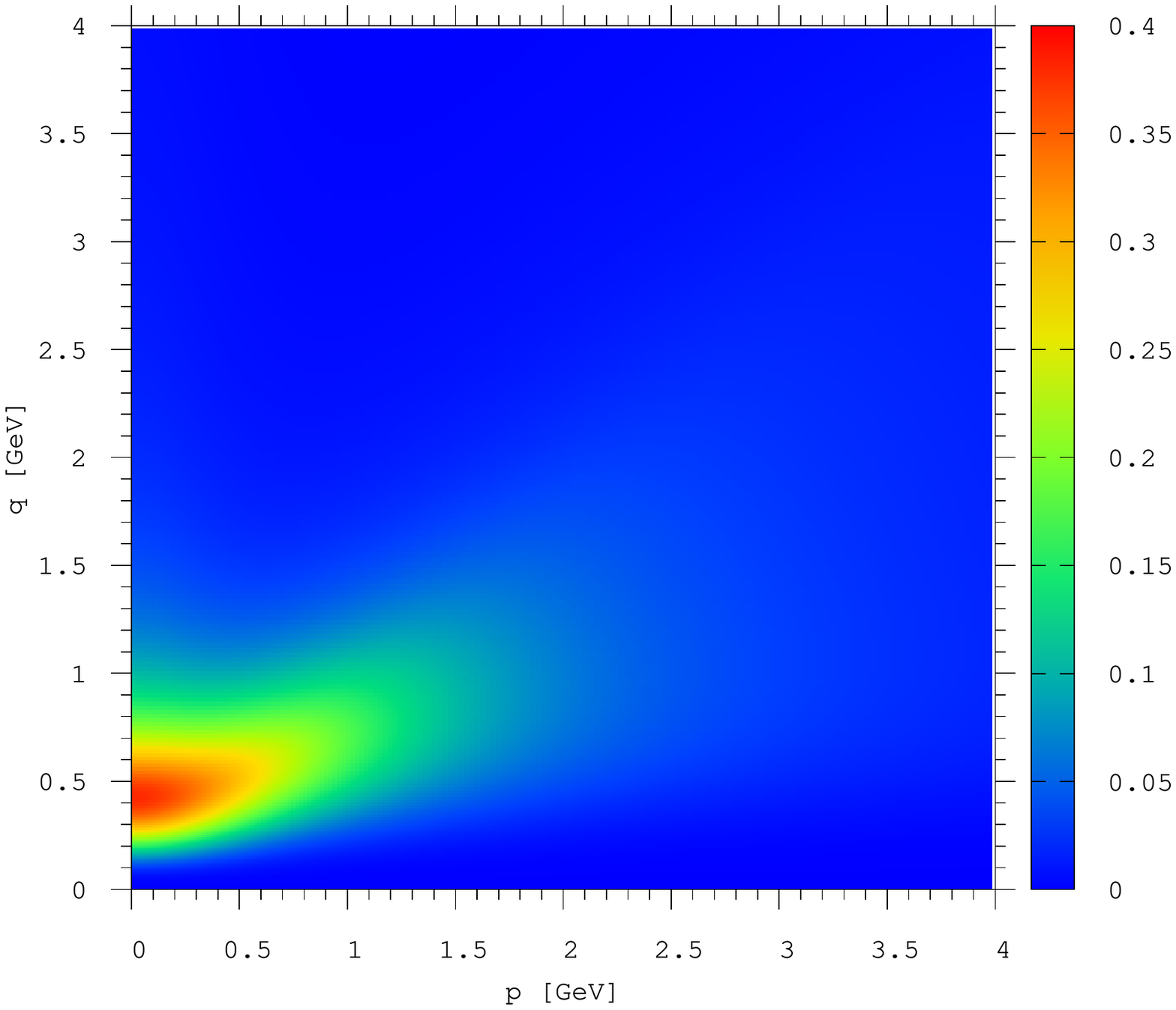}
      \vspace{-2cm}
   \caption{The scalar (top) $\mathcal{N}^{(0)}_{B} (p,q)$ and vector (bottom) $\mathcal{N}^{(0)}_{A} (p,q) / p^2$ kernel components of the Dyson-Schwinger equations.
                  For comparison with the kernels shown in~\cite{Rojas:2013tza}, the above kernels do not include the Gauss-Legendre quadrature weights required
                  to perform the integration over the gluon momentum $q$. This applies also to Figs. \ref{fig:NKernels_1}, \ref{fig:NKernels_3}, \ref{fig:NKernelsProp_2}
                  and \ref{fig:NKernelsProp_3}.}
   \label{fig:NKernels_0}
\end{figure}

\begin{figure}[t] 
   \centering
   \vspace{-2cm}
      \includegraphics[width=3.5in]{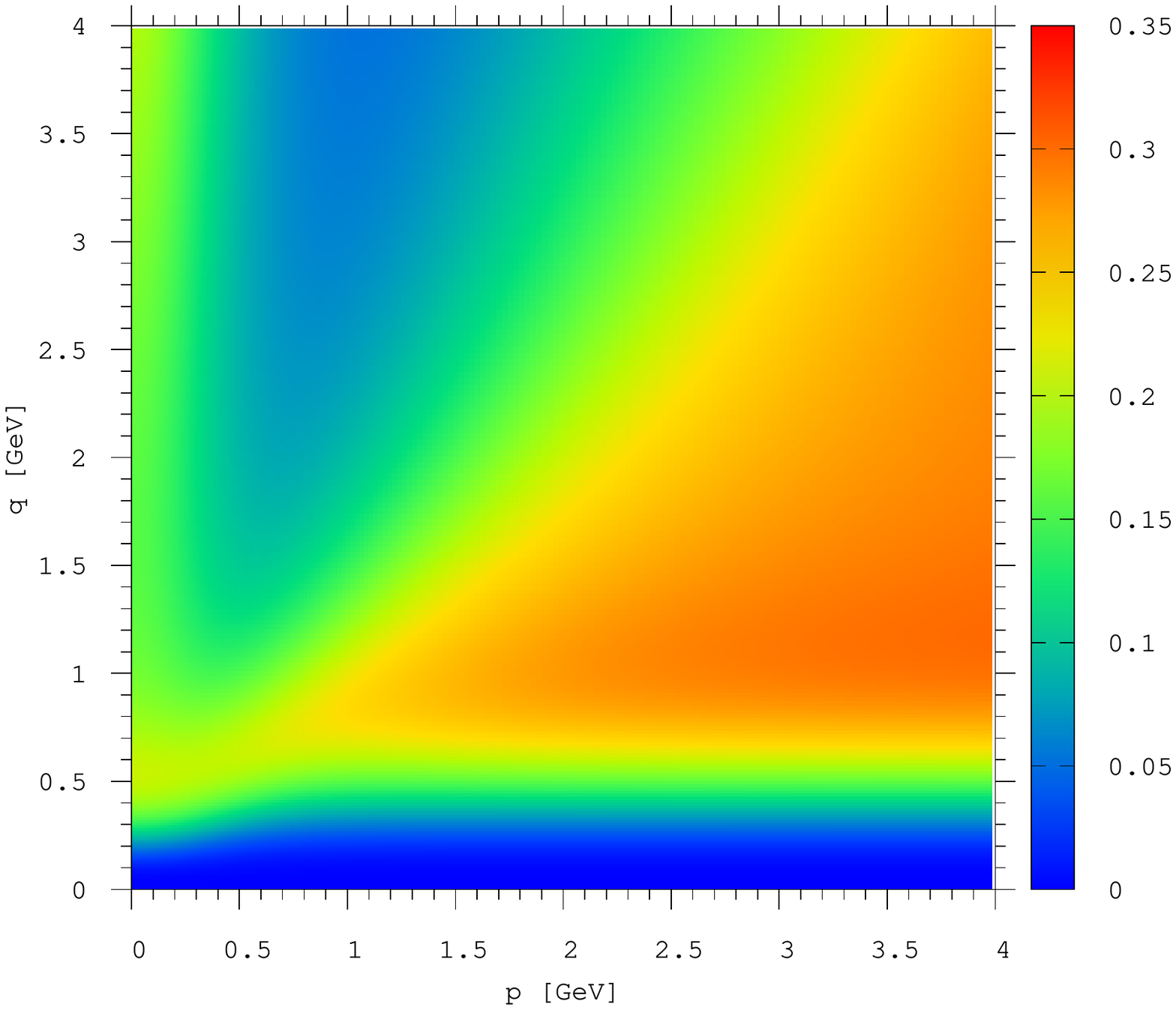} \\
      \vspace{-4cm}
      \includegraphics[width=3.5in]{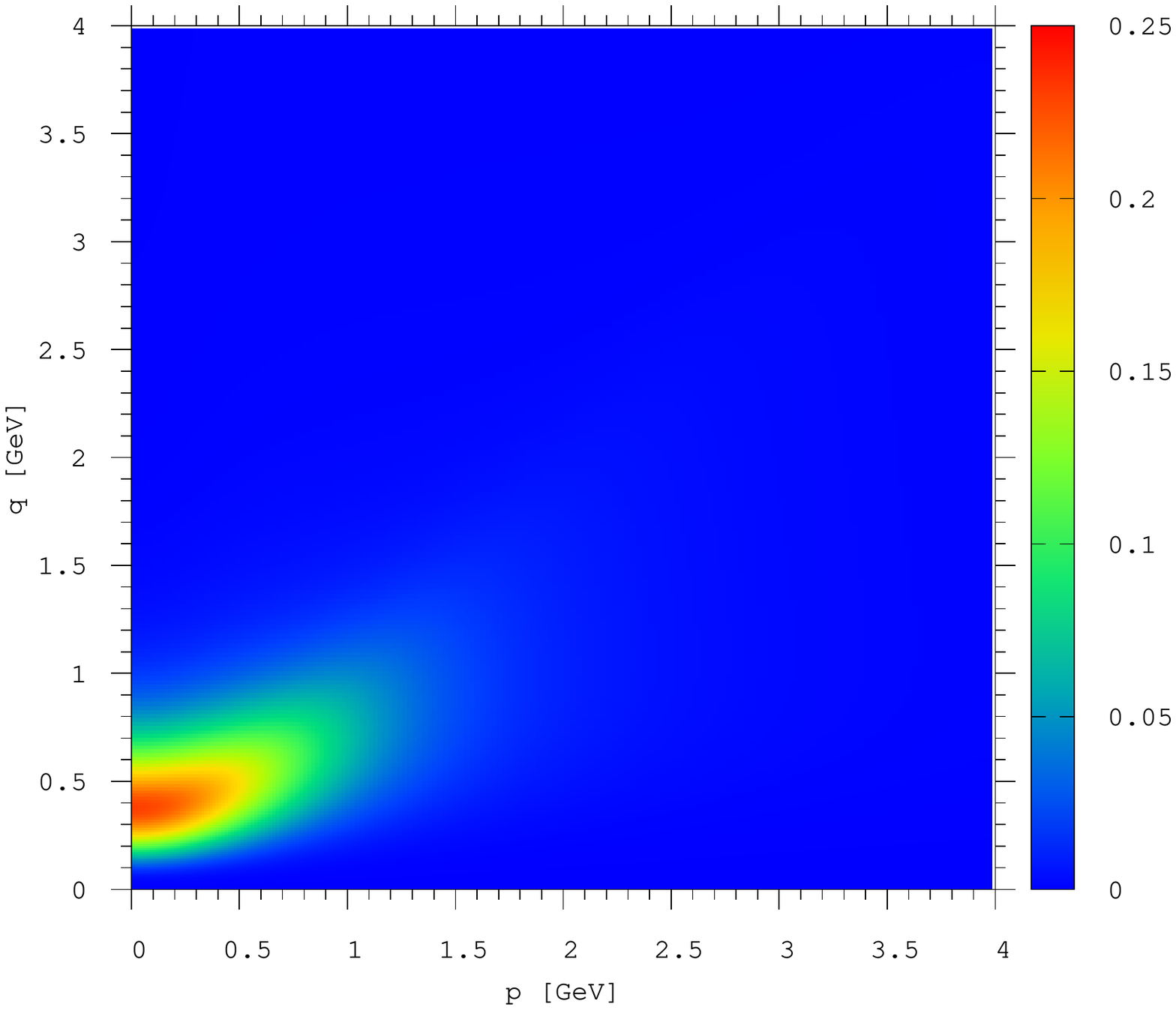}
      \vspace{-2cm}
      \caption{The scalar (top) $\mathcal{N}^{(1)}_{B} (p,q)$ and vector (bottom) $\mathcal{N}^{(1)}_{A} (p,q)/ p^2$ kernel components of the Dyson-Schwinger equations.
                    See also the caption of Fig.~\ref{fig:NKernels_0}.}
   \label{fig:NKernels_1}
\end{figure}

\begin{figure}[t] 
   \centering
   \vspace{-2cm}
      \includegraphics[width=3.5in]{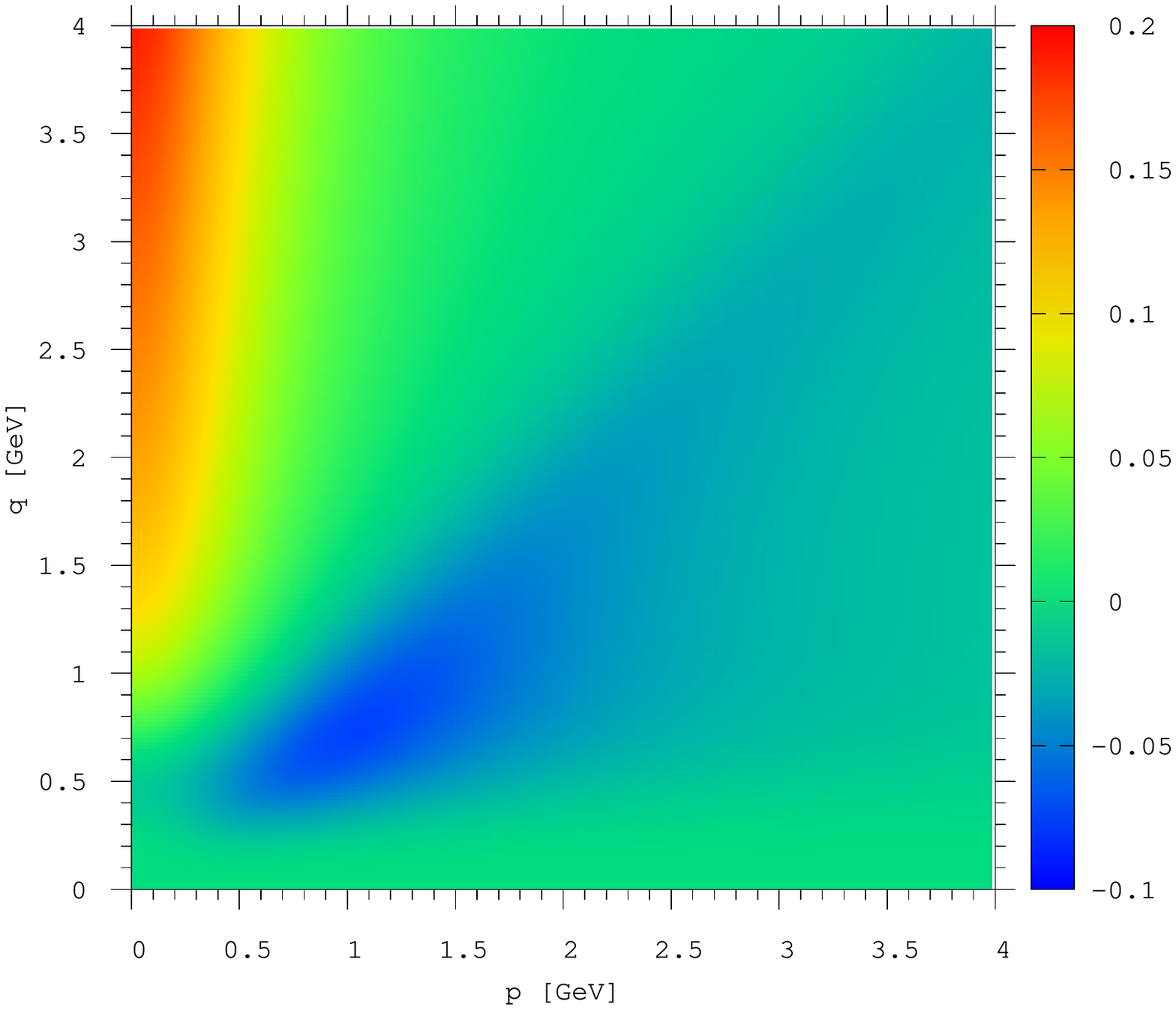} \\
      \vspace{-4cm}
      \includegraphics[width=3.5in]{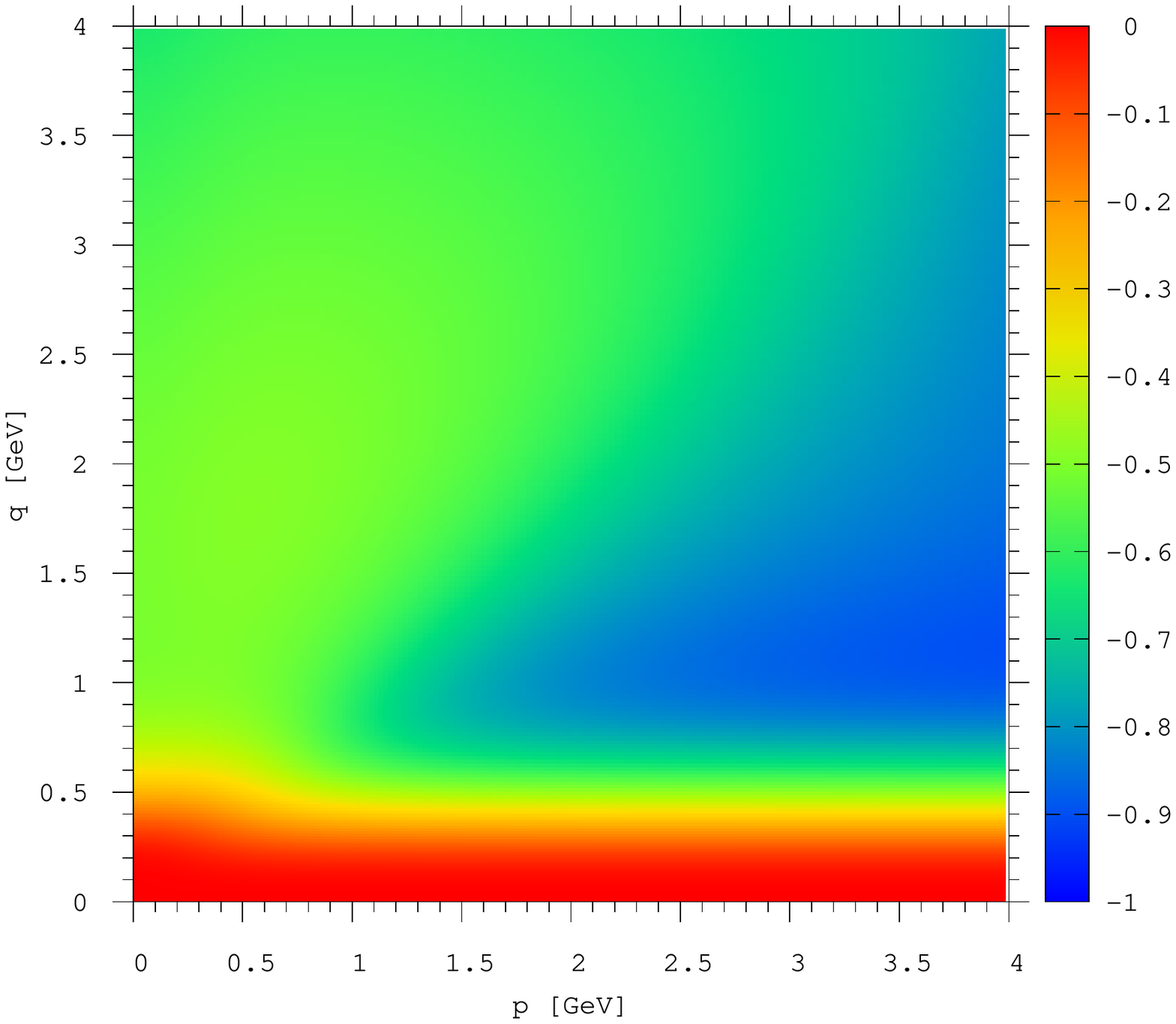} 
   \vspace{-2cm}
   \caption{The scalar (top) $\mathcal{N}^{(3)}_{B} (p,q)$ and vector (bottom) $\mathcal{N}^{(3)}_{A} (p,q)/ p^2$ kernel components of the Dyson-Schwinger equations.
                 See also the caption of Fig.~\ref{fig:NKernels_0}.}
   \label{fig:NKernels_3}
\end{figure}

Let us now discuss the solutions of the Euclidean space Dy\-son-Schwinger equations (\ref{Eq:DSEMinkScalarXEuclidean}) and (\ref{Eq:DSEMinkVectorXEuclidean}) 
for the quark-ghost kernel, i.e. for the quark-gluon vertex. The momentum integration will be performed as described in \ref{Sec:integracao}, i.e.
by introduction an hard cutoff $\Lambda$, and 
with all integrations performed with Gauss-Legendre quadrature. For the angular integration we consider 500 Gauss-Legendre points as in~\cite{Rojas:2013tza}.
After angular momentum integration, one is left with the kernels 
\begin{equation}
   \mathcal{N}^{(0,1,3)}_{A,B} (p,q) =  \frac{ q^3 ~D(q^2) ~ F(q^2)}{\left[ A(k^2) \right]^2  \, k^2 + \left[ B(k^2) \right]^2} ~ \mathcal{K}^{(0,1,3)}_{A,B} (p,q)
\end{equation}
that can be seen on Figs.~\ref{fig:NKernels_0},~\ref{fig:NKernels_1} and~\ref{fig:NKernels_3}, without taking into account the Gauss-Legendre
weights associated to the integration over the gluon momentum. 
The inclusion of the Gauss-Legendre weights associated to the $q$ momentum integration does not change the outcome reported
on Figs.~\ref{fig:NKernels_0},~\ref{fig:NKernels_1} and~\ref{fig:NKernels_3} and the main difference being that the associated numerical values are considerably smaller.

The major contributions of the $\mathcal{N}^{(0)}_{B}$ and $\mathcal{N}^{(0)}_{A}$ kernels occur
in a well defined momentum region where $p \lesssim 1.5$ GeV and $q \lesssim 1.5$ GeV. Further, for $p$, $q \gtrsim 2$ GeV the kernels become marginal. 
The results for $\mathcal{N}^{(0)}_{B}$ and $\mathcal{N}^{(0)}_{A}$  reproduce the corresponding behaviour observed in~\cite{Rojas:2013tza}.
It follows that the integration over momentum in Eqs. (\ref{Eq:DSEMinkScalarXEuclidean}) and (\ref{Eq:DSEMinkVectorXEuclidean}) associated to the kernels 
$\mathcal{N}^{(0)}_{B}$ and $\mathcal{N}^{(0)}_{A}$ kernels, that  are coupled to $X_0(q^2)$, is finite.

\begin{figure}[t] 
 \vspace{-2cm}
   \centering
      \includegraphics[width=3.5in]{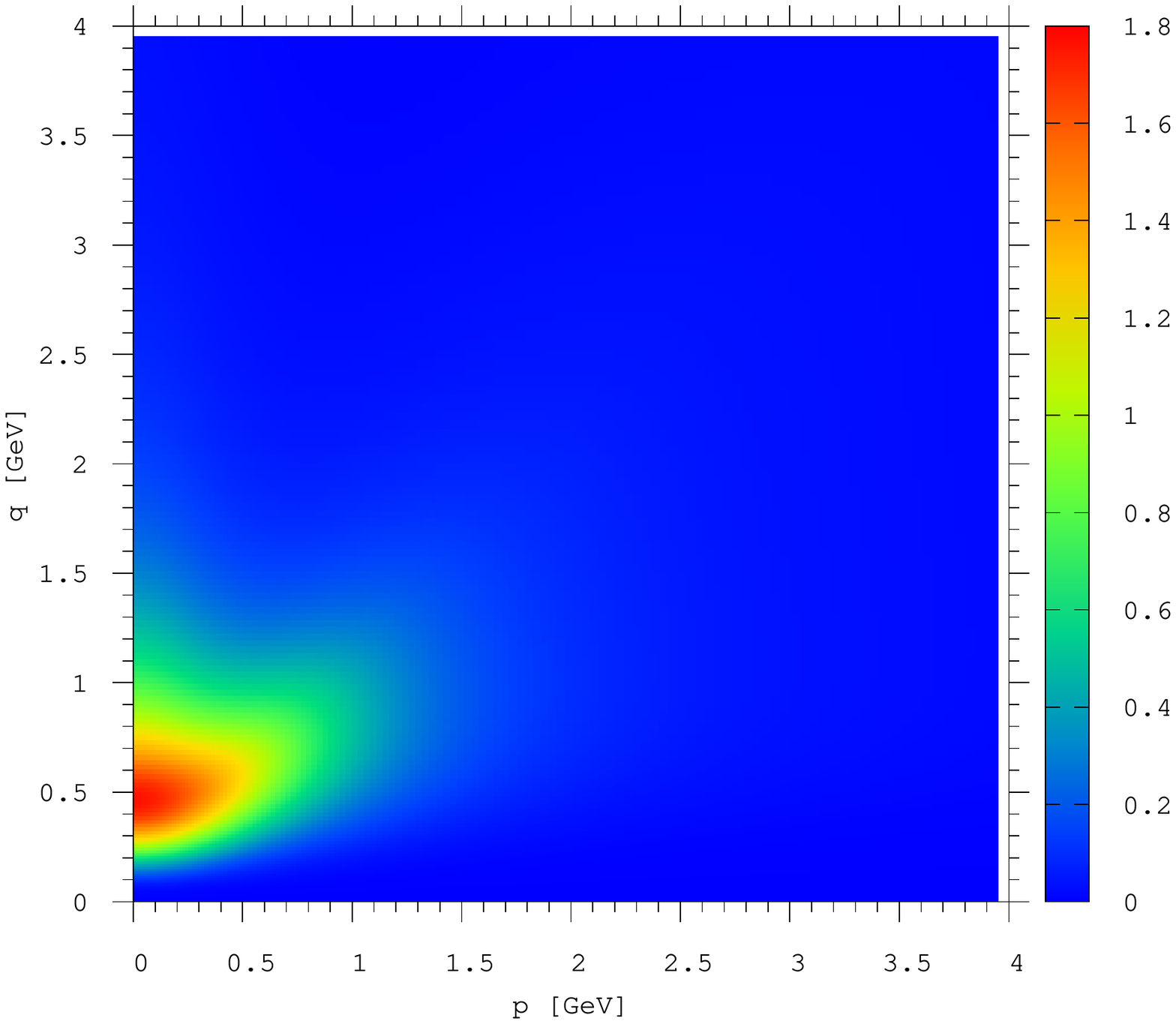} \\
      \vspace{-4cm}
      \includegraphics[width=3.5in]{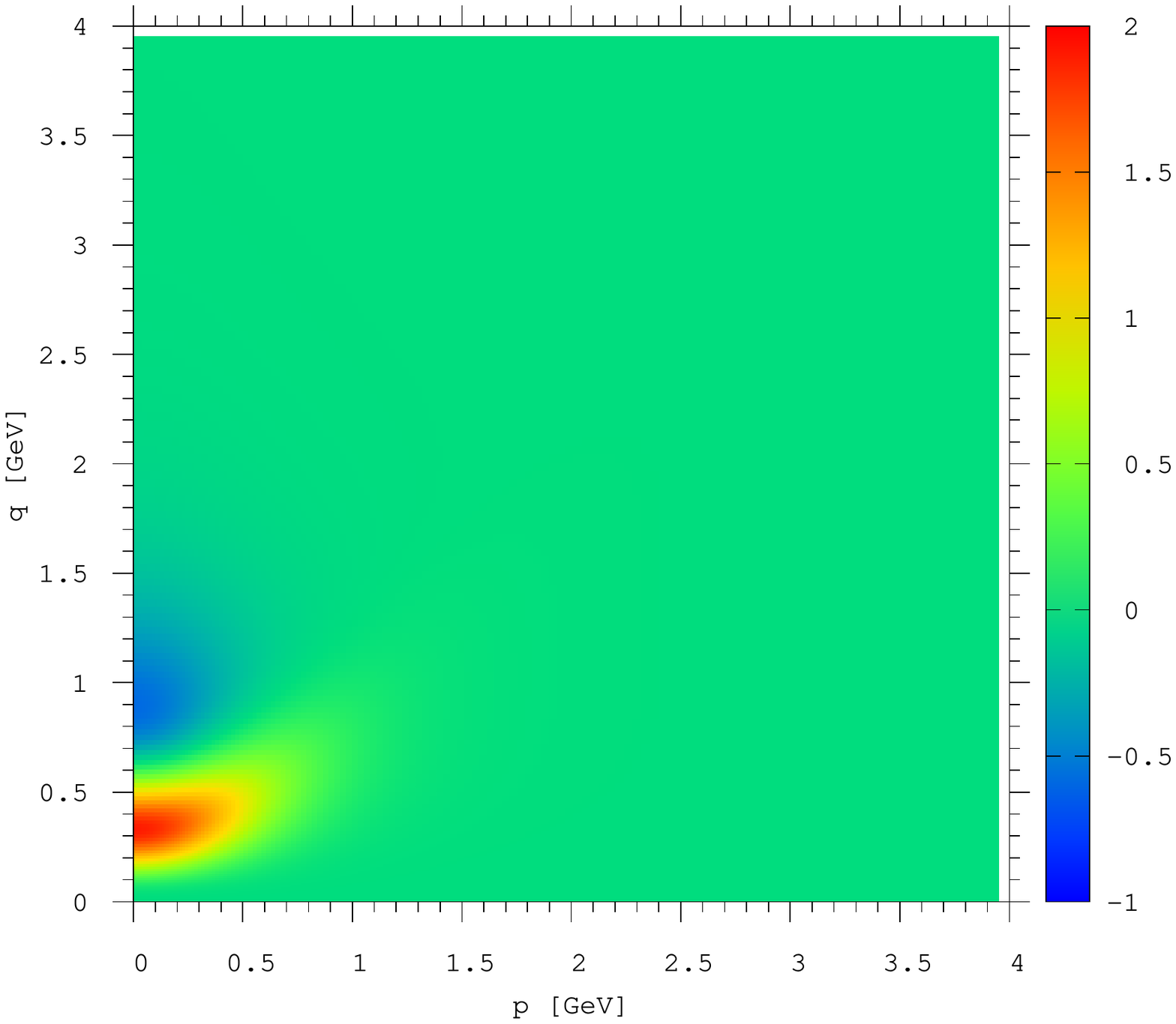} 
   \vspace{-2cm}
   \caption{The scalar (top) $\mathcal{N}^{(1)}_{B} (p,q)$ and vector (bottom) $\mathcal{N}^{(1)}_{A} (p,q)/ p^2$ kernels including the term of the gluon propagator as 
                 defined in (\ref{Eq:AnsatzX1}). See also the caption of Fig.~\ref{fig:NKernels_0}.}
   \label{fig:NKernelsProp_2}
\end{figure}

\begin{figure}[t] 
 \vspace{-2cm}
   \centering
      \includegraphics[width=3.5in]{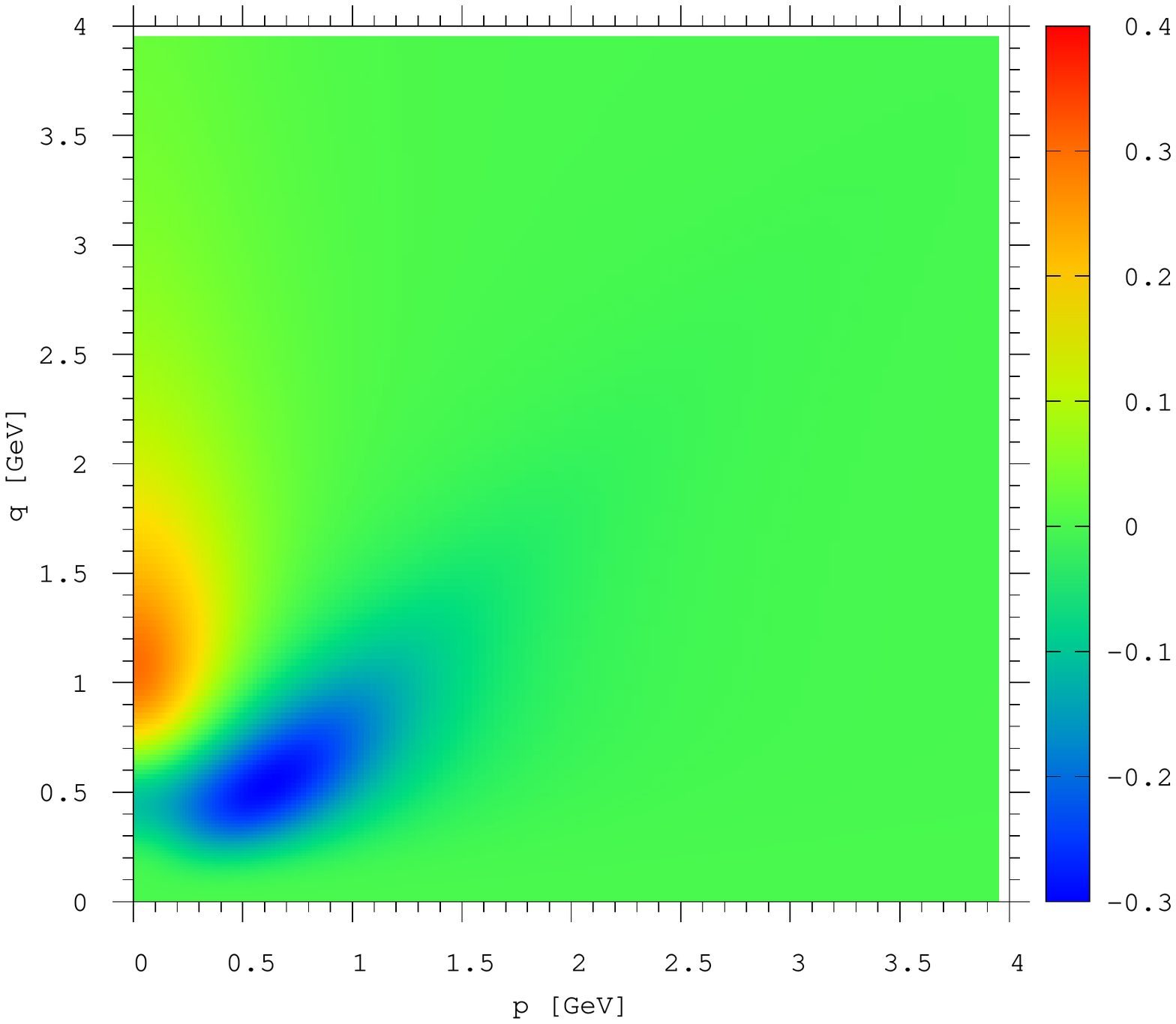} \\
            \vspace{-4cm}
      \includegraphics[width=3.5in]{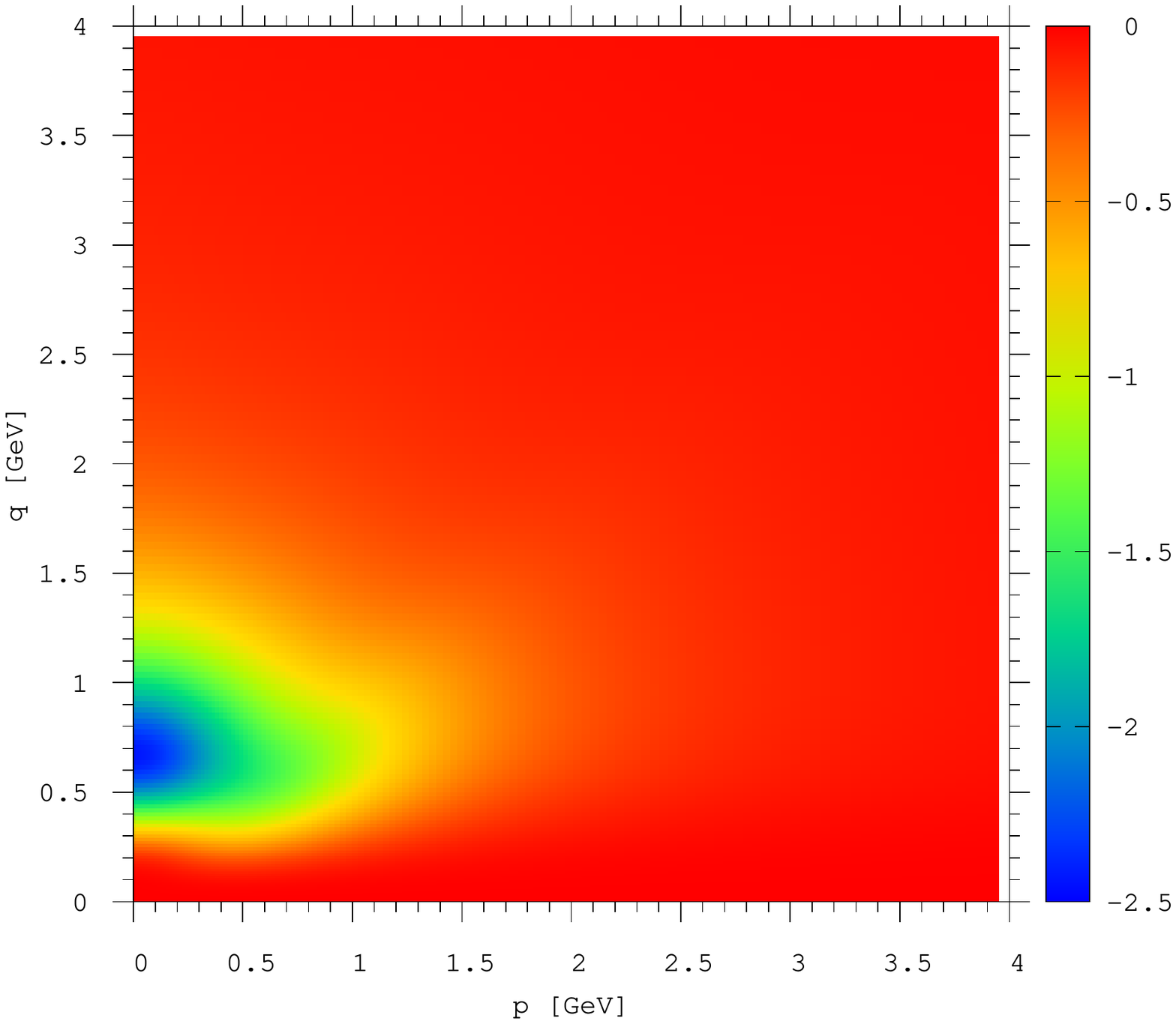} 
   \vspace{-2cm}
   \caption{The scalar (top) $\mathcal{N}^{(3)}_{B} (p,q)$ and vector (bottom) $\mathcal{N}^{(3)}_{A} (p,q)/ p^2$  kernels including the term of the gluon propagator as 
                 defined in (\ref{Eq:AnsatzX3}). See also the caption of Fig.~\ref{fig:NKernels_0}.}
   \label{fig:NKernelsProp_3}
\end{figure}

The function $\mathcal{N}^{(1)}_{A} (p, q)$ displays a similar pattern and, again, the integration over the gluon momentum associated with 
$\mathcal{N}^{(1)}_{A}$ is expected to be well behaved. 
On the other hand the remaining kernels, i.e.  $\mathcal{N}^{(1)}_{B} (p, q)$, $\mathcal{N}^{(3)}_{B}  (p, q)$ and $\mathcal{N}^{(3)}_{A}  (p, q)$,
are all increasing functions of $q$. The requirement of a finite integration over $q$ demands that $X_1$ and $X_3$ should approach zero fast enough
to compensate the increase with $q$ of these kernel functions; see the discussion of the kernels ultraviolet limit in Sec. \ref{Sec:DSEeuclidean}. 
The ansatz (\ref{Eq:AnsatzX0}) -- (\ref{Eq:AnsatzX3}) adds a multiplicative gluon propagator term that is just enough to regularize the 
high momentum associated to $\mathcal{N}^{(1)}_{B} (p, q)$, $\mathcal{N}^{(3)}_{B}  (p, q)$ and $\mathcal{N}^{(3)}_{A}  (p, q)$.
Indeed if one takes into account the multiplicative gluon propagator contribution to the kernels, those who are divergent become well behaved.
This can be seen on Figs.~\ref{fig:NKernelsProp_2} and~\ref{fig:NKernelsProp_3} where the kernels, now including the multiplicative gluon propagator term,
are reported. The new versions of $\mathcal{N}^{(1)}_{B} (p, q)$, $\mathcal{N}^{(3)}_{B}  (p, q)$ and $\mathcal{N}^{(3)}_{A}  (p, q)$
mimic the pattern observed for $\mathcal{N}^{(0)}_{B}$, $\mathcal{N}^{(0)}_{A}$ and $\mathcal{N}^{(1)}_{A} (p, q)$ and, once more, their main contribution to the integral
equations happens for $p \lesssim 2$ GeV and $q \lesssim 2$ GeV. The inclusion of the gluon propagator in the kernels makes the integration over $q$ finite.

The Dyson-Schwinger equations are solved using a hard cutoff that is set to $\Lambda = 20$ GeV. All quantities are renormalised in the MOM scheme,
using the same renormalisation scale as in~\cite{Rojas:2013tza}, i.e $\mu = 4.3$ GeV, so that one can compare easily the results of the two works.
The renormalized quantites satisfy the identities
\begin{equation}
   Z ( \mu^2 )  =  \frac{1}{A( \mu^2 )} = 1 \ , \quad D(\mu^2) =  \frac{1}{\mu^2} \ , \quad F( \mu^2 ) = 1 \ .
\end{equation}
The bare quark mass quoted in the lattice simulation for the ensemble used here reads $m^{bm} = 8$ MeV~\cite{Oliveira:2016muq}.
In the following we set $Z_1 = 1$, take the value for $Z_2$ from the vector component of the gap equation at the cutoff and ''measure`` the bare quark
mass using the scalar component of the gap equation at the cutoff momenta. In this way $m^{b.m.}$ does not coincide with the value quoted
in the simulation but, as can be seen below, its value is close to  the 8 MeV quoted above.

The results shown on Secs.~\ref{Sec:OneLoopSolution},~\ref{Sec:solveall},~\ref{Sec_X0one} and~\ref{Sec:compareX} were computed using the same
value for $\alpha_s( \mu ) = 0.295$ as in~\cite{Rojas:2013tza}. In Sec.~\ref{Sec:tune_alpha} we allow $\alpha_s( \mu )$ to deviate from this value
and provide a ``best value''. From Sec.~\ref{Sec:tune_alpha} onwards, the results reported use the optimal value for the strong coupling constant.

\subsection{One-Loop Dressed Perturbation Theory for $X_0(q^2)$ \label{Sec:OneLoopSolution}}

\begin{figure}[t] 
   \centering
      \includegraphics[width=3in]{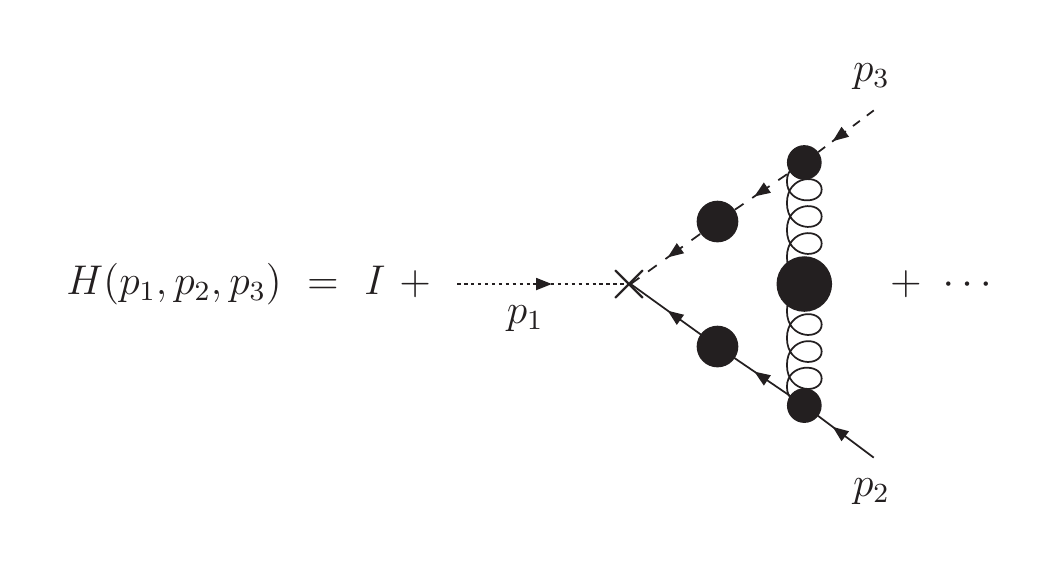} 
   \caption{One-loop dressed perturbation theory }
   \label{fig:Feynquarkghostkernel}
\end{figure}

The four longitudinal quark-gluon form factors were pa\-ra\-me\-tri\-sed in terms of the three quark-ghost kernel form factors $X_0$, $Y_1$ and $Y_3$. 
However, the quark gap equation provides only two independent equations and, therefore, it is not possible to compute all the form factors at once 
for the full range of momenta.

A first look at the quark-ghost kernel form factors is possible if one computes $X_0$ within one-loop dressed perturbation theory with a simplified
version of a quark-ghost kernel where one sets $Y_1 = Y_3 = 0$ and, then, solve the gap equation to estimate $Y_1$ and $Y_3$. The way 
the solutions of the Dyson-Schwinger equations for $Y_1$ and $Y_3$ are built also illustrates the numerical procedure used to solve the integral equations.

The one-loop dressed approximation to the quark-ghost kernel is represented on Fig.~\ref{fig:Feynquarkghostkernel} that, in the simplified version of
kernel, translates into the following integral equation
\begin{eqnarray}
& &    X_0(p^2)  =  1 \nonumber \\
 & & + \frac{C_F \, g^2}{8}  \int_k^\Lambda  \, \left[ p^2 - \frac{(k \cdot p)^2}{p^2} \right] \, \frac{ D(k^2) \, F ( (k + p/2)^2 ) \, F(k^2)}{(k + p/2)^2} \times
    \nonumber \\
    & & \hspace{0.5cm} \times \frac{  A( (k+p)^2 ) \, \left[  A( (k+p)^2 ) + A(p^2) \right] \, H_1( (k+p)^2 ) }{A^2( (k+p)^2 ) \, (k+p)^2  + B^2( (k+p)^2 ) } \times \nonumber \\
    & & \hspace{1cm} \times ~
    X_0 (k^2)  \,  \label{Eq:X0oneloop}
\end{eqnarray}
with ~$H_1 (q^2)$ representing the ghost-gluon vertex. When solving this equation we consider two version of the ghost-gluon vertex, namely its tree level
version where $H_1(q^2) = 1$ and an enhanced dressed vertex as given in~\cite{Dudal:2012zx} where
\begin{equation}
  H_1(q^2) =  c \left( 1 + \frac{ a^2 q^2}{q^4 + b^4} \right) + ( 1 - c ) \frac{w^4}{w^4 + q^4} \ ,
  \label{Eq:ghostgluonenh}
\end{equation}  
for $c = 1.26$, $a = 0.80$ GeV, $b = 1.3$ GeV and $w = 0.65$ GeV. 
For a recent analysis of the quark-ghost vertex see~\cite{Aguilar:2018csq}.
Equation (\ref{Eq:X0oneloop}) was solved after introducing a cutoff $\Lambda = 100$ GeV, after performing
the angular integration using 1000 Gauss-Legendre points and considering 2000 Gauss-Legendre points for the integration over $k$. The introduction of a
Gauss-Legendre quadrature reduces the integral equation (\ref{Eq:X0oneloop}) to a linear system of equations that was solved using the QR decomposition of the matrix appearing
in the linear system. 

\begin{figure}[t] 
   \centering
      \includegraphics[width=3.5in]{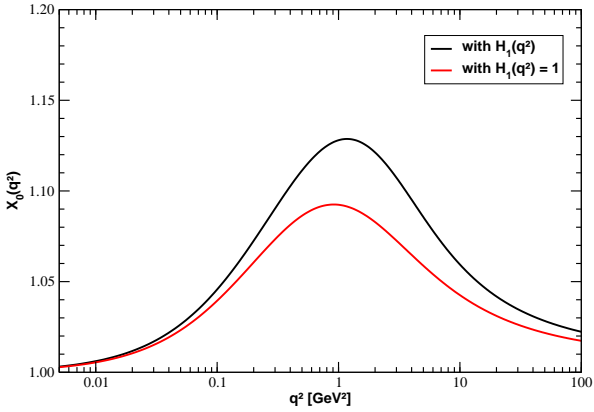} 
   \caption{Simplified one-loop dressed perturbation theory estimation for $X_0(q^2)$}
   \label{fig:X0onelooop}
\end{figure}

The numerical solutions for $X_0$ can be seen on Fig.~\ref{fig:X0onelooop} and are, essentially, those reported in~\cite{Rojas:2013tza}.
According to one-loop dressed perturbation theory, the deviations of $X_0 (q^2)$ from its tree level value are, at most, of the order of 15\%.
We have also looked at the iterative solutions for Eq. (\ref{Eq:X0oneloop}) but no convergence was observed, therefore, the solutions reported  on
Fig.~\ref{fig:X0onelooop} are those computed from a single iteration.

The estimation of $X_0$ allows to solve the gap equation for $Y_1$ and $Y_3$. In order to solve the Dyson-Schwinger equations, after
the angular integration, the scalar and vector components of the equations are rewritten in the form of the larger linear system
\begin{eqnarray}
& &\left( \begin{array}{c}   B(p) - Z_2 m^{b.m.} -  \mathcal{N}^{(0)}_B X_0 (p)  \\  \\ A(p) - Z_2 -  \mathcal{⁄N}^{(0)}_A X_0 (p) \end{array} \right) = \nonumber \\
& & \hspace{2.5cm} = 
\left( \begin{array}{c@{\hspace{0.5cm}}c}   \mathcal{N}^{(1)}_B   & \mathcal{N}^{(3)}_B \\ & \\
                                      \mathcal{N}^{(1)}_A    &   \mathcal{N}^{(3)}_A \end{array} \right) ~
\left( \begin{array}{c}   Y_1   \\  \\ Y_3 \end{array} \right)   \ ;                                
\end{eqnarray}
from now on we will adopt the short name version $B = \mathcal{N} X$ to refer to this linear system of equations. Note that 
$Y_1$ has mass dimensions, while $X_0$ and $Y_3$ are dimensionless. Note that the kernels in $\mathcal{N}$
also have different dimensions and it is only after multiplication that we recover the proper dimensionful equation.

A direct solution of $B = \mathcal{N} X$ results in a meaningless result, with the components of $X$ oscillating over very large values due to the presence of very small
eigenvalues of the matrix $\mathcal{N}$, that translates the ill defined problem in  hands. 
The linear system can be solved using the Tikhonov regularisation~\cite{Tikhonov95} that  replaces the original linear system by a minimisation 
of the functional $|| B - \mathcal{N} X ||^2 + \epsilon ||X||^2$, where $\epsilon$ is a small parameter to be determined in the inversion. 
This functional favours solutions that solve approximately the linear system but whose norm is small. 
For real symmetric matrices, Tikhonov regularisation replaces the original
system by  its normal form 
$ \mathcal{N}^T B = ( \mathcal{N}^T \mathcal{N}  + \epsilon )X$. 
Although in our case $\mathcal{N}$ is not a symmetric matrix, we will solve the system as given in its normal form.

\begin{figure}[t] 
   \centering
      \includegraphics[width=3.5in]{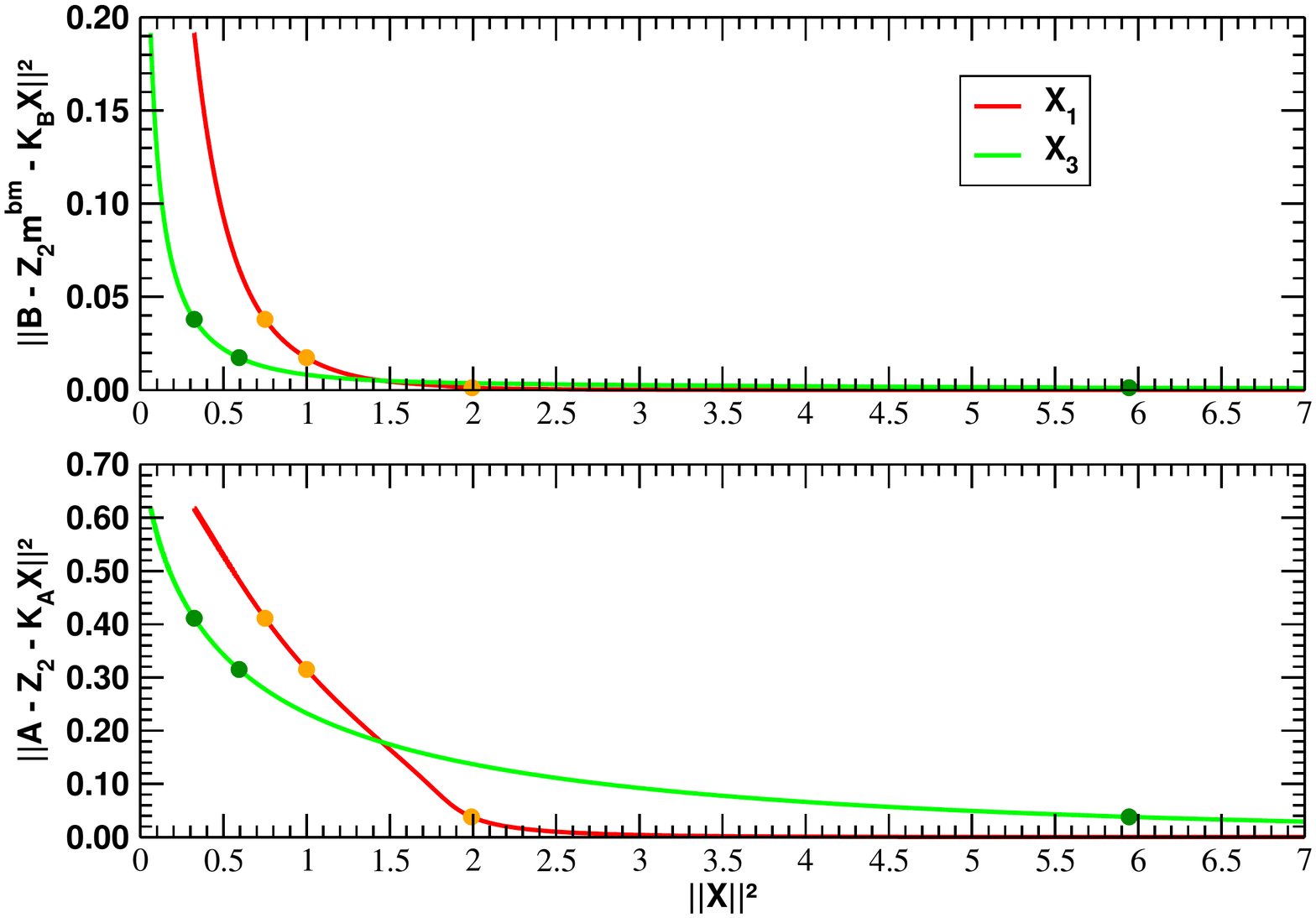} \\
      \vspace{-0.5cm}
      \includegraphics[width=3.5in]{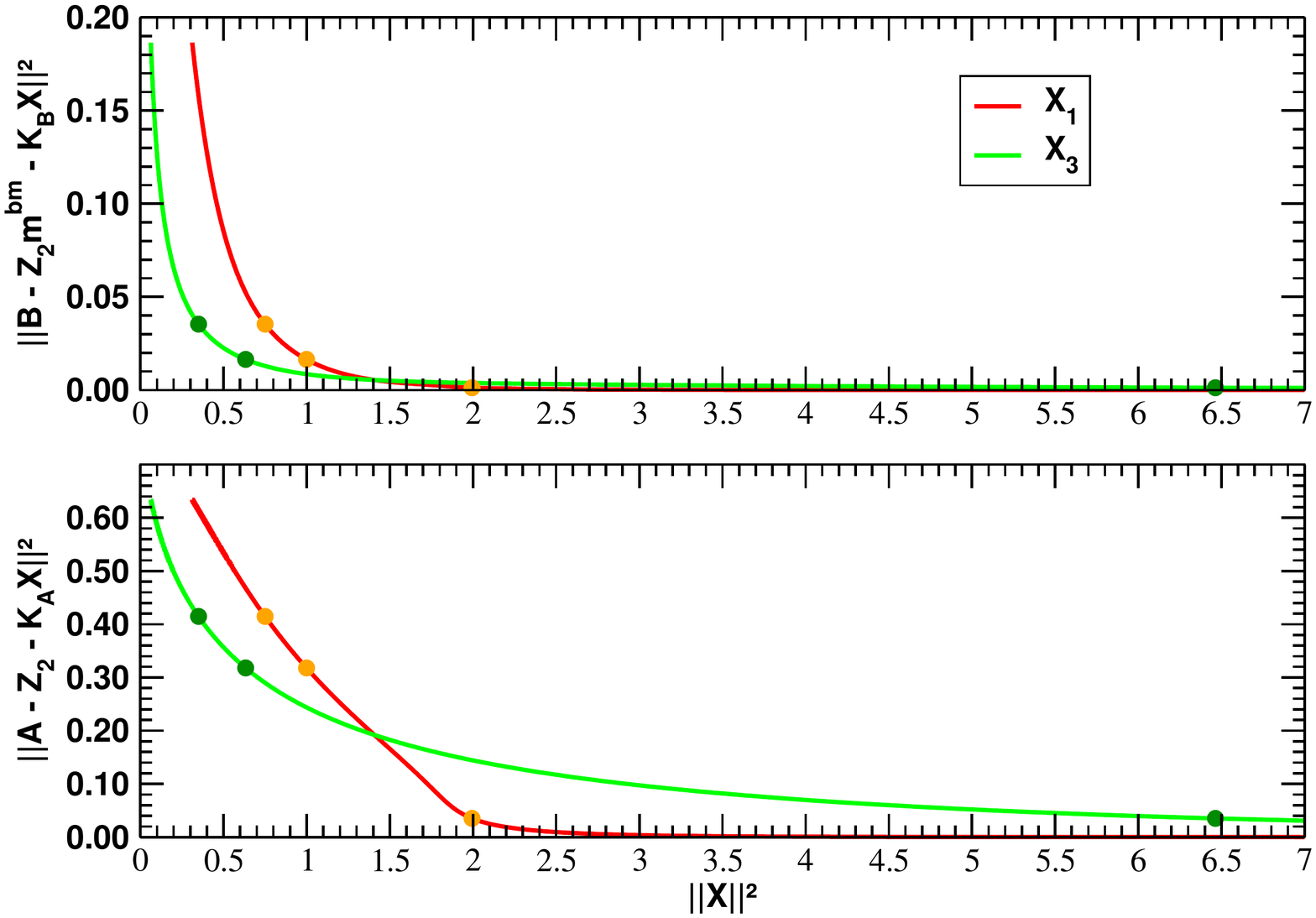}
   \caption{Residuum versus norm for the scalar and vector equation when solving the gap equation for $X_1$ and $X_3$ with $X_0$ as given by one-loop
   dressed perturbation theory. The left plot refers to the inversion using $H_1(q^2) = 1$, while the right plot are the results for the inversion using the improved
   gluon-ghost vertex. Smaller values of the regularising parameter $\epsilon$ are associated to solutions with larger norms, while larger values of $\epsilon$
   produce $Y_1$ and $Y_3$ with smaller norms. Recall that $Y_1$ has mass dimensions, while $Y_3$ is dimensionless.}
   \label{fig:X0onelooopnormresiduo}
\end{figure}

The determination of the optimal $\epsilon$ is done by solving $ \mathcal{N}^T B = ( \mathcal{N}^T \mathcal{N}  + \epsilon )X$ for various values of
$\epsilon$ and look at how $|| B - \mathcal{N} X ||^2$ and $||X||^2$ behave as a function of the regularisation parameter $\epsilon$. 
The outcome of the inversions for different $\epsilon$
can be seen on Fig.~\ref{fig:X0onelooopnormresiduo}. 
For smaller values of $\epsilon$, i.e. when one is closer to the original ill defined problem, the corresponding
solution of the linear system results on $Y_1$ and $Y_3$ with larger norms.
The larger values of the regularisation parameter $\epsilon$ are associated to solutions of the modified linear system with smaller $Y_1$ and $Y_3$ norms.
The optimal value of $\epsilon$ is given by the solution whose residuum, i.e. the difference between the lhs and the rhs of the original equations,
is among the smallest values just before the norms of $Y_1$ and $Y_3$ start to grow but without changing the residuum. 
On the above figure we point out three solutions in the region where $\epsilon$ takes approximately its optimum value.

Our first comment on Fig.~\ref{fig:X0onelooopnormresiduo} being that both the scalar and vector components of the Dyson-Schwinger equations
can be resolved with the ansatz considered, i.e. setting $X_0(p^2)$ to its one-loop dressed perturbative result and getting $Y_1(p^2)$ and $Y_3(p^2)$
from solving the modified gap equations, provided we let the norm of $Y_1$ and $Y_3$ to be large enough. Of course, for large norms $Y_1$ and $Y_3$
are free to vary over a large range of values and the solutions with smaller norms are preferred.

From Fig.~\ref{fig:X0onelooopnormresiduo}  three typical solutions close to the optimal solution, as defined previously, are identified.
For the $X_0$ perturbative solution using the tree level (TL) ghost-gluon vertex, the characteristics of these solutions are

\begin{center}
\begin{small}
   \begin{tabular}{llllll } 
             &   $||Y_1||^2$  & $||Y_3||^2$  & $||\Delta\mbox{Sca}||^2$  & $||\Delta\mbox{Vec}||^2$  & $\epsilon$   \\ 
   I (TL)  &    1.991         &  5.945          & 0.00125                           & 0.03811                            & 0.0095         \\
   II (TL)  &   0.999         &  0.594          &  0.01739                            & 0.3151                            & 0.175           \\
   III (TL)  &   0.749      &  0.324          &  0.03796                            & 0.4113                             & 0.291           
   \end{tabular}
\end{small}   
\end{center}

\noindent
for $m^{b.m.} = 6.852$ MeV, $Z_2 = 1.0016$, where $\Delta\mbox{Sca} = B - Z_2 m^{b.m.} - K_B X$, $||Y_2||$ is given in GeV,
and $\Delta\mbox{Vec} = A - Z_2 - K_A X$ with $||Y_3||$ being dimensionless. For all these solutions $||X_0 - 1||^2 = 0.24214$.
On the other hand, the characteristics of the solutions computed with the enhanced (Enh) ghost-gluon vertex given by Eq. (\ref{Eq:ghostgluonenh}) are

\begin{center}
\begin{small}
   \begin{tabular}{llllll } 
                  &   $||Y_1||^2$  & $||Y_3||^2$  & $||\Delta\mbox{Sca}||^2$  & $||\Delta\mbox{Vec}||^2$  & $\epsilon$   \\ 
   I (Enh)    &    1.994        &  6.464          &  0.001257                         & 0.03526                           & 0.0085           \\
   II (Enh)   &    0.999        &  0.633          &  0.01644                           & 0.3178                            & 0.17          \\
   III (Enh)  &    0.745       &  0.350          &  0.03536                            &0.4146                            & 0.28           
   \end{tabular}
\end{small}   
\end{center}

\noindent
have the same $m^{b.m.}$ and $Z_2$ as the previous ones and $||X_0 - 1||^2 = 0.45283$. In both cases, the norms of $X_0$, $Y_1$ and $Y_3$ are the norms of the 
corresponding part of the vector that appear in the linear system .

\begin{figure}[t] 
   \centering
   \vspace{-0.3cm}
      \includegraphics[width=3.5in]{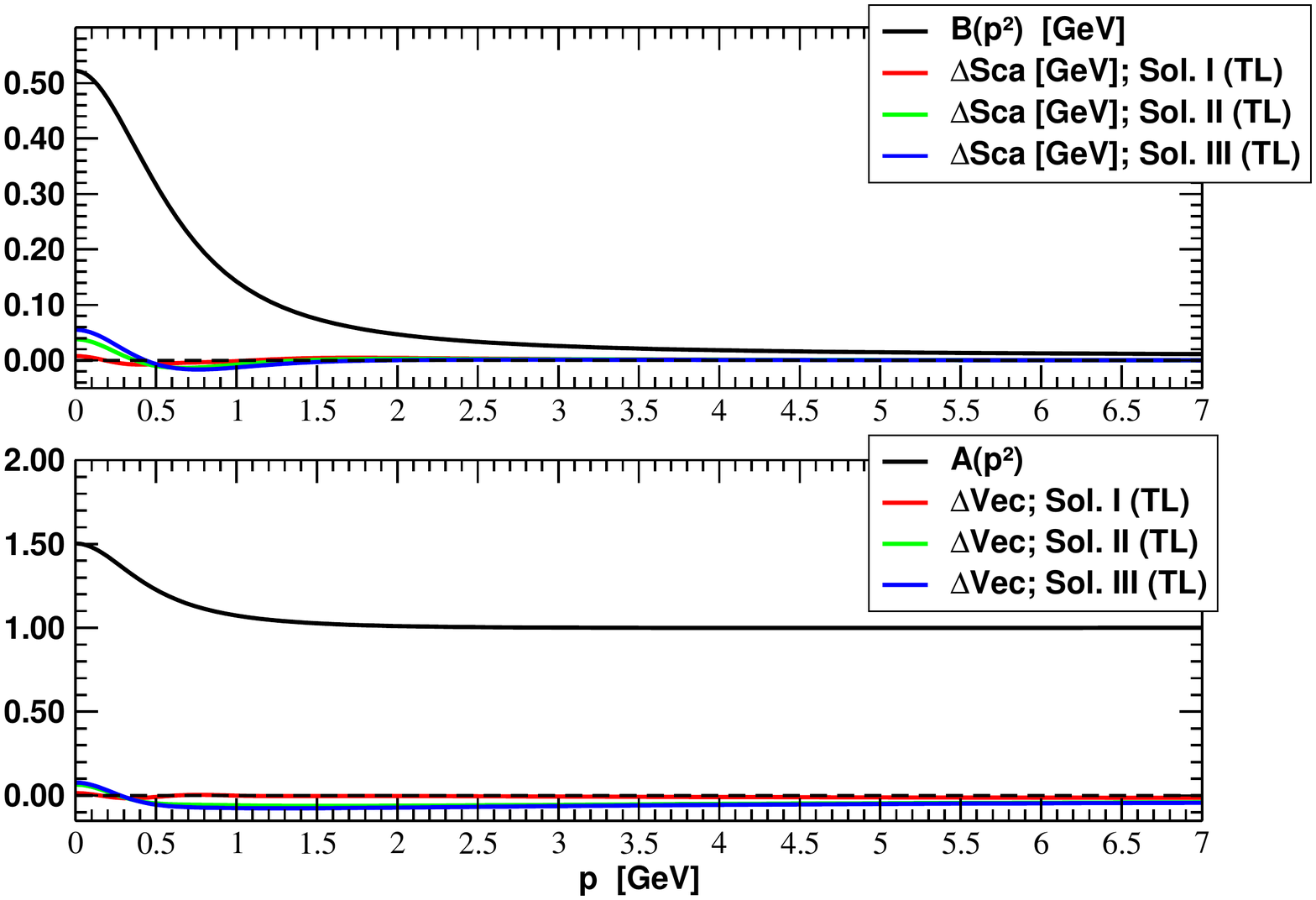} \\
      \vspace{-0.5cm}
      \includegraphics[width=3.5in]{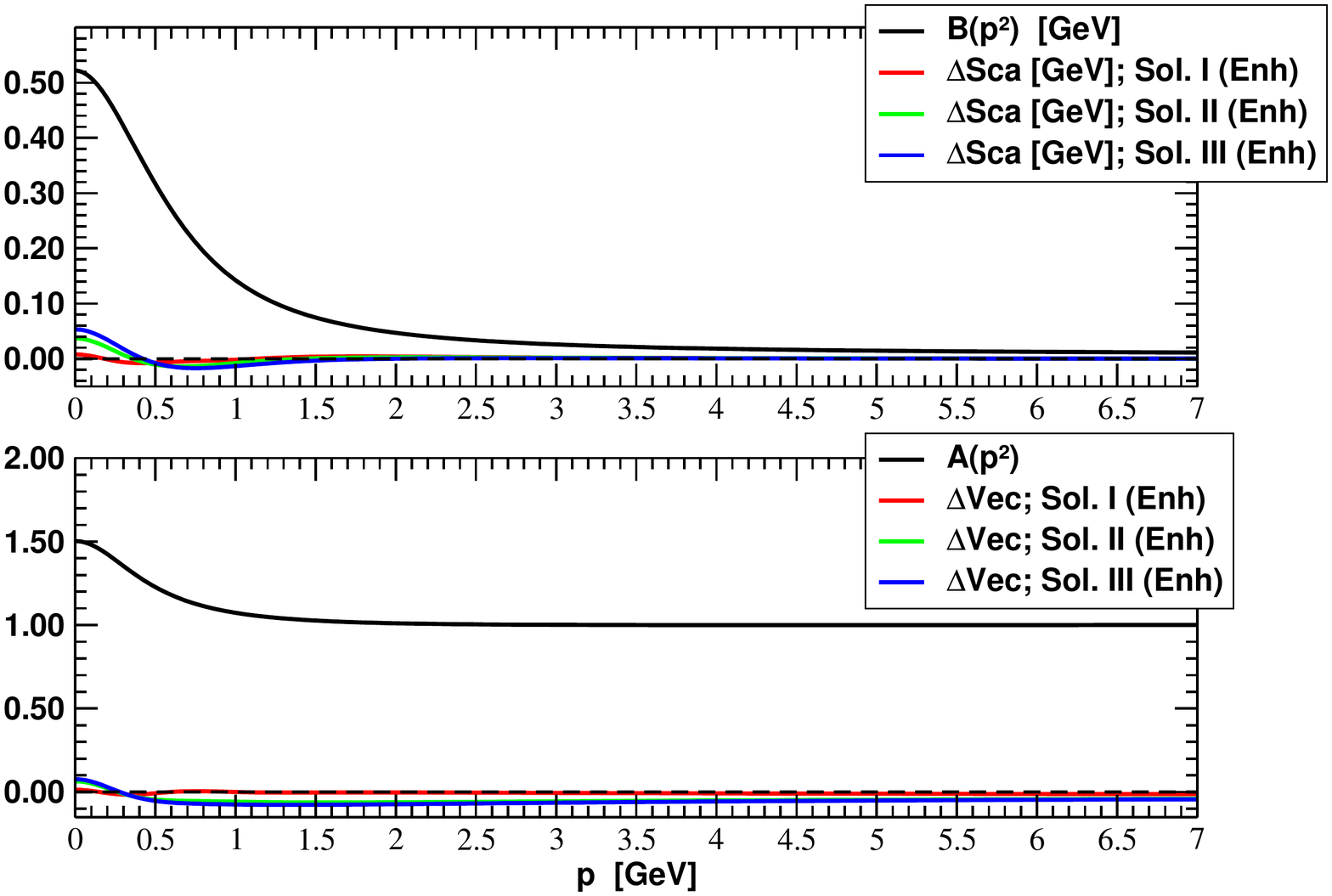}
   \caption{Scalar (top) and Vector (bottom) Dyson-Schwinger equations and the differences between its lhs and rhs when using the tree level ghost-gluon vertex (top)
   and the enhanced gluon-ghost vertex (bottom).}
   \label{fig:X0onelooopDSE}
\end{figure}

\begin{figure}[t] 
   \centering
   \vspace{-0.3cm}
      \includegraphics[width=3.5in]{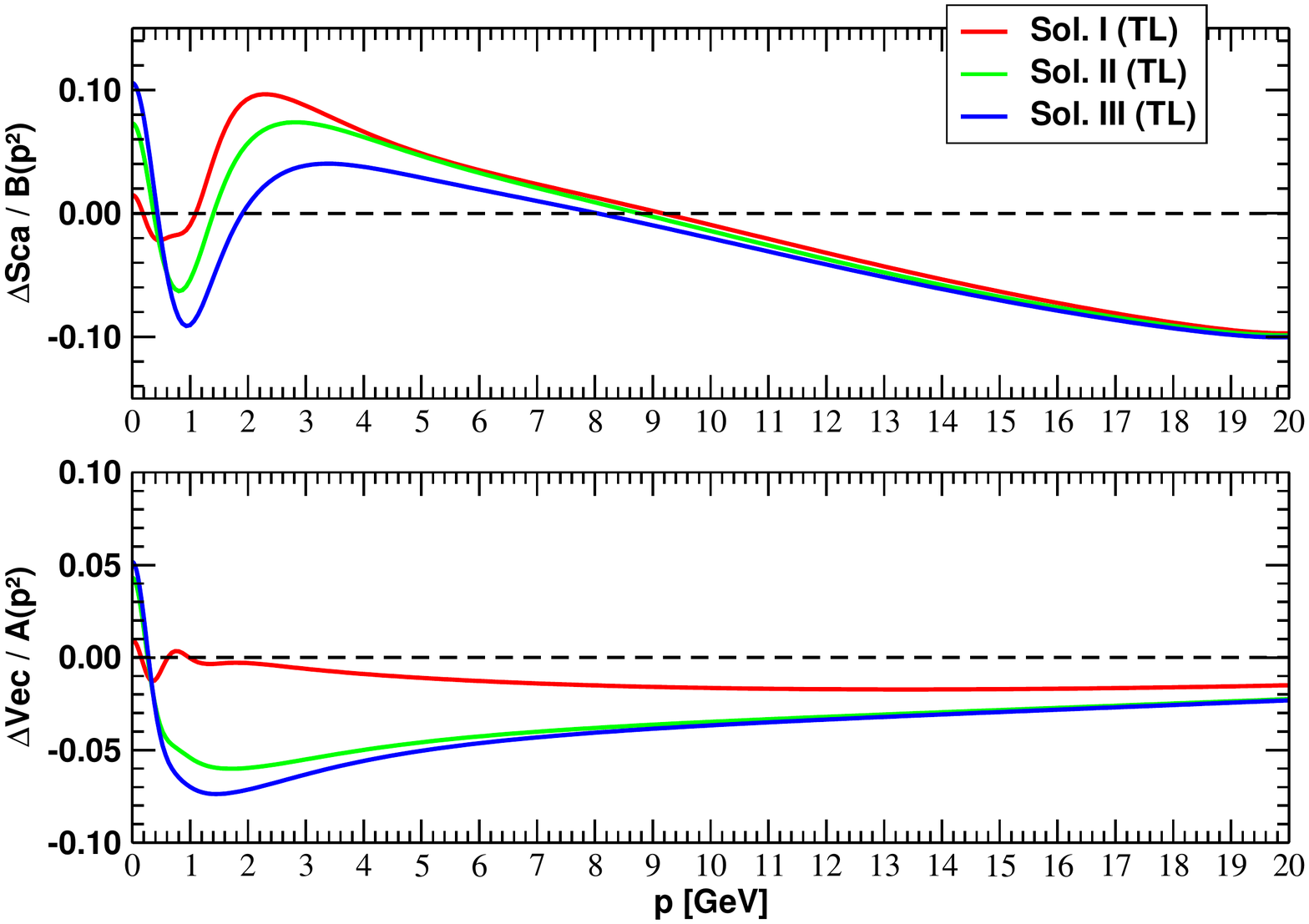} \\
      \vspace{-0.5cm}
      \includegraphics[width=3.5in]{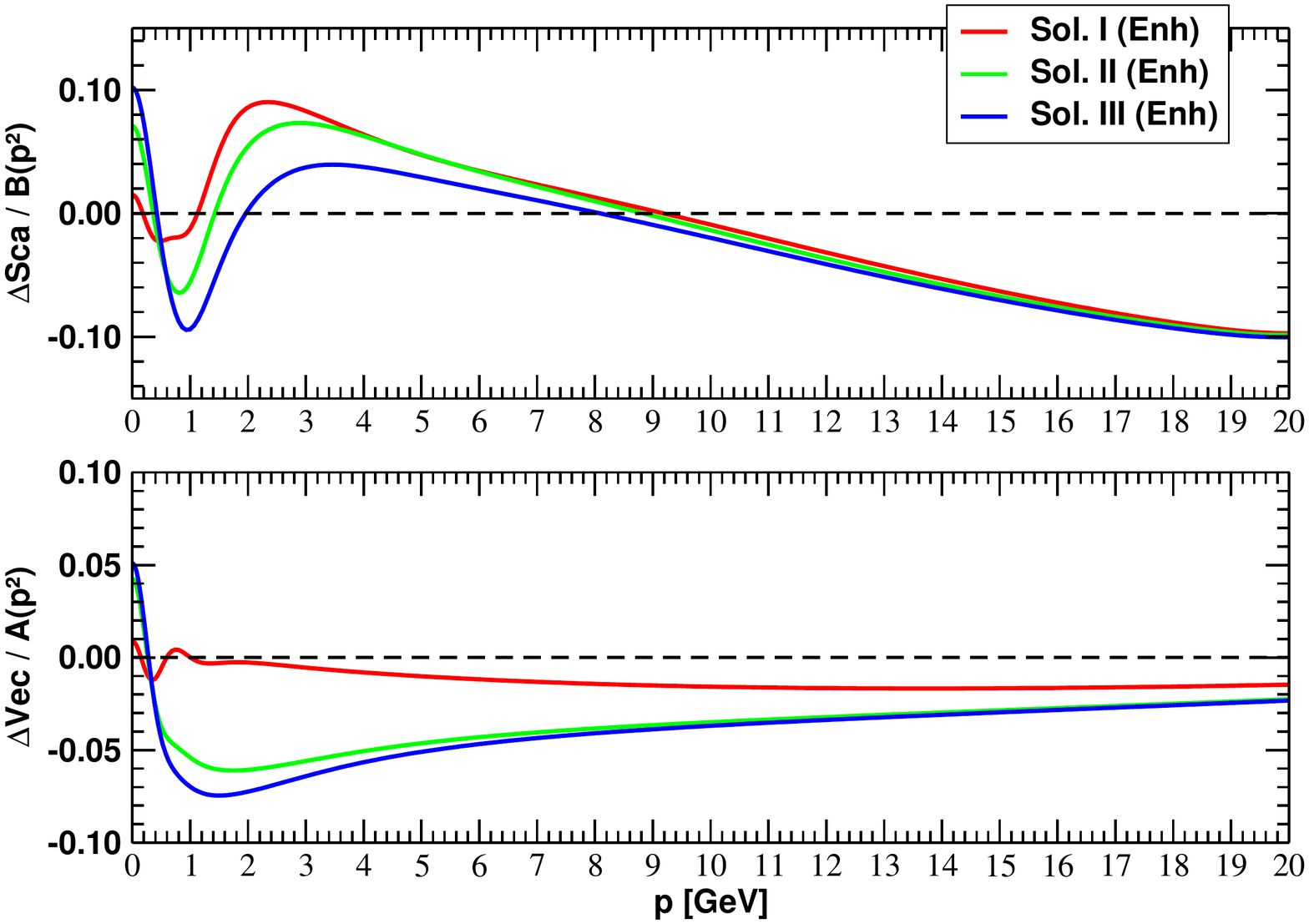}
   \caption{Relative error on the solution of the Scalar (top) and Vector (bottom) Dyson-Schwinger equation when using the tree level ghost-gluon vertex (top)
   and the enhanced gluon-ghost vertex (bottom).}
   \label{fig:X0onelooopDSEErrRel}
\end{figure}

The quality of the solutions can be appreciated on
Fig.~\ref{fig:X0onelooopDSE} where we show both the l.h.s. of the scalar and vector components of the gap equation, 
together with the difference between the l.h.s. and the computed r.h.s. using the $X_0$ from one-loop perturbation theory and $Y_1$ and $Y_3$ that solve the modified linear system.
The relative error both for the scalar and vector components of the Dyson-Schwinger equations are shown on Fig.~\ref{fig:X0onelooopDSEErrRel}. On the figures we have defined
\begin{eqnarray}
\Delta \mbox{Sca} & = & B - (Z_2 m^{b.m.} + K^{(0)}_B X_0 + K^{(1)}_B Y_1 + K^{(3)}_B Y_3 ) \ ,  \label{Eq:DefSca} \\
\Delta \mbox{Vec} & = & A - (Z_2 + K^{(0)}_A X_0 + K^{(1)}_A Y_1 + K^{(3)}_A Y_3 ) \ . \label{Eq:DefVec} 
\end{eqnarray}
$||\Delta\mbox{Sca}||^2$  and $||\Delta\mbox{Vec}||^2$  should be understood as the sum of the squares of the components of the linear systems
(\ref{Eq:DefSca}) and (\ref{Eq:DefVec}), respectively, over the Gauss-Legendre points.
As Fig.~\ref{fig:X0onelooopDSEErrRel} shows, the relative error on the DSE equations is below 10\% for the scalar equation and below 8\% for the vector
equation. Surprisingly, despite the larger values of $||\Delta\mbox{Vec}||^2$ relative to $||\Delta\mbox{Sca}||^2$, the vector component of the gap equation
is better resolved. This is also due to the fact that $A(p^2)$ spans a narrower range of values relative to $B(p^2)$. We have tried to rescale the linear system
by $1/A(p^2)$ for the vector equation and by $1/B(p^2)$ for the scalar equation to try to improve the quality of the solutions, specially at large momentum.
However, the numeric solutions of the rescaled linear systems produced $Y_1$ and $Y_3$ that don't seem reasonable and, for example, result in
a $Y_1$ at the cutoff that is far away from zero. Further, for the rescaled systems the $\Delta \mbox{Sca}$ and $\Delta \mbox{Vec}$ are larger than
the ones obtained without rescaling the linear system. For all these reasons we disregard the rescaled linear system solutions.

\begin{figure}[t] 
   \centering
   \vspace{-0.3cm}
      \includegraphics[width=3.5in]{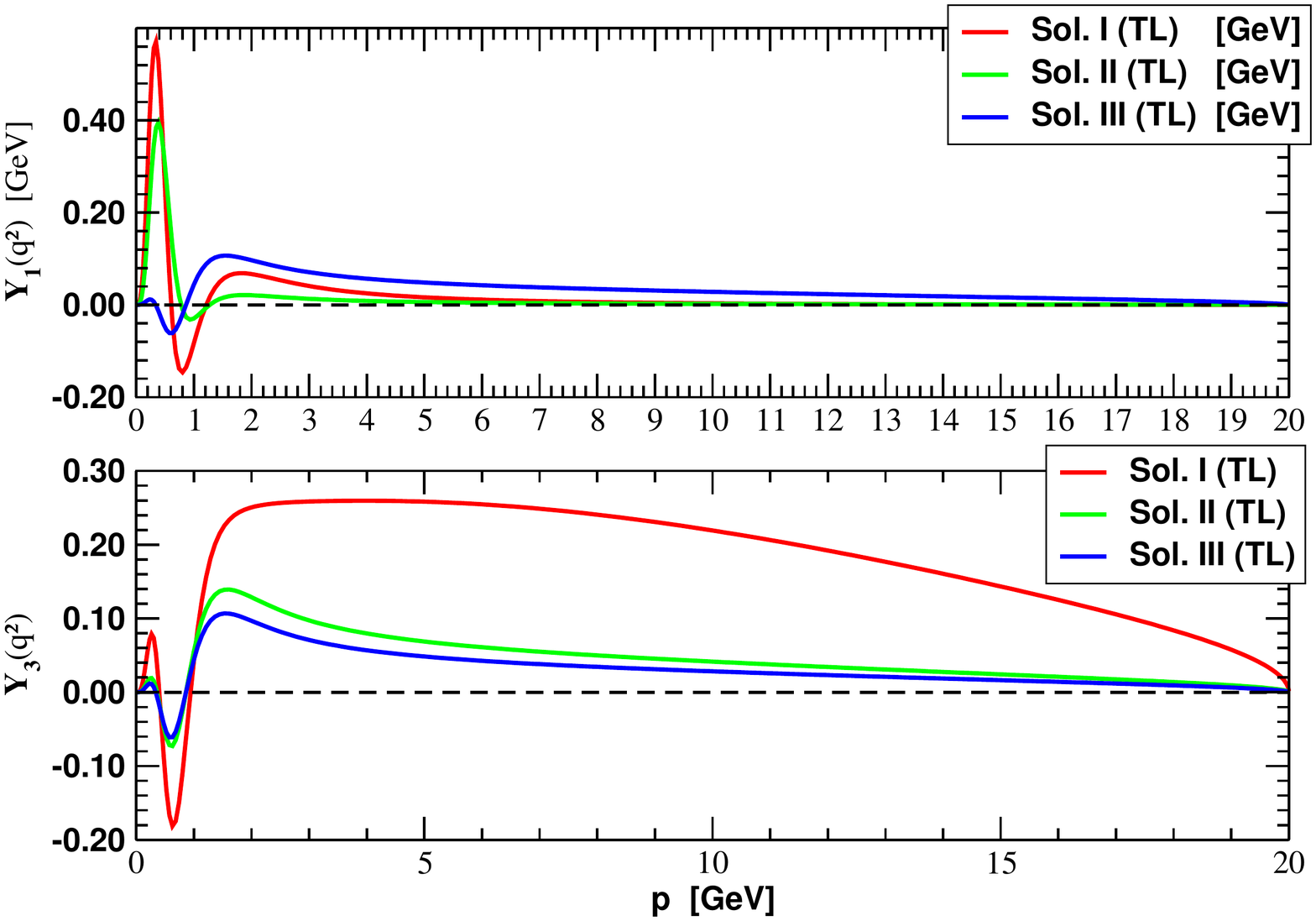} \\
      \vspace{-0.5cm}
      \includegraphics[width=3.5in]{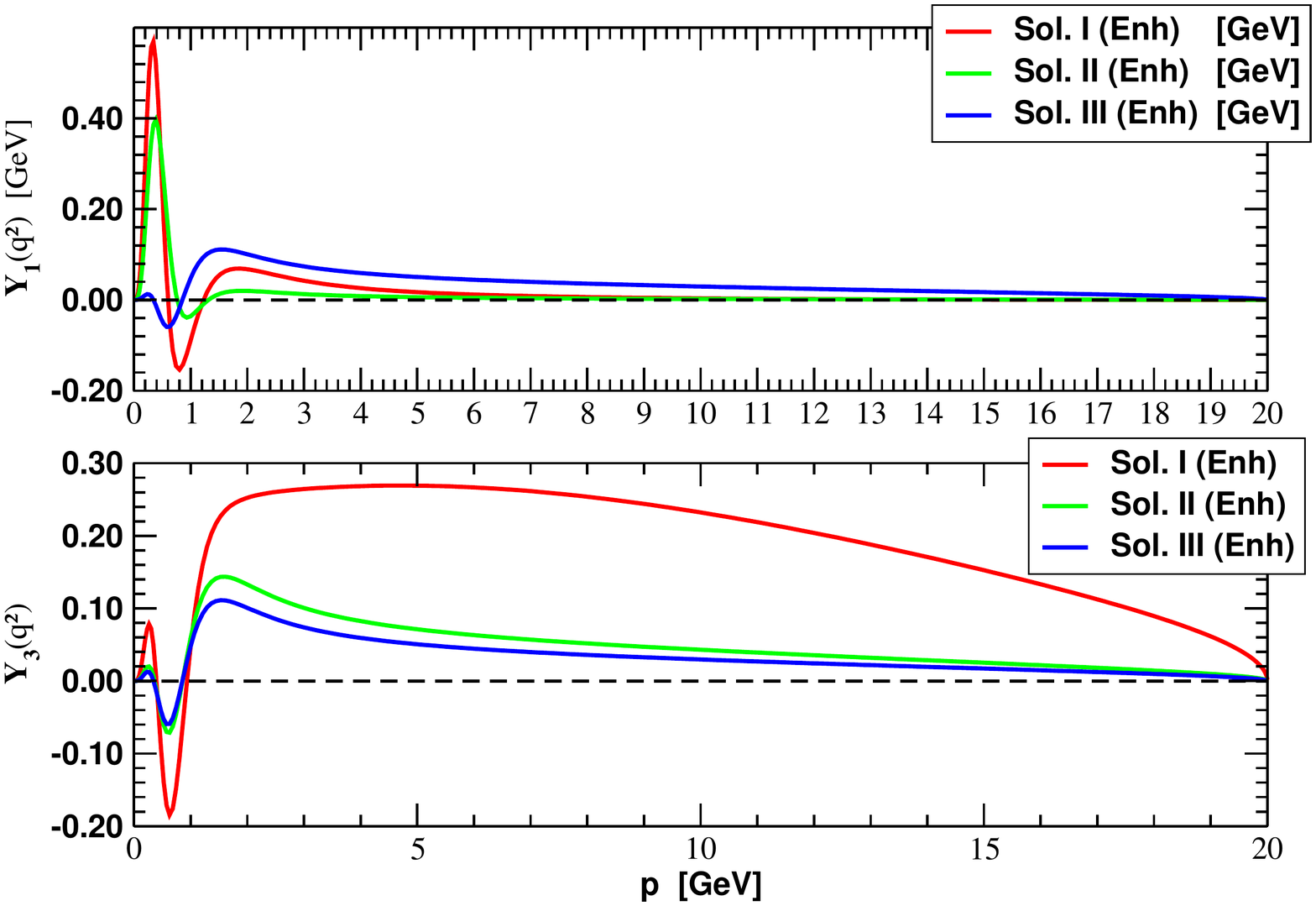}
   \caption{The quark-ghost kernel form factors $X_1$ and $X_3$ computed for the tree level ghost-gluon vertex (top) and the enhanced gluon-ghost vertex (bottom).}
   \label{fig:X0onelooopX1X3}
\end{figure}

The quark-ghost kernel form factors $Y_1$ and $Y_3$ computed for the various $\epsilon$ associated to the solutions I (TL) -- III (TL) and I (Enh) -- III (Enh)
can be seen on Fig~\ref{fig:X0onelooopX1X3}. For $Y_1$ the outcome of resolving the integral equations using either the tree level or the enhanced
ghost-gluon vertex result on essentially the same function. Further, the various solutions provide essentially the same $Y_1(p^2)$, with the exception of
III (TL) and III (Enh) that return a suppressed form factor relative to the other solutions. For $Y_3$ the situation is similar, with the form factor associated
to the solutions I (TL) and I (Enh) being enhanced at momentum above 2 GeV. Looking at Fig.~\ref{fig:X0onelooopDSEErrRel}, one can observe that
solutions II (TL) and II (Enh) are those with smaller relative errors over the full range of momentum considered. So, from now on we will take these
solutions as our best solutions associated to the perturbative $X_0$ form factor. Note that the scalar equation is solved with a relative error
$\lesssim$ 8\% and the vector equation is solved with a relative error $\lesssim$ 6\%.

The quark-ghost kernel
form factors $X_0$, $X_1$ and $X_3$ were also computed in~\cite{Aguilar:2018epe}, see their Fig. 4, combining the quark gap equation with a one-loop
dressed perturbation theory for the quark-ghost kernel. The comparison between the results of the two calculations is not straightforward. 
Indeed, if in our calculation $X_0$ is assumed to be a function only of the gluon momentum, in~\cite{Aguilar:2018epe} the authors take into account 
its full momentum dependence and evaluate how it changes with the quark momentum, the gluon momentum and the angle between these two momenta. 
There calculation results on a $X_0$ that is always close to its tree level value $X_0 = 1$ and whose values are in the range $ [1 , 1.12 ]$.
A qualitative comparison with the here reported form factor, shows that, from the point of view of the dynamical range of values, our $X_0$ computed using the 
enhanced ghost-gluon vertex is closer to that reported in~\cite{Aguilar:2018epe}. In both cases $X_0$ is always close to its tree level value and differs from unit by, at most,
10\%.

The comparison between the remaining form factors is slightly more involved. Indeed, the $X_1$, $X_2$ and $X_3$ referred in~\cite{Aguilar:2018epe}
compared with the expressions given in Eqs. (\ref{Eq:AnsatzX1}) and (\ref{Eq:AnsatzX3})  and not directly with the $Y_1$ and $Y_3$ reported in 
Fig.~\ref{fig:X0onelooopX1X3}. Note that there are signs differences on the definition of the various quark-ghost kernel form factors between~\cite{Aguilar:2018epe}
and the current work.
Due to the presence of the gluon propagator in (\ref{Eq:AnsatzX1}) and (\ref{Eq:AnsatzX3}) one expects some angular dependence of the quark-ghost kernel form factors.
Further, due to the parameterisation used for the gluon propagator, the form factors are expected to be enhanced when, simultaneous, the quark and gluon momenta 
become smaller, i.e. in the infrared limit.
This is precisely what is observed for the $X_1$ and $X_2$ reported in~\cite{Aguilar:2018epe}.

\begin{figure}[t] 
   \centering
   \vspace{-0.3cm}
      \includegraphics[width=3.5in]{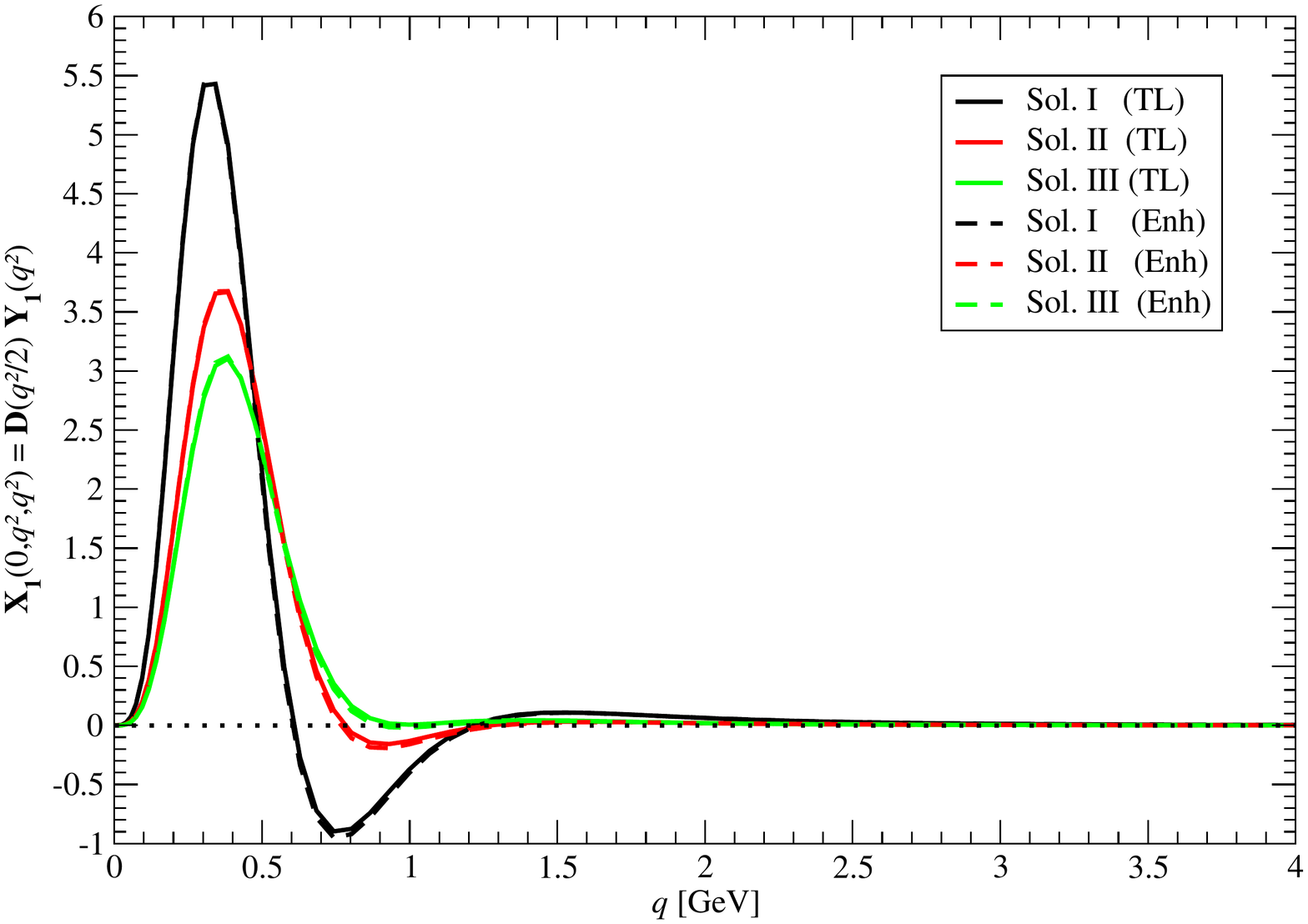} \\
      \vspace{-0.5cm}
      \includegraphics[width=3.5in]{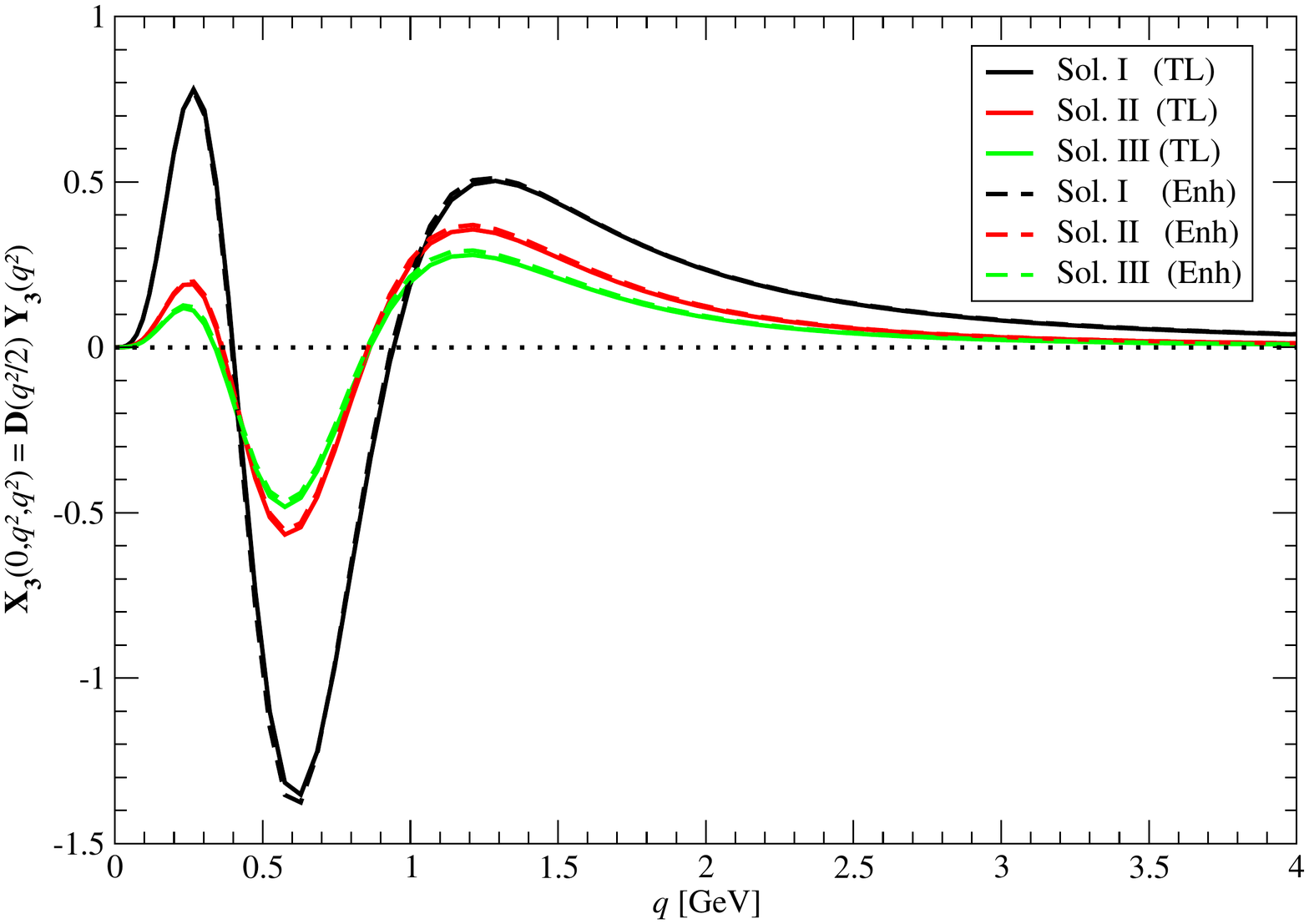}
   \caption{The quark-ghost kernel form factors $X_1$ (top) and $X_3$ (bottom) as defined in Eqs. (\ref{Eq:AnsatzX1}) and (\ref{Eq:AnsatzX3}) for the soft quark limit defined by
                  a vanishing quark momentum.}
   \label{fig:TrueX1X3Pert}
\end{figure}

Our estimations for $X_1$ and $X_2$ for the solutions referred previously and for the particular kinematics $p = 0$, sometimes called the soft quark limit, can be seen 
on Fig.~\ref{fig:TrueX1X3Pert}. The first comment about these form factors is that they seem to be independent of the ghost-gluon vertex and, indeed,
the form factors associated to the solution using a tree level ghost-gluon vertex, named (TL) in the figure, are essentially indistinguishable from those computed
using the enhanced ghost-gluon vertex, named (Enh) in the figure.
The quark-ghost kernels form factors $X_1$ and $X_3$ differ from there perturbative values at low momenta, i.e. for $q \lesssim 1$ GeV for $X_1$ and
for $q \lesssim 2$ GeV for $X_3$. According to~\cite{Aguilar:2018epe}, these two form factors increase (in absolute value) for momenta
below $\sim 1$ GeV, a result that is in good qualitative agreement with our estimations. 
From their Fig. 4, it is not clear if $X_3 \ne 0$ extends over a wider range of momenta, when compared to $X_1$. Our calculation returns a $X_3$ that differ from zero on
larger range of momenta, when compared with $X_1$. 
Furthermore, the $X_1$ and $X_3$ computed in~\cite{Aguilar:2018epe} are monotonic increase functions (absolute values) when the zero momentum
limit is approached. The form factors reported on Fig.~\ref{fig:TrueX1X3Pert} show a pattern of maxima, with $X_1$ having a single maximum for $q \sim 350$ MeV for $X_1$ and
$X_3$ showing several maxima (in absolute value) at momenta $q \sim 250$ MeV, $\sim 600$ MeV and $\sim 1$ GeV. Note that the zero crossing for $X_3$ occur
for momenta that are of the same order of magnitude of $\Lambda_{QCD}$, the gluon mass (or twice $\Lambda_{QCD}$) and the usual considered as a non-perturbative
mass scale (1 GeV). It is not obvious why the zero crossing of $X_3$ occur for such mass scales and why the crossing is not
seem for $X_1$. If the form factors reported on Fig. 4 of~\cite{Aguilar:2018epe} never cross the zero value, that is not the case of the form factors represented on Fig ~\ref{fig:TrueX1X3Pert}. $X_1$ shows various zeros that, curiously, seem to disappear for the solution with the smaller norm.
On the other hand, $X_3$ cross  zero  for $q \approx 363$ MeV and $835$ MeV for solutions II and III. This is a major difference between the two sets of
solutions under discussion. Another important difference being the dynamical range of values. Our $X_1$ is within the range of values $[ 0 , 3.5]$ GeV$^{-1}$, while the same
form factor computed in~\cite{Aguilar:2018epe} is within $[ 0 , 0.2]$ GeV$^{-1}$ which represents a factor of $\sim 20$ smaller than our estimation.
However, the two calculations report a $|X_3|$ within the range $[ 0 , 0.5]$ GeV$^{-2}$. Our result overestimates $X_1$ relative to~\cite{Aguilar:2018epe} but
returns a $|X_3|$ within the same dynamical range of values. Another major difference being that the maxima of the form factors does not occur at vanishing momenta
as in~\cite{Aguilar:2018epe} but at finite and small momenta, $\sim \Lambda_{QCD}$ for $X_1$ and $\sim 2 \, \Lambda_{QCD}$ for the absolute maxima of $|X_3|$.

The estimations of $X_0$, $X_1$ and $X_3$ suggest that the quark-gluon vertex is dominated, at the infrared by those terms that are associated with
$X_1$. If this is the case, then, given the definitions (\ref{Vertex:L1}) -- (\ref{Vertex:L4}) one expects that the dominant contributions to the quark-gluon vertex
to be associated with the form factors
\begin{eqnarray}
& & \lambda_1 (-p, \, p-q, \, q )  \cong  F(q^2)  \, \bigg[ B(p^2)  + B( k^2 ) \bigg] \times \nonumber \\
 & & \hspace{3cm} \times ~ D\left( \frac{p^2 + (p-q)^2}{2} \right) \, Y_1(q^2) ,  \\
& & \lambda_3 (-p, \, p-q, \, q )  \cong  \nonumber \\
& & \hspace{0.25cm}  \cong  F(q^2) \, \frac{ A( k^2 ) \, \bigg( q^2 -   (p q) \bigg) - A(p^2) \, ( p q ) }{q^2 - 2 (p \cdot q)}  \times \nonumber \\
& & \hspace{3cm} \times ~D\left( \frac{p^2 + (p-q)^2}{2} \right) \, Y_1(q^2)  , 
\\
& & 
 \lambda_4 (-p, \, p-q, \, q ) \cong  \nonumber \\
 & & \hspace{0.25cm} \cong \frac{F(q^2)}{2} ~ ~  \Delta A ~ ~ D\left( \frac{p^2 + (p-q)^2}{2} \right)  \, Y_1(q^2)
\end{eqnarray}
and since $Z \sim 1$ and $B \propto M(p^2)$, one expects the dominant form factor for the quark-gluon vertex to be associated with the tree level operator, i.e. with $\lambda_1$.

\subsection{Solving the Dyson-Schwinger Equations \label{Sec:solveall}}

Let us now discuss the simultaneous computation of $X_0$, $X_1$ and $X_3$ from the modified linear system of equations that replace the 
original Dyson-Schwinger integral equations. The procedure to build the linear system as well as the regularisation of the
corresponding linear system of equations follow the steps described in the previous section.
First the angular integration is performed using 800 Gauss-Legendre points. 
Then, for the momentum integration the cutoff $\Lambda = 20$ GeV is introduced and  we consider 200 Gauss-Legendre points to perform 
the integration over the loop momentum. Further, to determine the solutions of the Dyson-Schwinger equations, now already in the form of a 
linear system of equations, the scalar component and the vector component of the gap equation are grouped into a large linear system as follows
\begin{equation}
\left( \begin{array}{c}   B(p) - Z_2 m^{b.m.}  \\  \\ A(p) - Z_2 \end{array} \right) =
\left( \begin{array}{c@{\hspace{0.5cm}}c@{\hspace{0.5cm}}c}   
                             \mathcal{N}^{(0)}_B     &   \mathcal{N}^{(1)}_B   & \mathcal{N}^{(3)}_B \\ 
                             & \\
                              \mathcal{N}^{(0)}_A     &   \mathcal{N}^{(1)}_A    &   \mathcal{N}^{(3)}_A \end{array} \right) 
\left( \begin{array}{c}   X_0 \\  Y_1  \\ Y_3 \end{array} \right)                              
\label{Eq:DSE_FullLinearSystem}
\end{equation}
that again we refer, as a short name, by $B = \mathcal{N} X$. The upper  component of the large vector $X$ contains
the form factor $X_0$ defined in all the set of Gauss-Legendre points used in the integration over the loop momenta.
The remaining components of the large $X$ vector are the form factors $Y_1$ and $Y_3$ defined at the lower first half set of Gauss-Legendre points 
used in the integration over the momentum. This means that the solution of the linear system (\ref{Eq:DSE_FullLinearSystem}) returns
$X_0(q^2)$ for $q \in [ 0 , \, \Lambda]$ and $Y_1(q^2)$ and $Y_3(q^2)$ for $q \in [ 0 , \, \Lambda / 2]$. 
For $Y_1$ and $Y_3$ and for $p > \Lambda/2$ the form factors will be assumed to vanish.
In order to fulfil the boundary conditions for $X_0 (q^2)$ we write $X_0 (q^2) = 1 + \tilde{X}_0(q^2)$ and solve the linear system for $\tilde{X}_0(q^2)$, 
rebuilding $X_0 (q^2)$ at the end.
The resulting linear system is then regularised using the Tikhonov regularisation  and the corresponding 
$\mathcal{N}^T B = ( \mathcal{N}^T \mathcal{N}  + \epsilon )X$ linear system is solved for various $\epsilon$.
The choice of the optimal regularisation parameter $\epsilon$ follows the criteria discussed in Sec.~\ref{Sec:OneLoopSolution}. 
We have checked that by interchanging the roles of $X_0$, $Y_1$ and $Y_3$ when building the large linear system the solutions are unchanged;
more on that below. The differences only occur for those functions calculated only for $q \in [ 0 , \, \Lambda / 2]$, 
compared to the version of the linear system were they are computed in the range $q \in [ 0 , \, \Lambda ]$.
In the first case, i.e. for the solutions computed only for $q \in [ 0 , \, \Lambda / 2]$, there appears a discontinuity at $q = \Lambda /2$ 
(recall that the form factors are set to zero for momenta above $\Lambda/2$) but for smaller $q$ the form factors of all versions of the linear system 
are indistinguishable.

\begin{figure}[t] 
   \centering
   \vspace{-0.3cm}
      \includegraphics[width=3.5in]{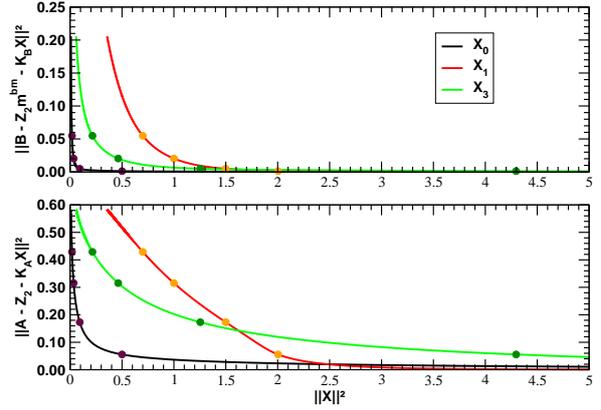}
   \caption{Residuum versus norm for the scalar and vector equation when solving the gap equation for $X_0$, $Y_1$ and $Y_3$.
                The smaller values of the regularising parameter $\epsilon$ are associated to solutions with larger norms, while larger values of $\epsilon$
                produce form factors with smaller norms. Recall that $Y_1$ has mass dimensions, while $X_0$ and $Y_3$ are dimensionless.}
   \label{fig:XFull_normresiduo}
\end{figure}

On Fig.~\ref{fig:XFull_normresiduo} the residuum squared of the scalar (top) and vector (bottom) components of the gap equation are shown against the norm
of various form factors. Smaller values of $\epsilon$ are associated to solutions with larger norms and appear at the right side of the plots, while larger values
of $\epsilon$ are associated to solutions with smaller norms that show up on the left side of the plots. As shown on the figure, the residuum for both equations
has a stronger dependence on $\epsilon$ and can take quite small values. For the solutions featured in the plot, the smallest residuum squared reaches values of
the order of $10^{-5}$ for the scalar equation and $10^{-4}$ for its vector component. Similar values for the minimum residuum were also observed for the 
solutions computed using the perturbative estimation of $X_0$.

On Fig.~\ref{fig:XFull_normresiduo} we identify four solutions associated to an $\epsilon$ around its optimal value and whose characteristics are

\begin{center}
\begin{small}
   \begin{tabular}{lllllll } 
             & \!\!\!\!\!\!   $||X_0 - 1||^2$ & \!\!\!\!\!\!  $||Y_1||^2$  & \!\!\! $||Y_3||^2$  & $||\Delta\mbox{Sca}||^2$  & $||\Delta\mbox{Vec}||^2$  & $\epsilon$   \\ 
   I         & \!\!\!\!\!\!   0.5003             & \!\!\!\!\!\!  2.0039         & \!\!\! 4.2966         & 0.001205                           & 0.05568                           & 0.012   \\
  II         & \!\!\!\!\!\!   0.09298           & \!\!\!\!\!\!  1.4994         & \!\!\! 1.2538          & 0.004890                           & 0.1730                             & 0.071   \\
  III        & \!\!\!\!\!\!   0.03408           & \!\!\!\!\!\!  1.0002         & \!\!\! 0.4634         & 0.02034                             & 0.3154                             & 0.191   \\
  IV        &\!\!\!\!\!\!    0.01811           & \!\!\!\!\!\!  0.7001         & \!\!\! 0.2150         & 0.05486                             & 0.4291                             & 0.365   \\
   \end{tabular}
\end{small}   
\end{center}

\noindent
for $m^{b.m.} = 6.852$ MeV, $Z_2 = 1.0016$, where $\Delta\mbox{Sca}$ and $||Y_1||$ are  given in GeV, while $\Delta\mbox{Vec}$, $||X_0||$ and $||Y_3||$
are dimensionless.

\begin{figure}[t] 
   \centering
   \vspace{-0.3cm}
      \includegraphics[width=3.5in]{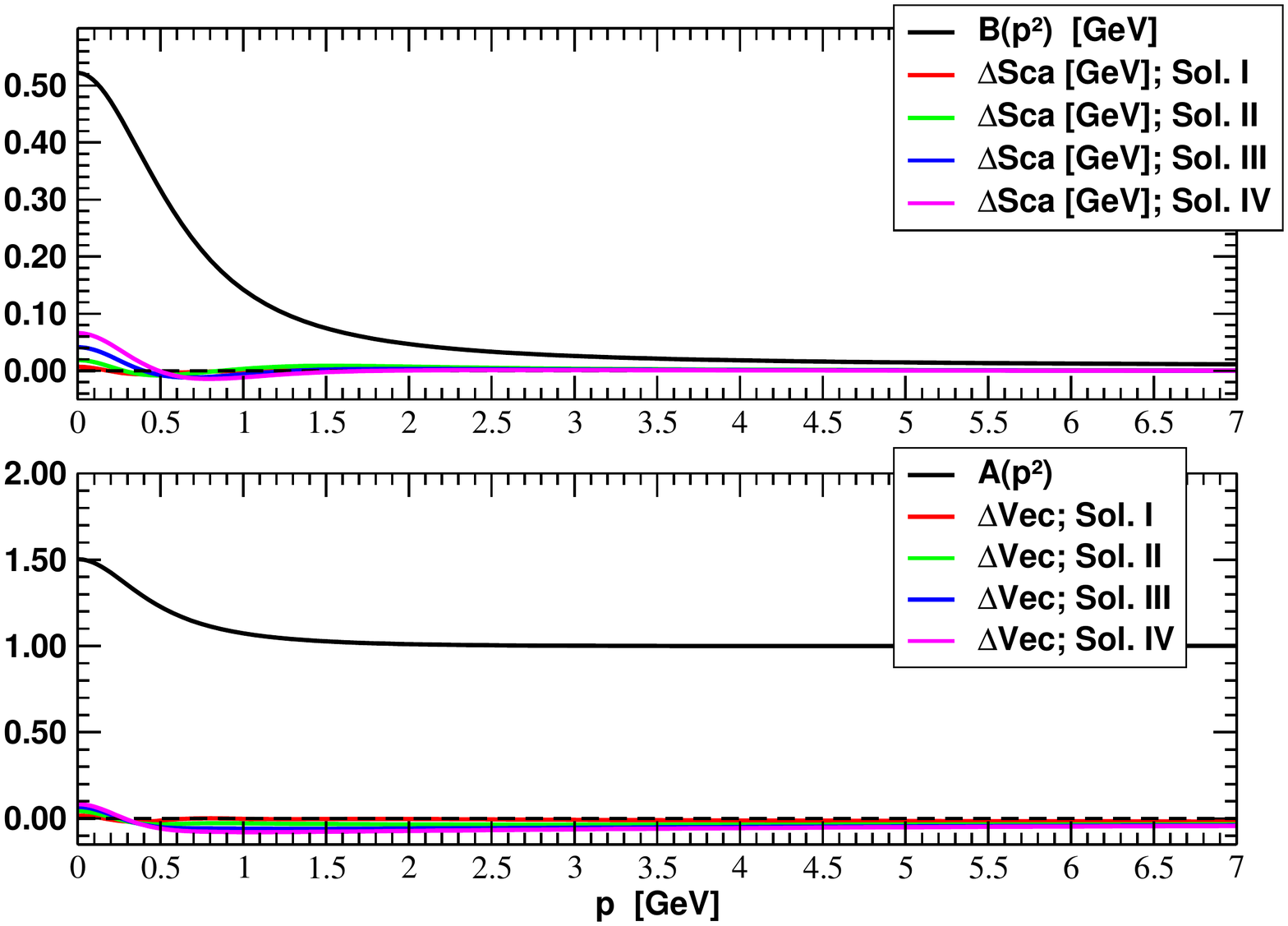} \\
      \vspace{-0.5cm}
      \includegraphics[width=3.5in]{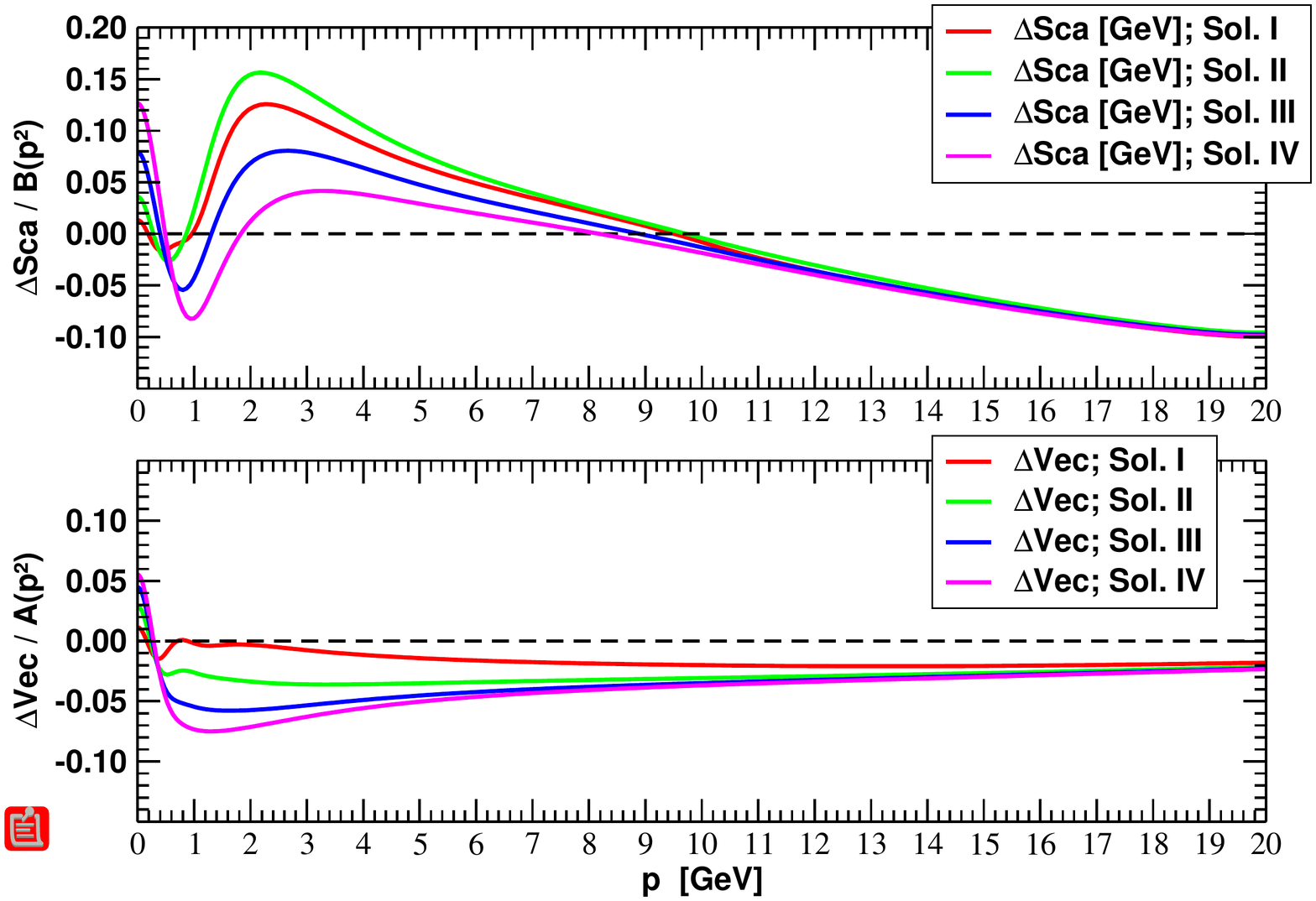}
   \caption{The lhs of the Dyson-Schwinger equations, scalar component on top and vector component at bottom, together with $\Delta\mbox{Sca}$ and
                 $\Delta\mbox{Vec}$ (top). The plots on the left are the relative error for the scalar equation (top) and vector equation (bottom). }
   \label{fig:XFull_DSE_scalar_vector}
\end{figure}

The relative errors for the solutions I -- IV of the regularised linear system are show on Fig.~\ref{fig:XFull_DSE_scalar_vector}.
In general the solution for the vector component of the equation is satisfactory, with the scalar component of the equation being more demanding and
not all of the solutions I -- IV resolve the scalar part of the gap equation with a relative error below 10\%. Only solutions III and IV resolve the DSE
equations with a relative error below 8\%. In particular for these solutions the value of $||X_0 - 1||$ is of the order of $10^{-2}$ suggesting that the
non-perturbative solution prefers having a $X_0 \simeq 1$ and, in this sense and for this form factor, are close to the result from perturbation theory
discussed in Sec.~\ref{Sec:OneLoopSolution}. The observed growth of the relative error for $p \gtrsim 10$ GeV is probably related also to the
missing components of $Y_1$ and $Y_3$ which are set to zero for this range of momenta.

\begin{figure}[t] 
   \centering
   \vspace{-0.3cm}
      \includegraphics[width=3.5in]{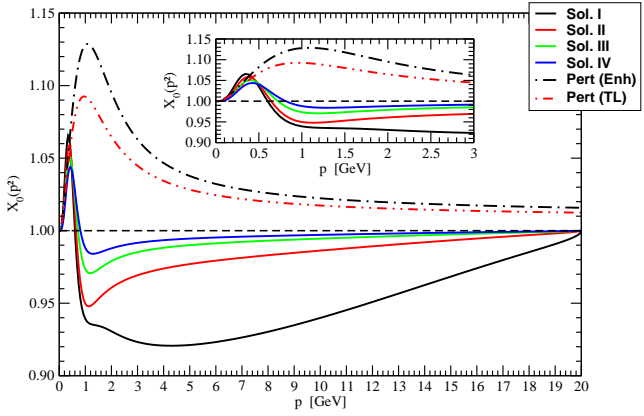}
   \caption{$X_0(p^2)$ from inverting the Dyson-Schwinger equations  together with its estimation using one-loop dressed perturbation theory by solving 
   exactly Eq. (\ref{Eq:X0oneloop}).}
   \label{fig:XFull_x0}
\end{figure}

\begin{figure}[t] 
   \centering
   \vspace{-0.3cm}
      \includegraphics[width=3.5in]{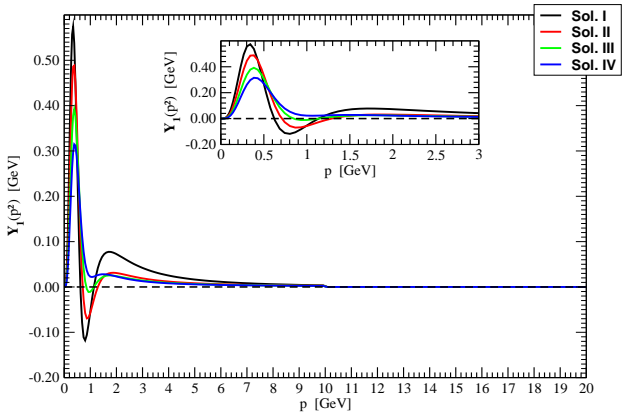}
   \caption{$Y_1(p^2)$ from inverting the Dyson-Schwinger equations.}
   \label{fig:XFull_x1}
\end{figure}

\begin{figure}[t] 
   \centering
      \includegraphics[width=3.5in]{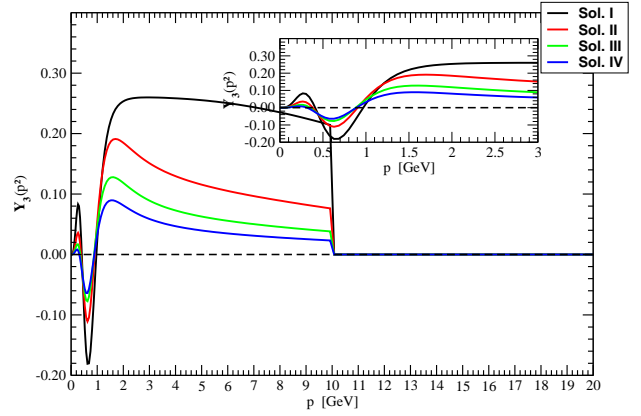}
   \caption{$X^{(3)}(p^2)$ from inverting the Dyson-Schwinger equations.}
   \label{fig:XFull_x3}
\end{figure}

The form factors $X_0$, $Y_1$ and $Y_3$ associated to the solutions I -- IV can be seen on Figs.~\ref{fig:XFull_x0} -- \ref{fig:XFull_x3}.
On Fig.~\ref{fig:XFull_x0} besides solutions I --IV we also show the perturbative $X_0(q^2)$ computed using one-loop dressed perturbation
theory with the tree level ghost-gluon vertex and its enhanced version. The perturbative solutions and those obtained solving the Dyson-Schwinger equations have
rather different structures, with perturbation theory providing larger $X_0(p^2)$ and predicting a relatively large tail. 
Indeed, the solutions of the regularised linear system recover their tree level value $X_0 (p^2) = 1$ from $p \gtrsim 10$ GeV onwards, while the perturbative solution 
only reproduces its tree level value at much larger momentum. 
The momentum scale associated to the absolute maxima of $X_0(p^2)$ occurs at essentially the same
$p \approx 400$ MeV, while the perturbative results points to a maximum of $X_0(p^2)$ at momenta slightly above the GeV scale.
Qualitatively, the non-perturbative solutions all have the same pattern for this form factor. The exception being Sol. I which clearly overestimates $|X_0|$
for $p \gtrsim 1$ GeV. The solutions III and IV  resolve the gap equation with the smaller relative errors that is below 8\%
-- see Fig.~\ref{fig:XFull_DSE_scalar_vector}.
 The non-perturbative solution of the DSE gives a $X_0(p^2)$ that differs from its tree level value by less than 5\%, that are above unit for momenta 
 $p \lesssim 1$ Gev. At this momenta scale the form factors take values below one, reaching a minimum for $p$ just
above 1 GeV, and approaching its tree level value at high momentum from below. The differences between the non-perturbative $X_0$ and its tree level value for 
$p \gtrsim 10$ GeV are rather small.

Our non-perturbative estimations for $Y_1(p^2)$ can be view\-ed on Fig.~\ref{fig:XFull_x1}. All the solutions I -- IV reproduce the same pattern for
this form factor, with a positive maxima around $p \simeq 400$ MeV and with $Y_1$ becoming small for $p \gtrsim 1.5$ GeV. In particular, for the solutions
III and IV, $Y_1(p^2)$ is particularly small ($\lesssim 0.4$ GeV) for $p \gtrsim 1.5$ GeV. One should not forget that the quark-ghost form factor appearing in the
quark-ghost kernel is not $Y_1$ but this function times the gluon propagator -- see Eq. (\ref{Eq:AnsatzX1}). The same applies to
$Y_3$ as can be seen on  Eq. (\ref{Eq:AnsatzX3}). Once more, as the norm of $Y_1$ decreases, the form factors seems to prefer to take only positive values.

The form factor $Y_3(p^2)$ is reported on Fig.~\ref{fig:XFull_x3}. It turns out that this function is positive for $p \lesssim 400$ MeV and for $p \gtrsim 1$ GeV,
takes negative values in between, with a maximum at $p \simeq 1.5$ GeV, and then slowly approaches its tree level value from above. Given the way
the solutions are computed, on Fig.~\ref{fig:XFull_x3} $Y_3(p^2)$ shows a jump at $p \simeq 10$ GeV that corresponds to $p = \Lambda/2$. A similar
behaviour can be seen on Fig.~\ref{fig:XFull_x1} for $Y_1(p^2)$. However, given that for $p \simeq 10$ GeV  one has a $Y_1(p^2) \simeq 0$, this sudden jump is not
so easily observed.

\begin{figure}[t] 
   \centering
   \vspace{-0.3cm}
      \includegraphics[width=3.5in]{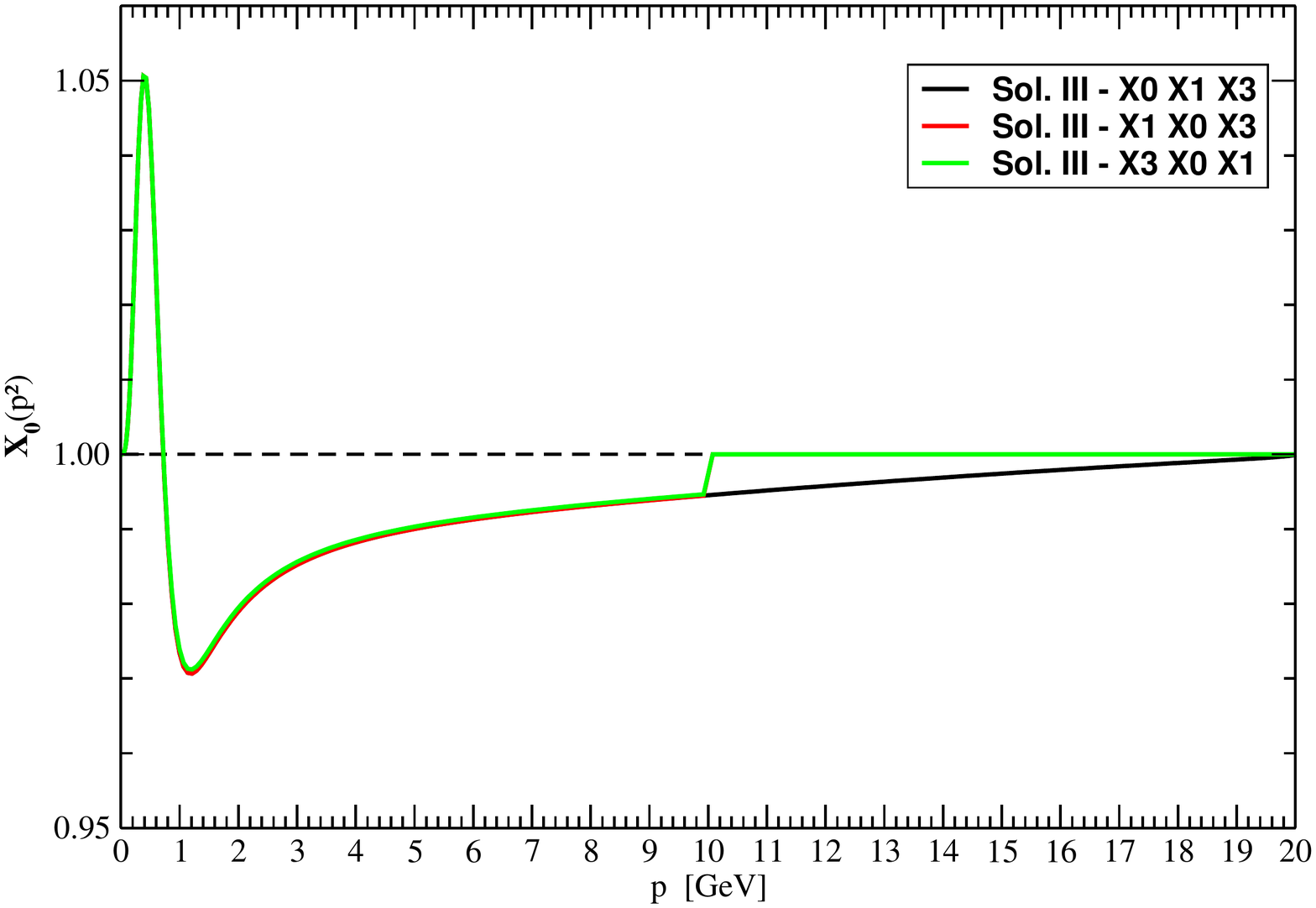} \\
      \vspace{-0.6cm}
      \includegraphics[width=3.5in]{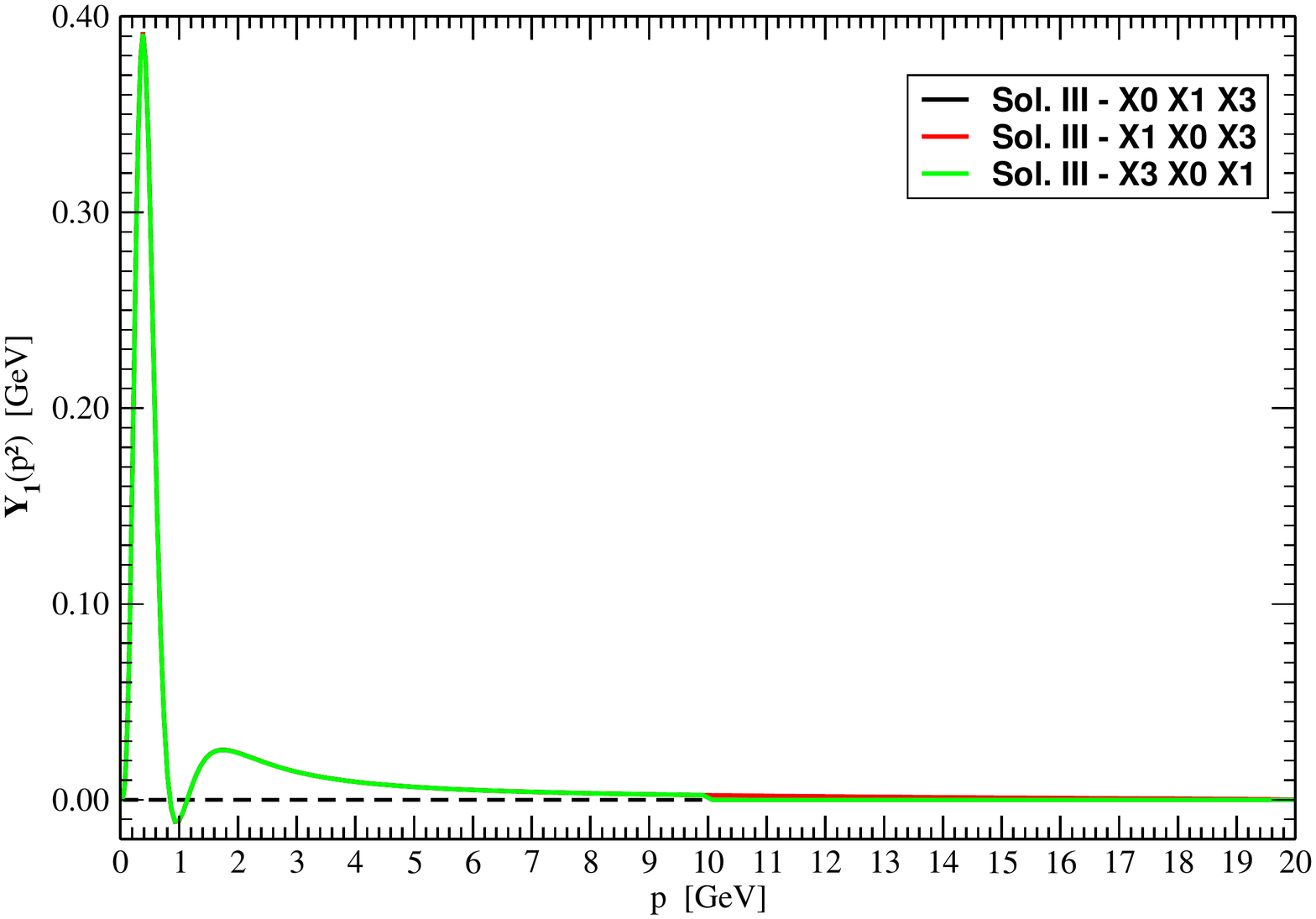} \\
      \vspace{-0.6cm}
      \includegraphics[width=3.5in]{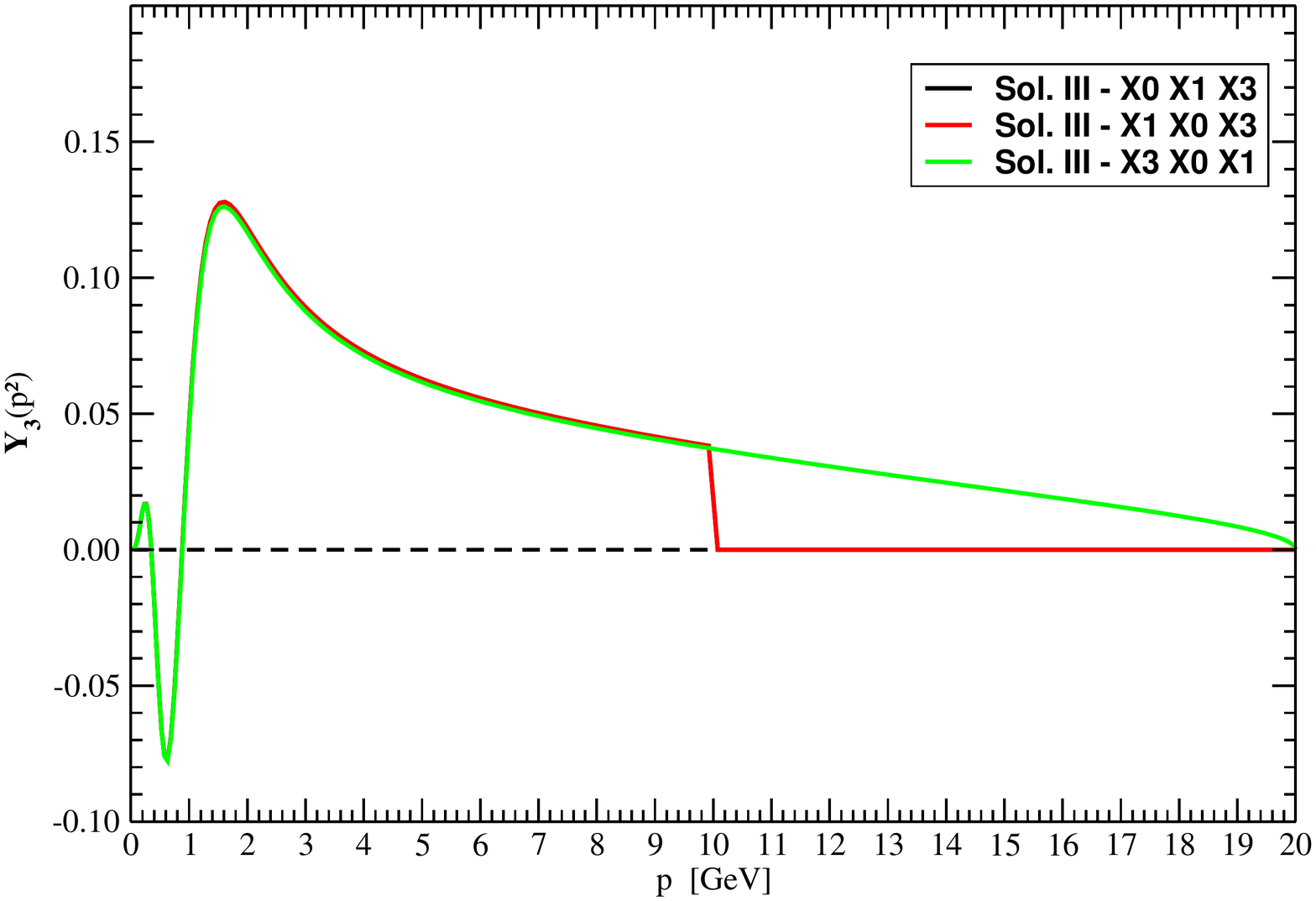}
   \caption{$X^{(0)}(p^2)$, $Y_1(p^2)$ and $Y_3(p^2)$ computed from inverting the Dyson-Schwinger equations (Sol. III)
   after build the large linear system (\ref{Eq:DSE_FullLinearSystem}) in all possible ways, i.e. by permuting the role of the form factors in $X$. See text for details.}
   \label{fig:XFull_compare}
\end{figure}

Finally, on Fig.~\ref{fig:XFull_compare} we provide the various solutions for $X_0(p^2)$, $Y_1(p^2)$ and $Y_3(p^2)$ after permuting the role
of the form factors when writing the extended vector $X$. For the so-called $X_0 \, X_1 \, X_3$ the extended vector included $X_0$ over the full set
of Gauss-Legendre points with $Y_1$ and $Y_3$ being obtained only in the range $p \in [ 0 \, , \, \Lambda/2]$.
For the so-called $X_1 \, X_0 \, X_3$ the extended vector included $Y_1$ over the full set
of Gauss-Legendre points with $X_0$ and $Y_3$ being obtained only in the range $p \in [ 0 \, , \, \Lambda/2]$.
For the so-called $X_3 \, X_0 \, X_1$ the extended vector included $Y_3$ over the full set
of Gauss-Legendre points with $X_0$ and $Y_1$ being obtained only in the range $p \in [ 0 \, , \, \Lambda/2]$. 
We call the readers attention to the stability of the solution of the various linear
systems. Furthermore, the comparison of Fig.~\ref{fig:X0onelooopX1X3} from Sec.\ref{Sec:OneLoopSolution} and Fig.~\ref{fig:XFull_compare}
show quite similar $Y_1$ and $Y_3$ suggesting, once again, that $X_0$ almost does not deviates from its tree level value.

\subsection{Solving the DSE for $X_0 = 1$ \label{Sec_X0one}}

\begin{figure}[t] 
   \centering
   \vspace{-0.3cm}
      \includegraphics[width=3.5in]{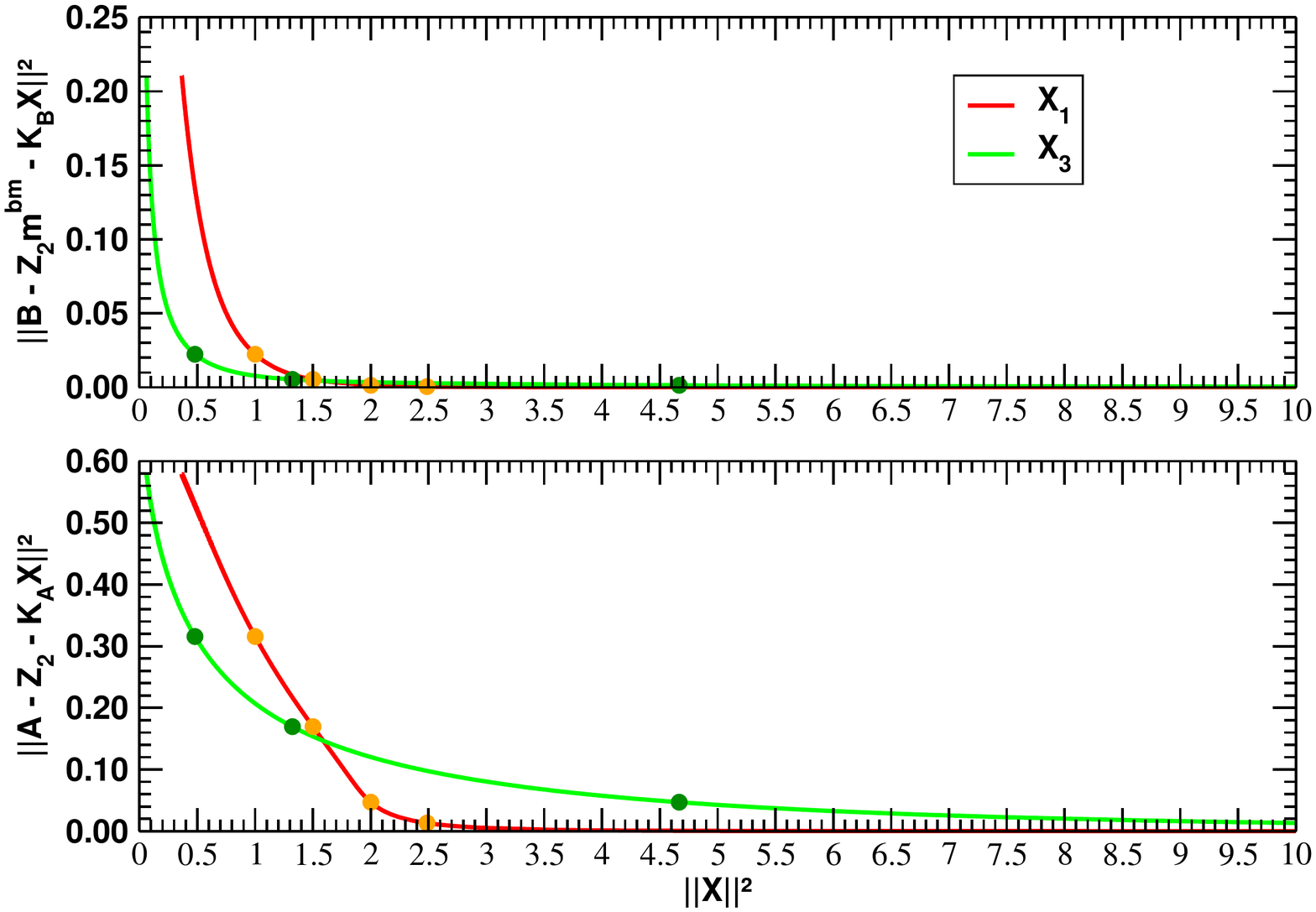}
   \caption{Residua versus norm of the form factors from inverting the Dyson-Schwinger equations with $X_0 = 1$.}
   \label{fig:ResdNormX0one}
\end{figure}

\begin{figure}[t] 
   \centering
   \vspace{-0.3cm}
      \includegraphics[width=3.5in]{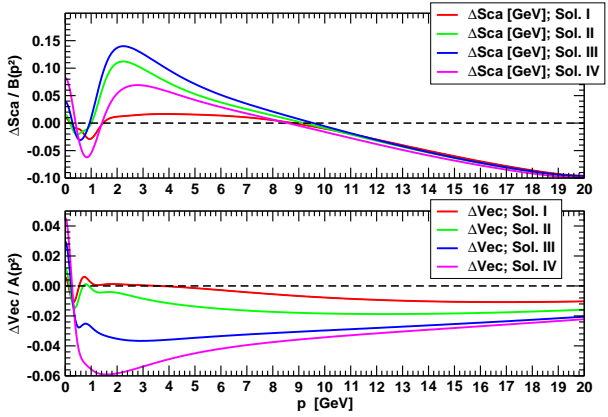}
   \caption{Relative error for the DSE associated to the highlighted solutions of the Dyson-Schwinger equations for $X_0 = 1$.}
   \label{fig:ErrRelDSEX0one}
\end{figure}

\begin{figure}[t] 
   \centering
   \vspace{-0.3cm}
      \includegraphics[width=3.5in]{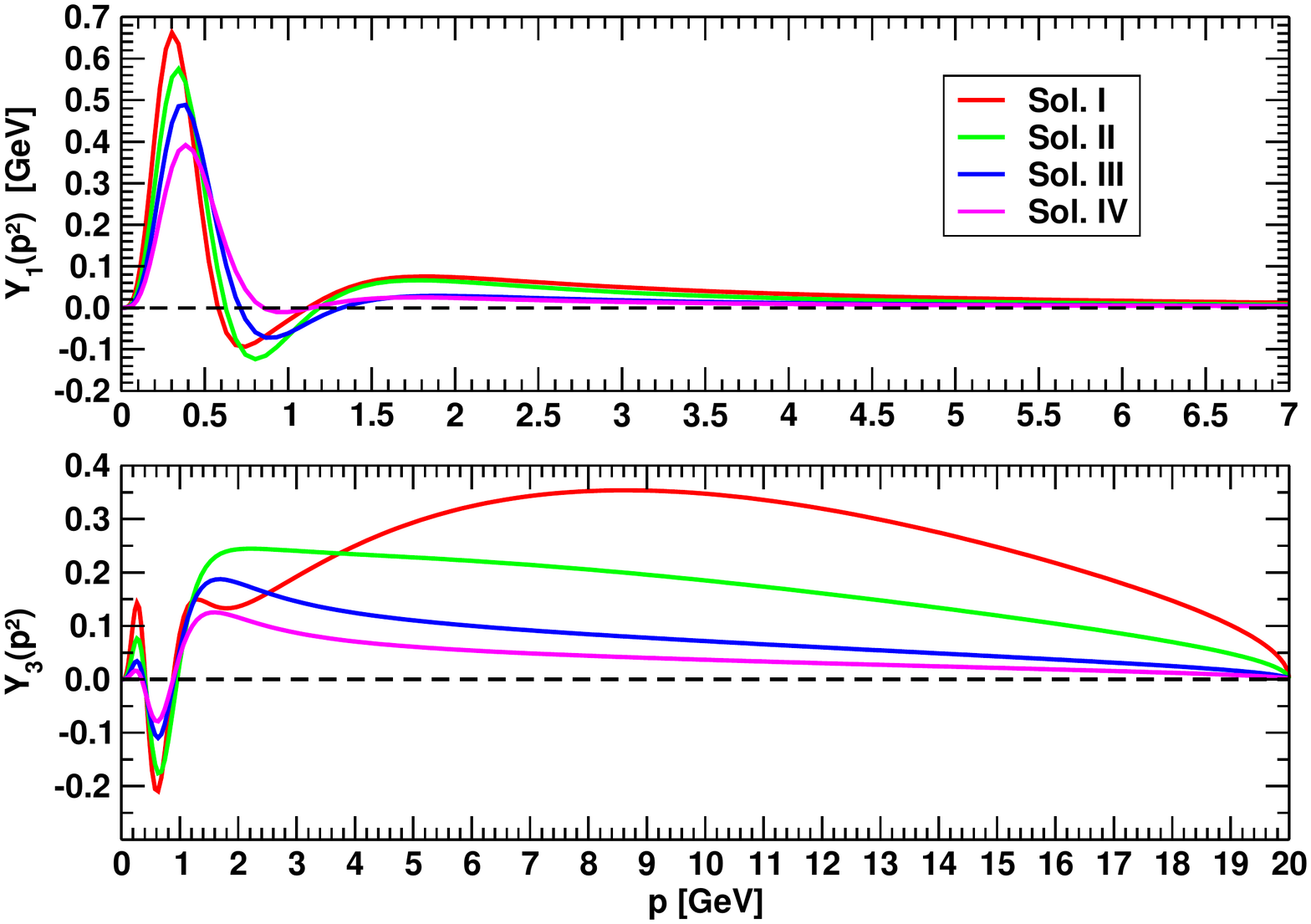}
   \caption{The solutions of the DSE for $Y_1$ and $Y_3$ when $X_0 = 1$.}
   \label{fig:X1X3X0one}
\end{figure}

The non-perturbative solutions of the Dyson-Schwinger equations discussed on the previous paragraph suggest that $X_0(p^2) \simeq 1$. Therefore,
herein we investigate the results by solving the DSE with $X_0(p^2) = 1$. The residua of the scalar and vector equations against the norm of the two remaining form factor $Y_1$ and $Y_3$ can be seen on Fig.~\ref{fig:ResdNormX0one}. The characteristics of the solutions
highlighted in the figure and associated to an $\epsilon$ close to its optimal value are

\begin{center}
\begin{small}
   \begin{tabular}{llllll } 
             & \!\!\!\!\!\!  $||Y_1||^2$  & $||Y_3||^2$  & $||\Delta\mbox{Sca}||^2$  & $||\Delta\mbox{Vec}||^2$  & $\epsilon$   \\ 
   I         & \!\!\!\!\!\!  2.4870         &  10.0286       &  0.0004739                      & 0.01314                            & 0.0025   \\
  II         & \!\!\!\!\!\!  2.0003         &  ~~4.6676     & 0.001351                         & 0.04691                           & 0.013   \\
  III        & \!\!\!\!\!\!  1.5006         &  ~~1.3225     & 0.005306                         & 0.1695                              & 0.0745   \\
  IV        & \!\!\!\!\!\!  1.0009         &  ~~0.4801     & 0.02232                           & 0.3157                              & 0.1975   \\
   \end{tabular}
\end{small}   
\end{center}

\noindent
for $m^{b.m.} = 6.852$ MeV, $Z_2 = 1.0016$, where $\Delta\mbox{Sca}$ and $||Y_1||$ are  given in GeV, while $\Delta\mbox{Vec}$, $||X_0||$ and $||Y_3||$
are dimensionless. The relative error on the Dyson-Schwinger equations for these solutions can be seen on Fig.~\ref{fig:ErrRelDSEX0one} which
shows that solution I resolves the DSE up to $p \simeq 10$ GeV with an error that is smaller than 3\% for the scalar equation and error of about 1\%
for the vector equation. However, the form factor $Y_3$ associated to solution I does not seem to be converged for momenta above 2 GeV.
The solution named IV solves the scalar equation with a relative error below 10\% and the vector equation with a relative error below 6\%.

The form factors $Y_1(p^2)$ and $Y_3(p^2)$ associated to the solutions I -- IV are reported on Fig.~\ref{fig:X1X3X0one} and reproduce the same
patterns as  the solutions computed in the previous sections.

\subsection{Full Form Factors and Comparison of Solutions \label{Sec:compareX}}

In Secs.~\ref{Sec:OneLoopSolution},~\ref{Sec:solveall} and~\ref{Sec_X0one} we have solved the Dyson-Schwinger equations assuming that the quark-ghost
kernel form factors are given by Eqs. (\ref{Eq:AnsatzX0}) -- (\ref{Eq:AnsatzX3}). So, besides, $X_0$, the full form factors appearing in $H$ and $\overline H$,
see Eqs. (\ref{Eq:quarkghostkernel}), should be multiplied by the gluon propagator at the proper kinematical configuration. Herein, we aim to compare
the various solutions found in previous section and, in this way, provide an estimation of the systematics associated to our ansatz, and also to provide
the form of the full functions appearing on $H$ and $\overline H$. Looking at the relative errors on the Dyson-Schwinger equations and at 
the convergence of the form factors at higher momenta, the comparison will
be done using Sol. II computed using the perturbative $X_0$ and the tree level ghost-gluon vertex,  Sol. III computed when the gap equation is solved for the full set of form factors and Sol. IV when the gap equation is solved
for $X_0 = 1$.

\begin{figure}[t] 
   \centering
   \vspace{-0.3cm}
      \includegraphics[width=3.5in]{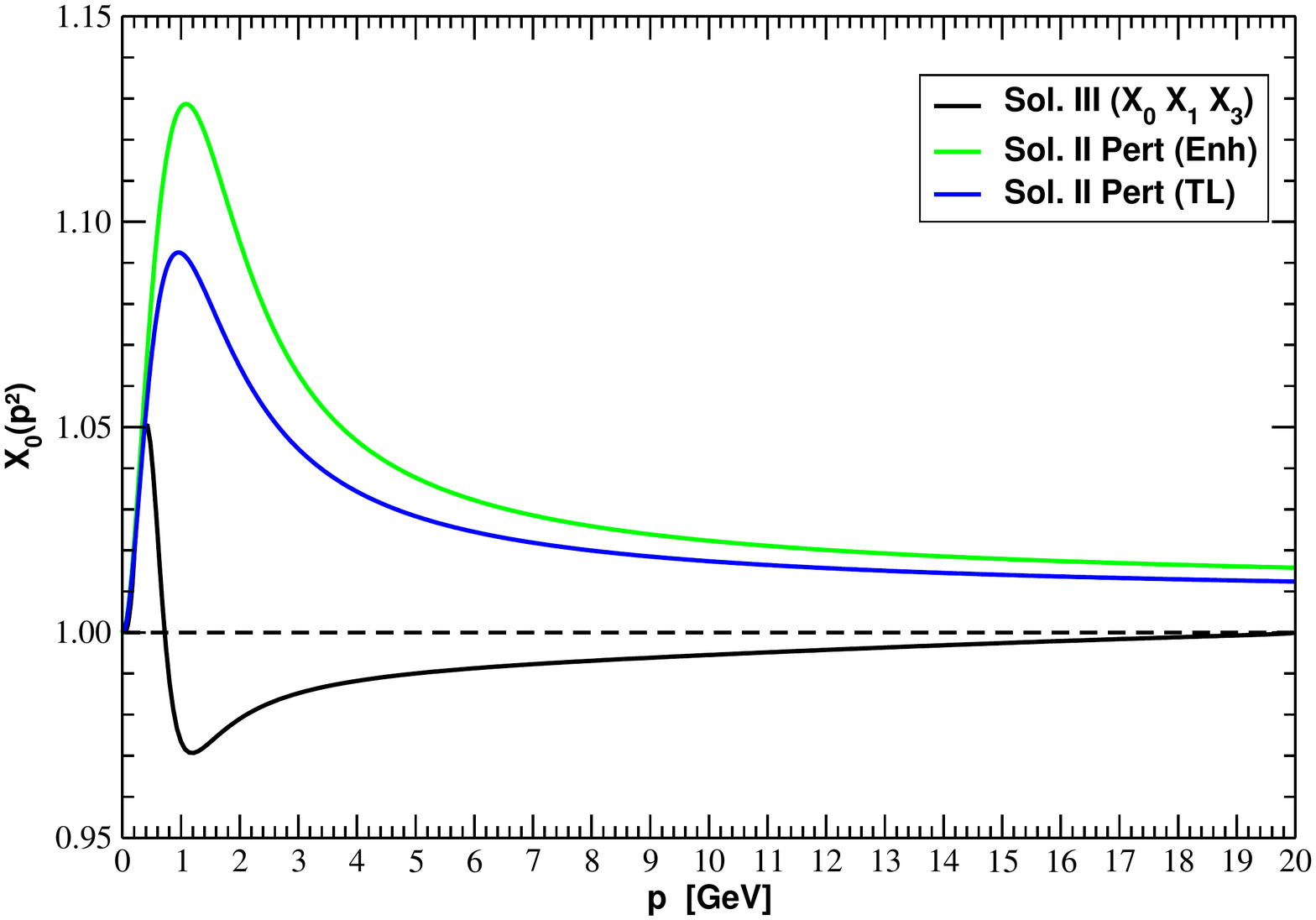}
   \caption{The quark-ghost kernel form factor $X_0(p^2)$.}
   \label{fig:X0comparesolutions}
\end{figure}

Let us start with the $X_0$ that we have assumed to be only a function of the gluon momenta. The perturbative solutions are compared with the solution
obtained inverting the Dyson-Schwinger equations for the full set of form factors used in our ansatz can be seen in Fig.~\ref{fig:X0comparesolutions}. This figure repeats 
partially Fig.~\ref{fig:XFull_x0} providing a clear view of the solutions. All solutions show a $X_0$ that essentially is close to  its tree level value, i.e.
$X_0 = 1$, with the perturbative solutions having the largest deviation from unit.

\begin{figure}[t] 
   \centering
   \vspace{-0.3cm}
      \includegraphics[width=3.5in]{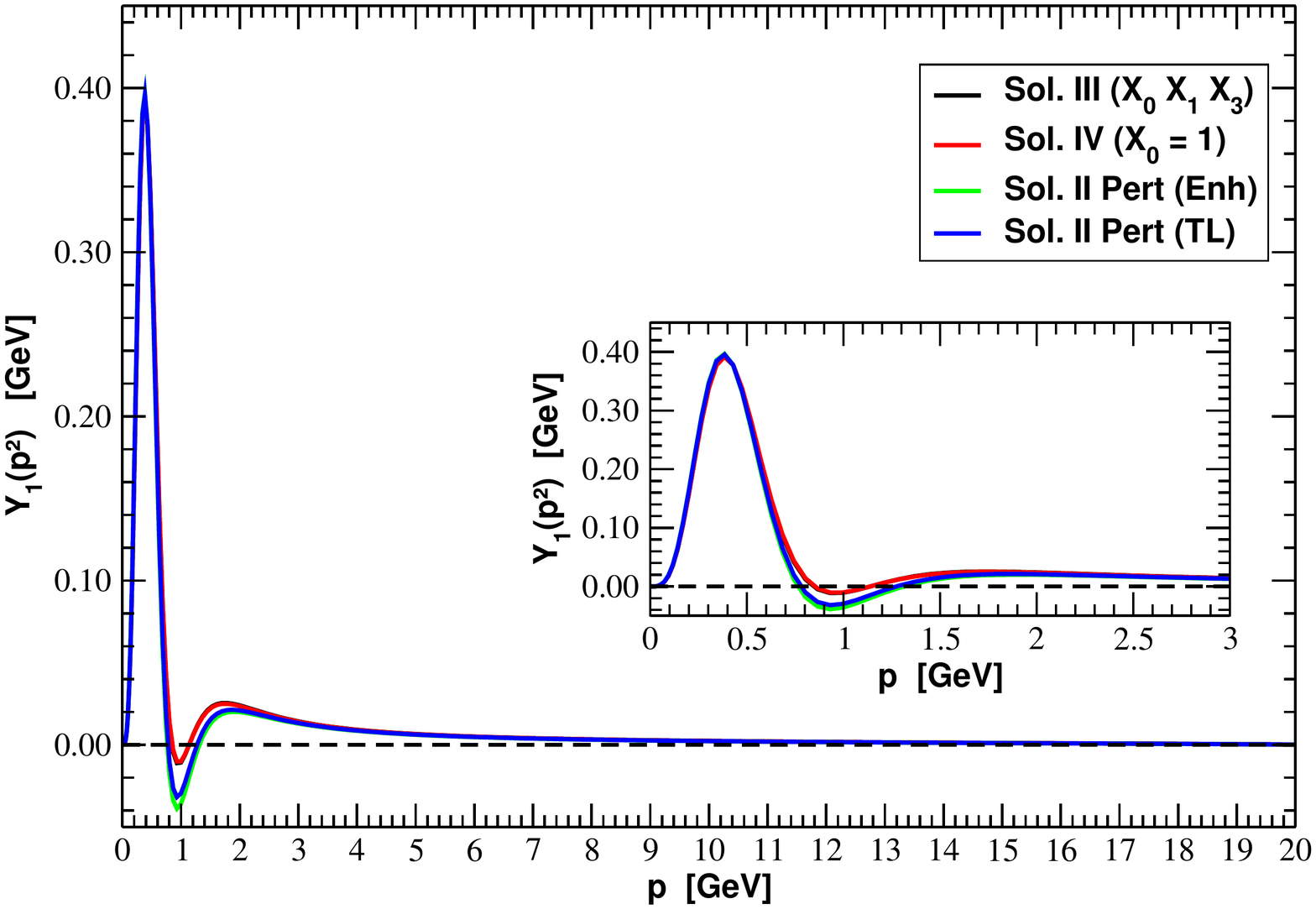}
   \caption{The quark-ghost kernel form factor $Y_1(p^2)$.}
   \label{fig:X1comparesolutions}
\end{figure}

The form factor $Y_1(p^2)$ can be seen on Fig.~\ref{fig:X1comparesolutions} for all the solutions. Note that all solutions reproduce essentially the same
function of the gluon momentum, with $Y_1(p^2)$ being small for $p\gtrsim 1.5$ GeV and showing a sharp peak at $p \simeq 400$ MeV. $Y_1(p^2)$
is positive defined except for a small range of momenta $p \in [ 0.75 \, , \, 1.4]$ GeV where it takes small negative values.

\begin{figure}[t] 
   \centering
   \vspace{-0.3cm}
      \includegraphics[width=3.5in]{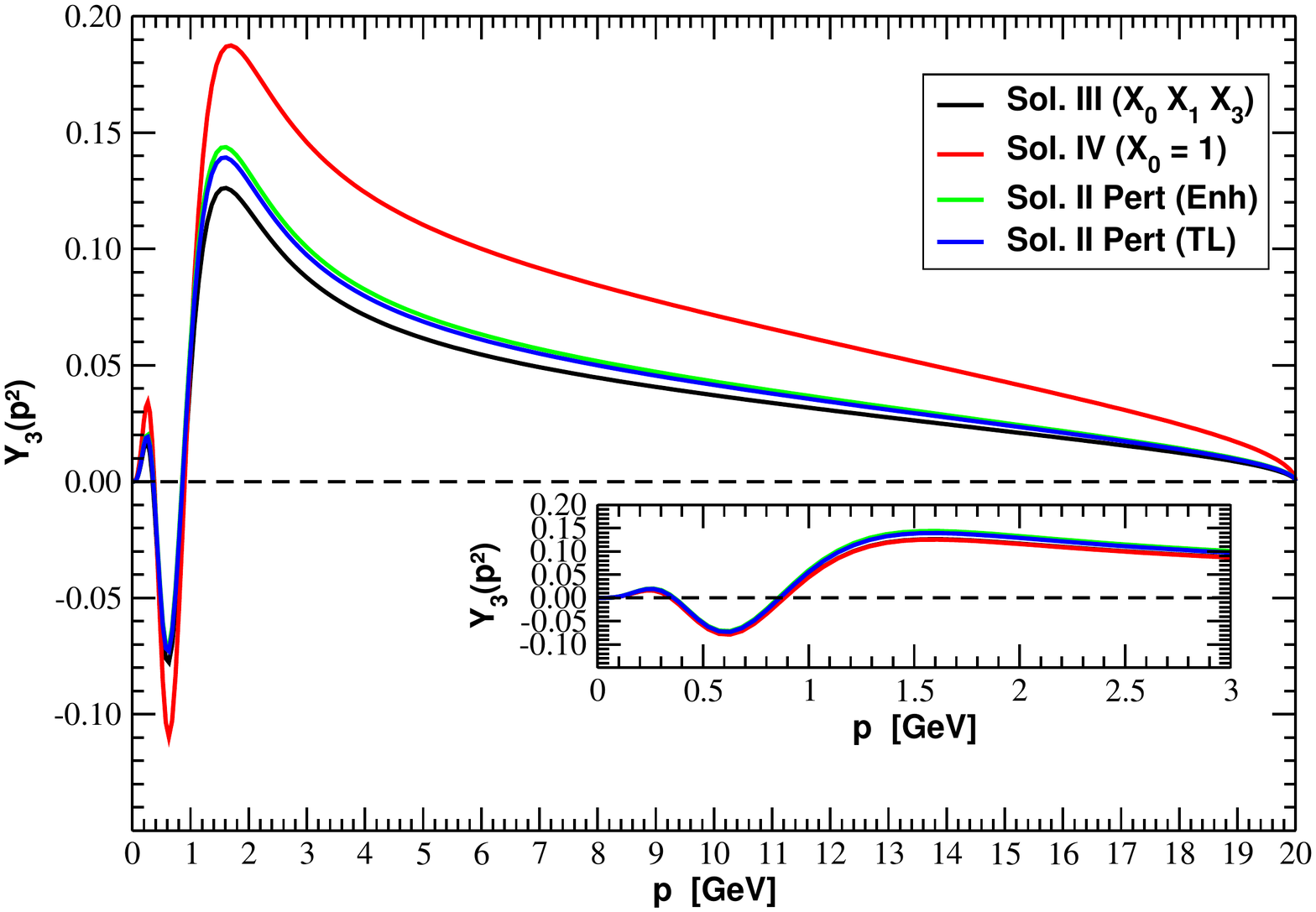}
   \caption{The quark-ghost kernel form factor $Y_3(p^2)$.}
   \label{fig:X3comparesolutions}
\end{figure}

The form factor $Y_3(p^2)$ can be seen on Fig.~\ref{fig:X3comparesolutions} for all the solutions. Surprisingly, $Y_3(p^2)$ seems to have a relative large tail that
appears in all the solutions. Up to momenta $p \simeq 3$ GeV the solutions reproduce essentially the same function. However for $p \simeq 3$ GeV the solution
associated to $X_0 = 1$ is enhanced relative to all the others, with the solutions associated to the one-loop perturbative $X_0$ being slightly enhanced
relative to the non-perturbative solution obtained from inverting the gap equation. $Y_3(p^2)$ shows a maxima at $p \simeq 200$ MeV, an absolute maxima at
$p \simeq 1.4$ GeV and an absolute minima at $p \simeq 650$ MeV. This form factor is positive defined at infrared momenta $p \lesssim 350$ MeV and
the high momenta $p \gtrsim 900$ MeV taking negative values in $p \in [ 0.35 \, , \, 0.9]$ GeV.

In summary, Figs.~\ref{fig:X1X3X0one} -- \ref{fig:X3comparesolutions} resume the computations of the quark-ghost kernel form factors performed so far.

\subsection{Tunning $\alpha_s$ \label{Sec:tune_alpha}}

The results for the relative errors on the scalar and vector components of the Dyson-Schwinger equation 
seen on Figs.~\ref{fig:X0onelooopDSEErrRel}, \ref{fig:XFull_DSE_scalar_vector}, \ref{fig:ErrRelDSEX0one} show a relative error that for $p \gtrsim 10$ GeV
grow with $p$ and take its maximum value $\sim 10$\% at the cutoff. This can be viewed in many ways and one of them being that our choice for the strong
coupling constant is not the best one. In our approach we mix quenched lattice results with dynamical simulations and, in order to be able to solve the
gap equation for the quark-ghost kernel, the renormalization constant $Z_1$, see Eq, (\ref{DSEselfenergy}), is set to identity\footnote{It can also be viewed as been
included in the definition of the various form factors.}. 
Although the original integral equation is linear on the form factors $X_0$, $Y_1$ and $Y_3$, the regularized system that is solved introduces an extra parameter 
that needs to be fixed in the way described above and, therefore, changing the strong coupling constant changes the balance between the regularizating parameter 
$\epsilon$ and the various form factors, allowing for adjustments on the solutions. Therefore, the relative errors on the integral equations can be adjusted by changing
the strong coupling constant. 

\begin{figure}[t] 
   \centering
   \vspace{-0.3cm}
      \includegraphics[width=3.5in]{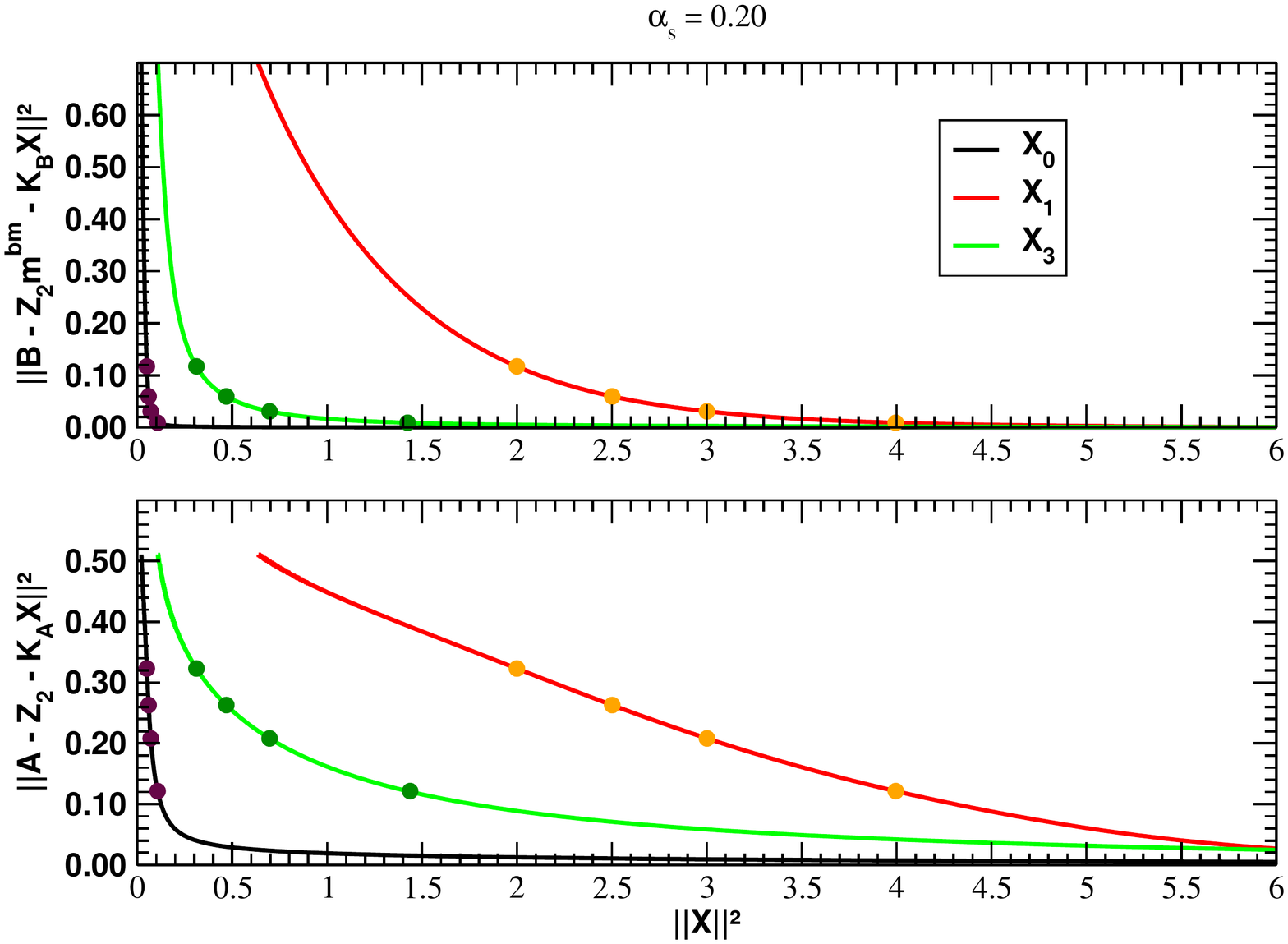} \\
      \vspace{-0.5cm}
      \includegraphics[width=3.5in]{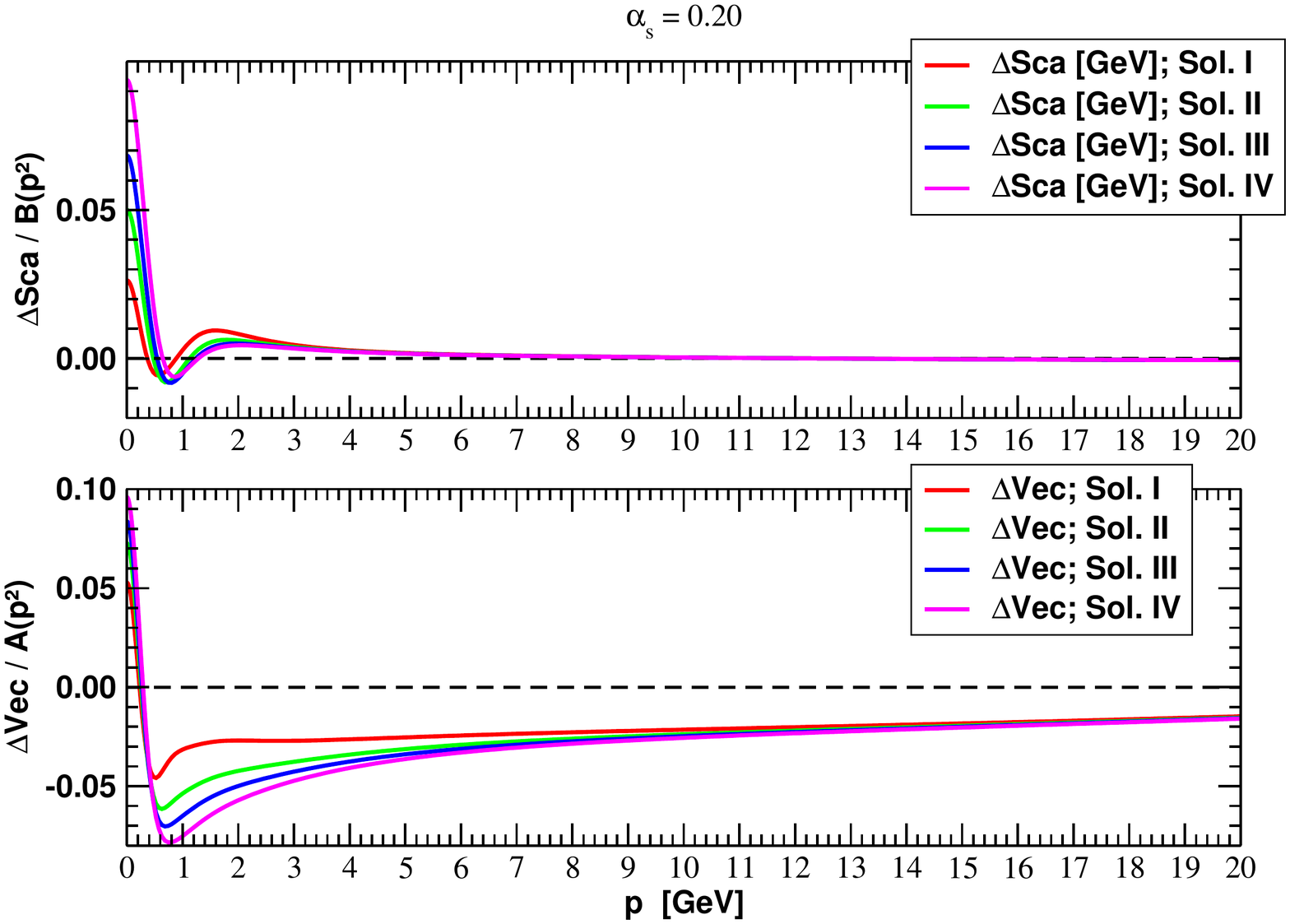}
   \caption{Norm versus Residuum (top) and relative error of the solutions of the gap equation for $\alpha_s(\mu) = 0.20$. The solutions on the right plot
    are those marked on the left plot with I being that associated to the most right mark.}
   \label{fig:changing_alpha_ErrRel_0p20}
\end{figure}

\begin{figure}[t] 
   \centering
   \vspace{-0.3cm}
      \includegraphics[width=3.5in]{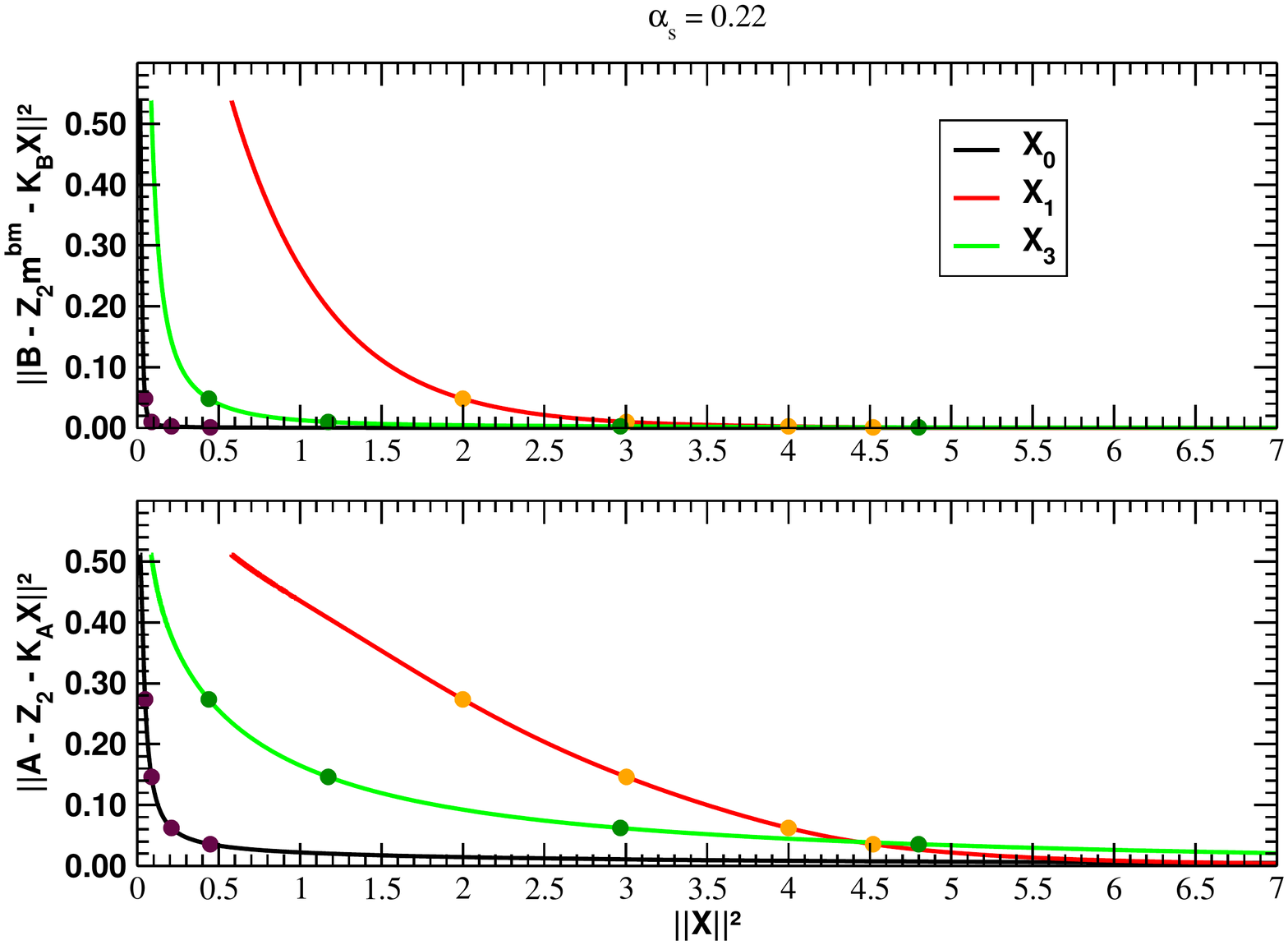} \\
      \vspace{-0.5cm}
      \includegraphics[width=3.5in]{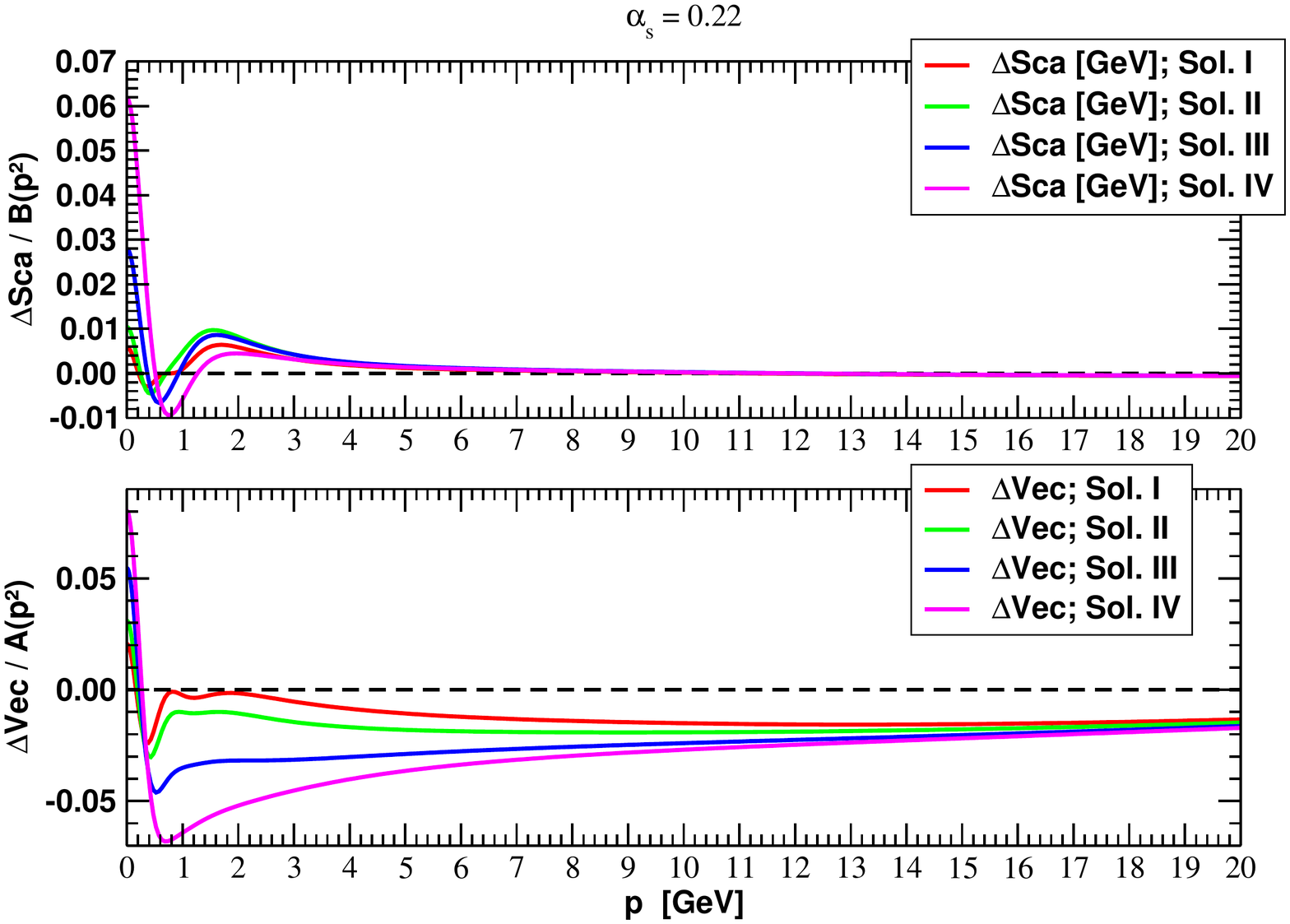}
   \caption{Norm versus Residuum (top) and relative error of the solutions of the gap equation for $\alpha_s(\mu) = 0.22$. The solutions on the right plot
    are those marked on the left plot with I being that associated to the most right mark.}
   \label{fig:changing_alpha_ErrRel_0p22}
\end{figure}

\begin{figure}[t] 
   \centering
   \vspace{-0.3cm}
      \includegraphics[width=3.5in]{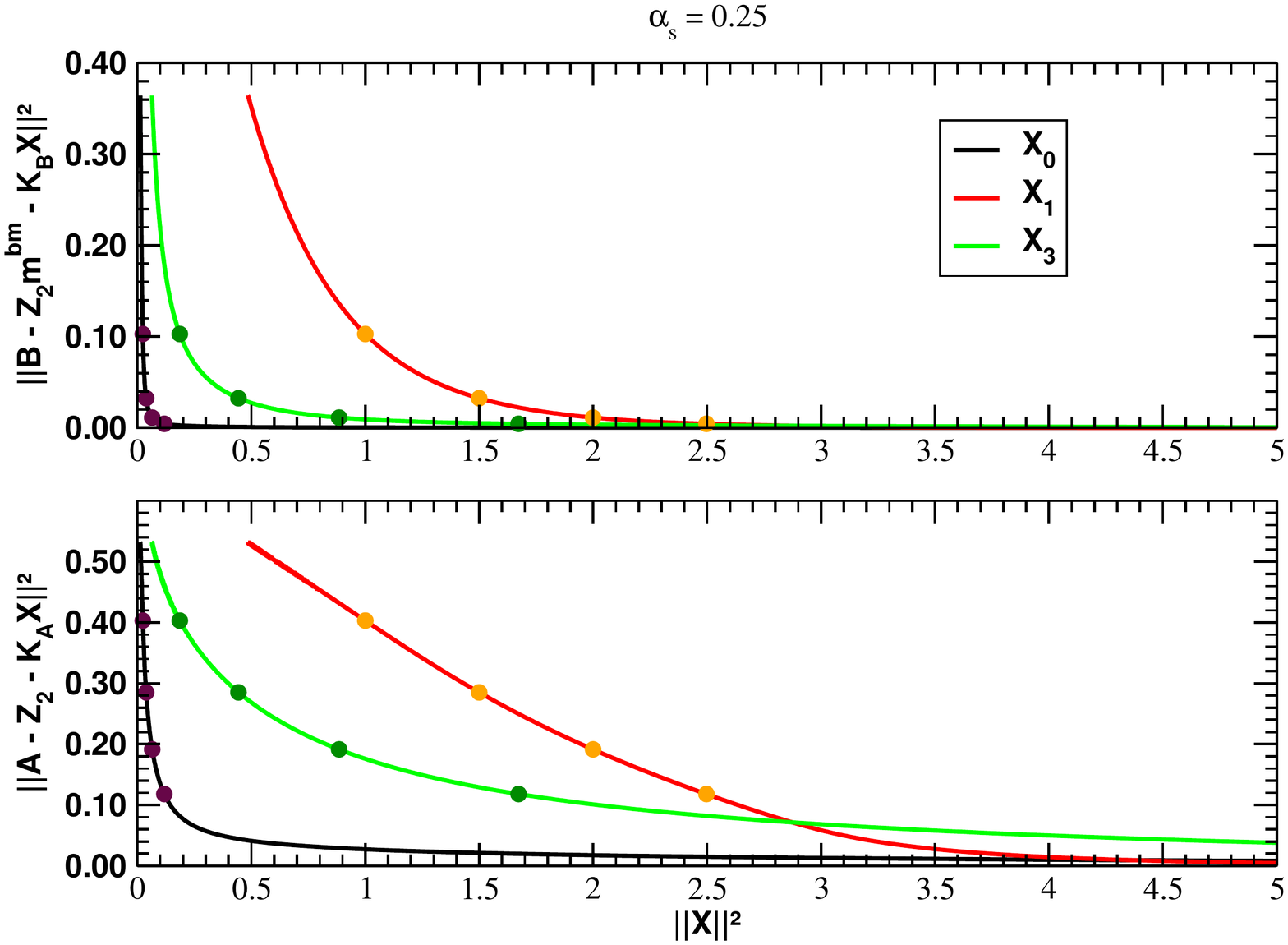} \\
      \vspace{-0.5cm}
      \includegraphics[width=3.5in]{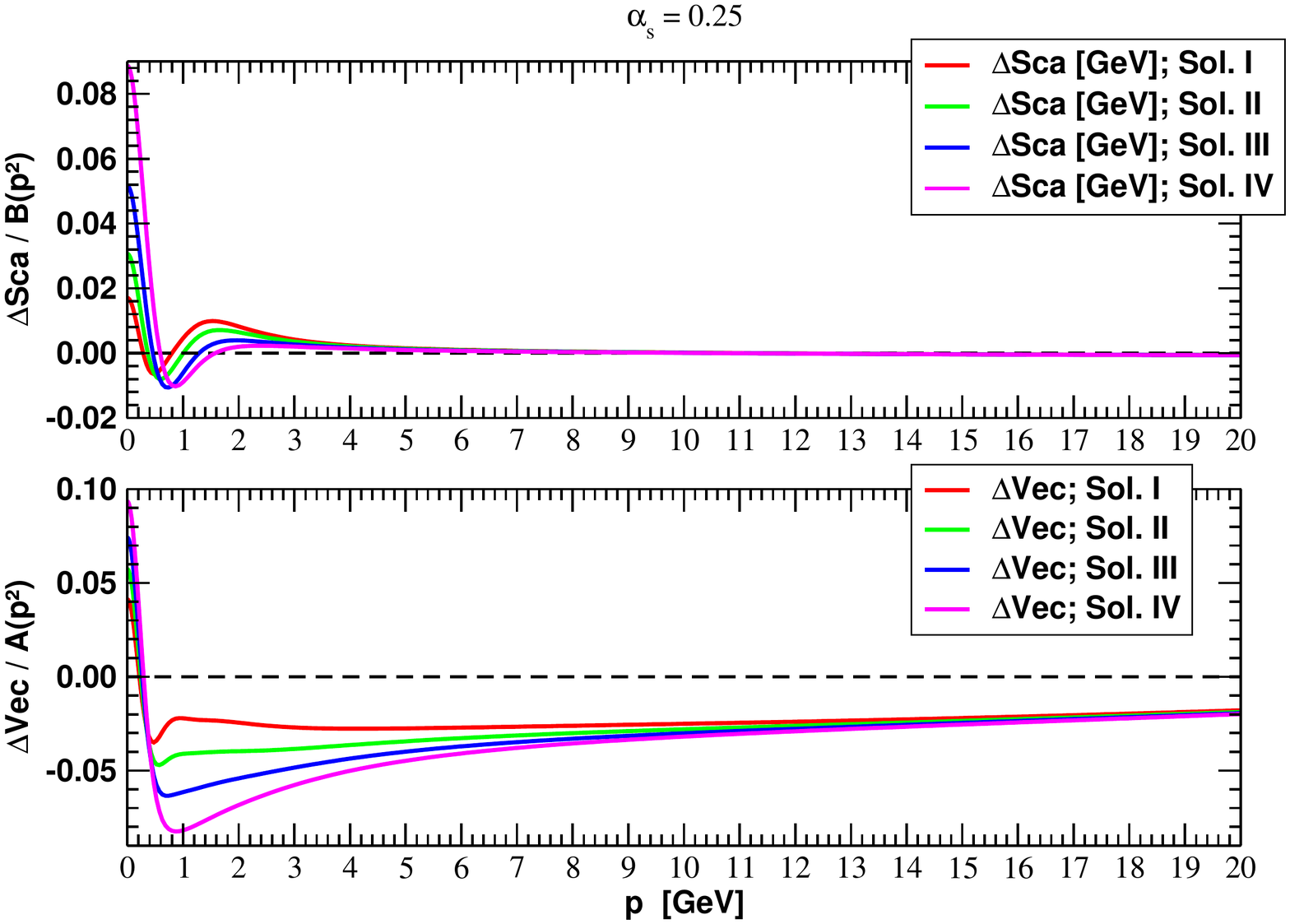}
   \caption{Norm versus Residuum (top) and relative error of the solutions of the gap equation for $\alpha_s(\mu) = 0.25$. The solutions on the right plot
    are those marked on the left plot with I being that associated to the most right mark.}
   \label{fig:changing_alpha_ErrRel_0p25}
\end{figure}

In this section, we report on the results of solving the regularised linear system of equations that replace the original equations
in the way it is described on Sec.~\ref{Sec:solveall} for $\alpha_s(\mu) = 0.20$, 0.22 and 0.25. The properties of the inversions of the regularised
system for the various values of the strong coupling constant can be found in Figs.~\ref{fig:changing_alpha_ErrRel_0p20},
\ref{fig:changing_alpha_ErrRel_0p22} and \ref{fig:changing_alpha_ErrRel_0p25} and should be compared with Fig.~\ref{fig:ErrRelDSEX0one} of the Sec.~\ref{Sec:solveall}.

\begin{figure}[t] 
   \centering
   \vspace{-0.3cm}
      \includegraphics[width=3.5in]{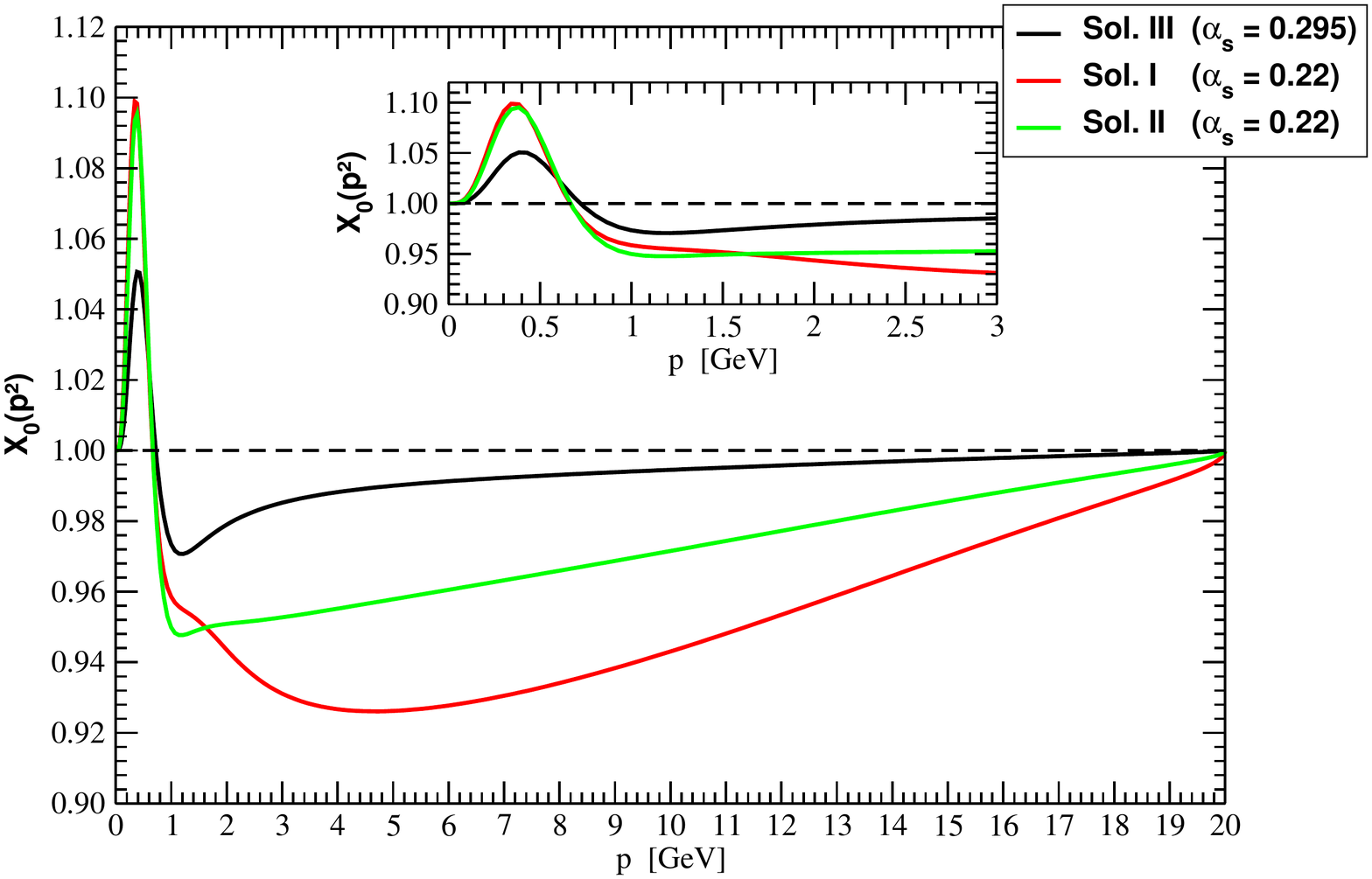}
   \caption{The quark-ghost kernel form factor $X_0(p^2)$ computed using $\alpha_s(\mu) = 0.22$.}
   \label{fig:X0_a0p22}
\end{figure}

\begin{figure}[t] 
   \centering
   \vspace{-0.3cm}
      \includegraphics[width=3.5in]{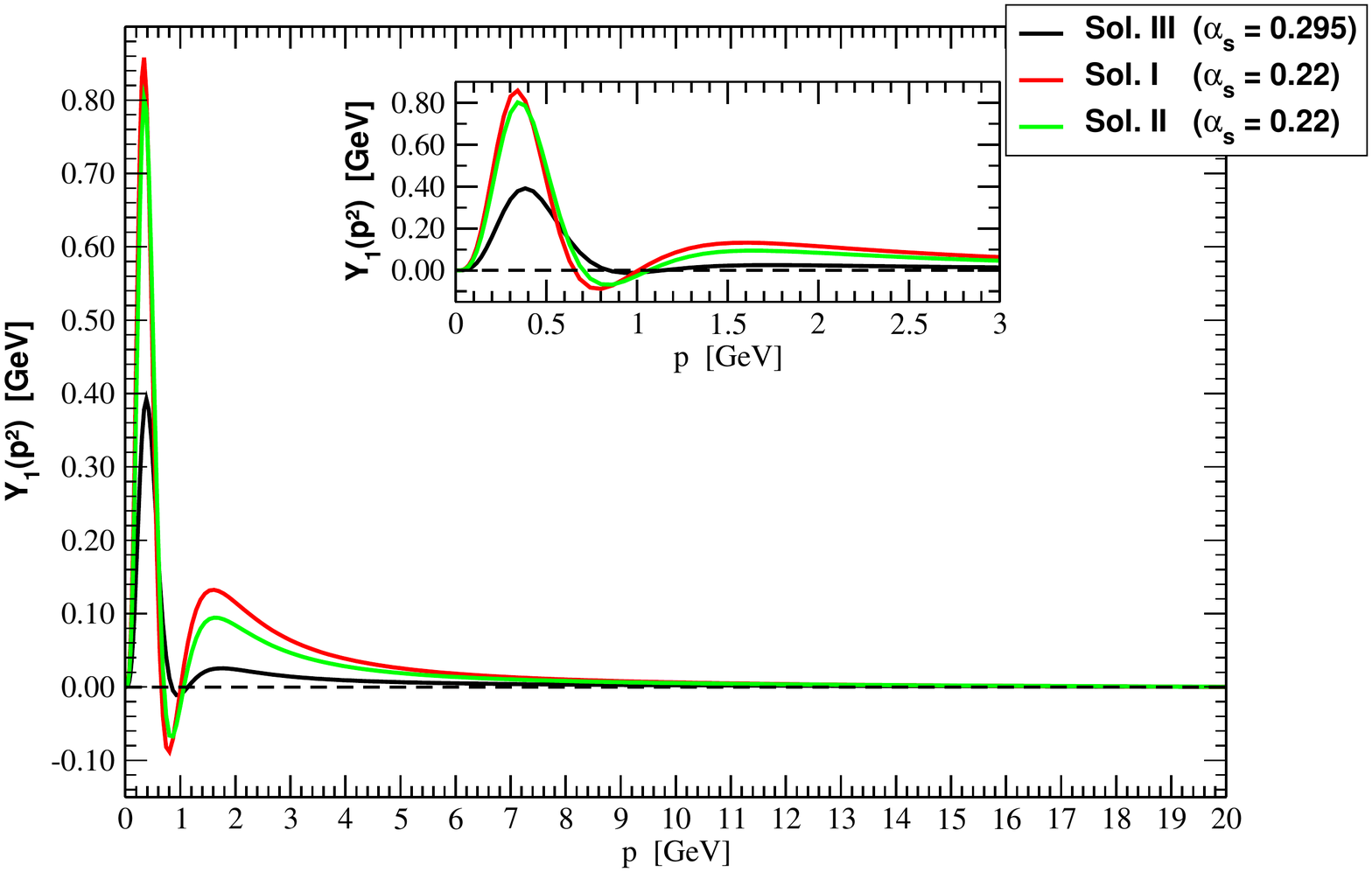}
   \caption{The quark-ghost kernel form factor $X_1(p^2)$ computed using $\alpha_s(\mu) = 0.22$.}
   \label{fig:X1_a0p22}
\end{figure}

\begin{figure}[t] 
   \centering
      \includegraphics[width=3.5in]{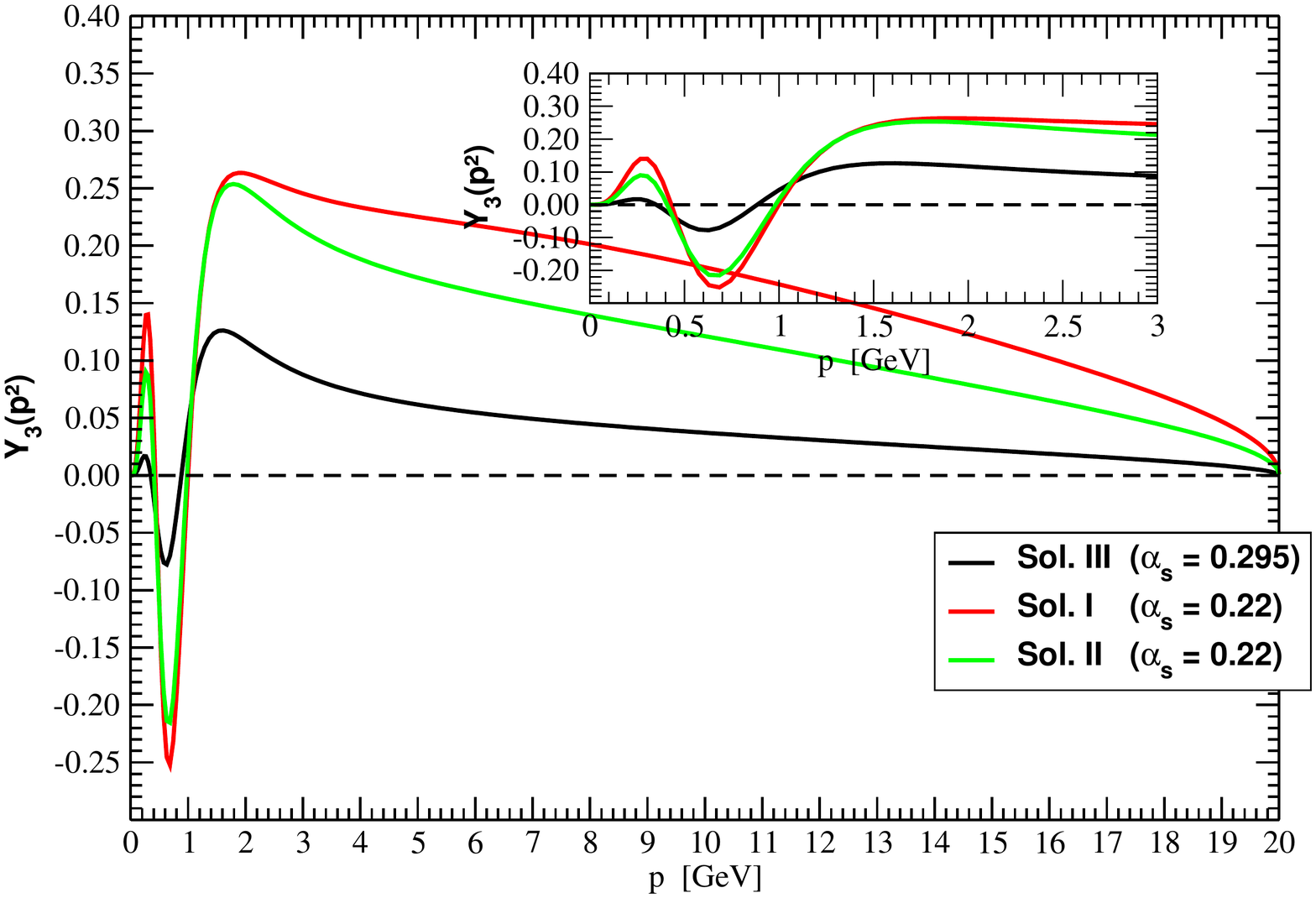}
   \caption{The quark-ghost kernel form factor $X_3(p^2)$ computed using $\alpha_s(\mu) = 0.22$.}
   \label{fig:X3_a0p22}
\end{figure}

As the figures shows, lowering the value of $\alpha_s( \mu )$ solves the problem of the increase of the relative error observed in Sec.~\ref{Sec:solveall}. Moreover,
of the various solutions considered, for $\alpha_s(\mu) = 0.22$ one can observe solutions whose relative error is of the order of $\sim 1$\% for the scalar equation
and $\sim 3$--4\% for the vector components, the solutions named Sol. I and II in Fig.~\ref{fig:changing_alpha_ErrRel_0p22}. The relative error associated to
the remaining solutions given on Figs.~\ref{fig:ErrRelDSEX0one},~\ref{fig:changing_alpha_ErrRel_0p20},~\ref{fig:changing_alpha_ErrRel_0p22} and 
\ref{fig:changing_alpha_ErrRel_0p25} are larger and, therefore, we take $\alpha_s(\mu) = 0.22$ as the optimal value for the strong coupling constant
within our approach. The corresponding quark-ghost kernel can be seen on Figs.~\ref{fig:X0_a0p22},~\ref{fig:X1_a0p22} and~\ref{fig:X3_a0p22}, together
with the corresponding solution computed using $\alpha_s(\mu) = 0.295$. The solutions for the two values of $\alpha_s$ are similar, although those
associated to the smaller value of $\alpha_s$ achieve higher values. If at momenta $p 
\gtrsim 1$ GeV Sol. I takes absolute values that are higher than those of Sol. II, at lower momenta the difference between the two solutions
is marginal.

\section{The Quark-Gluon Vertex Form Factors \label{Sec:lambda1_lambda4}}

\begin{figure}[t]
\vspace{-0.3cm}
        \includegraphics[width=3.5in]{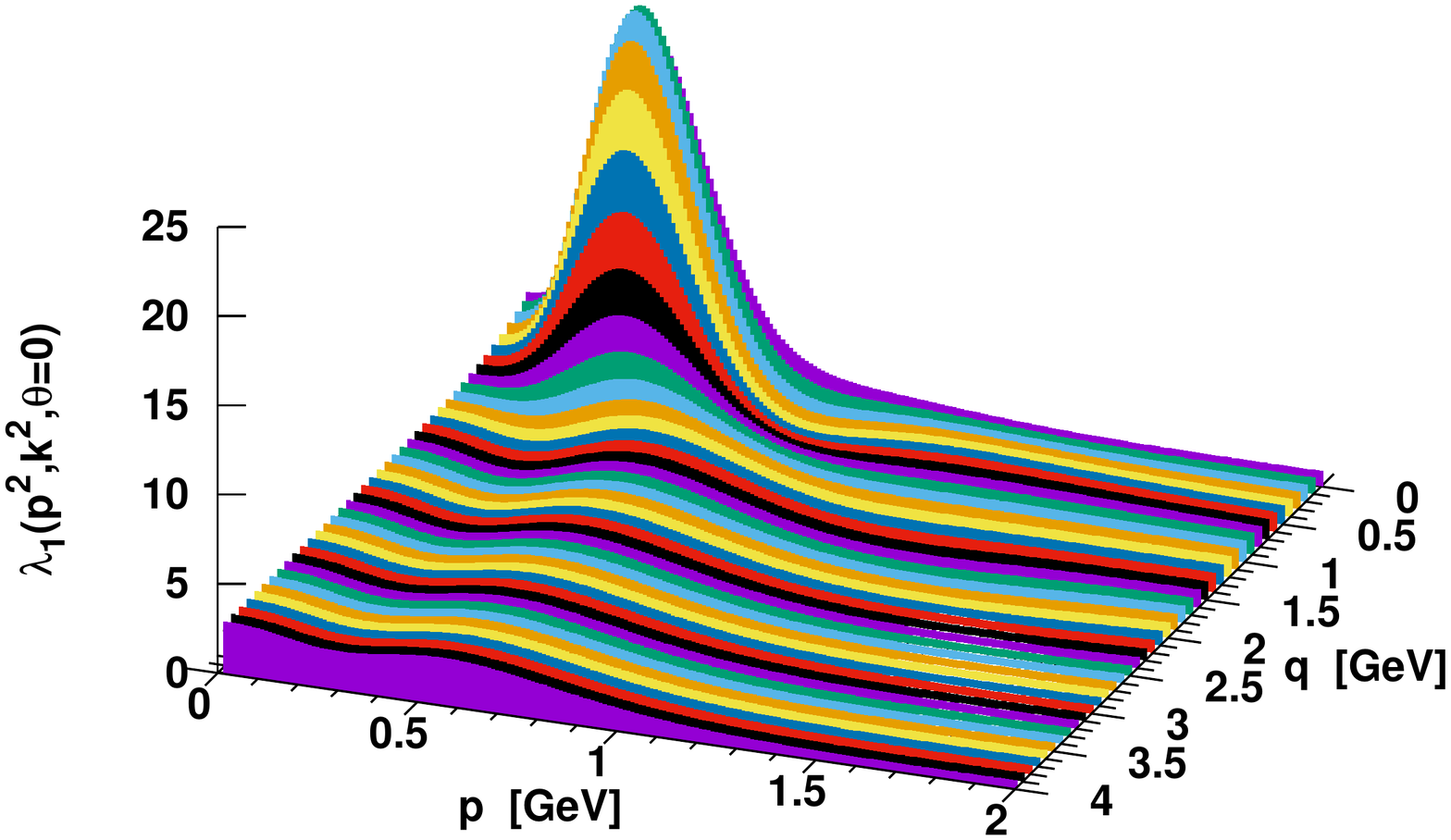}     
\vspace{-2cm} \\
        \includegraphics[width=3.5in]{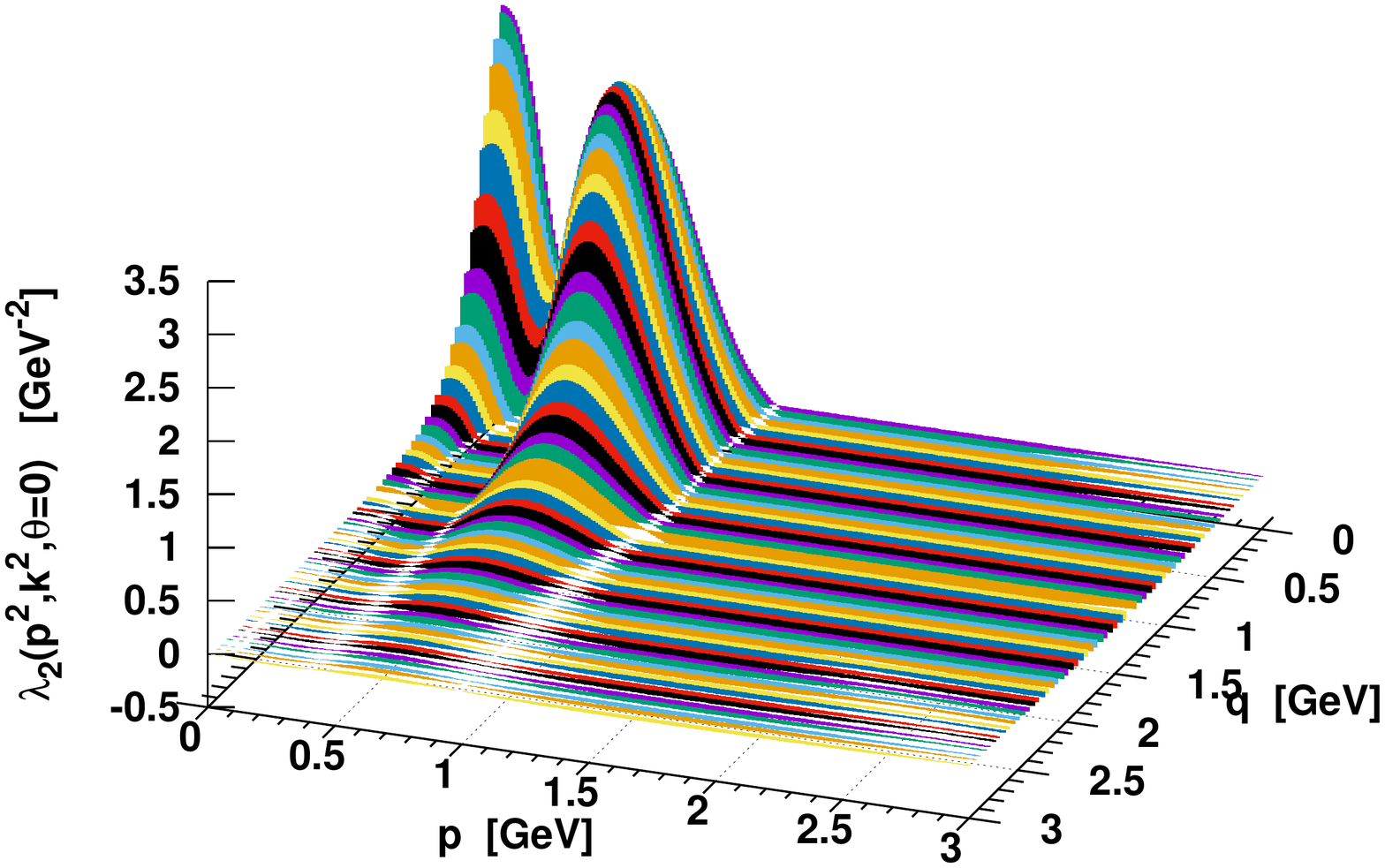} 
\vspace{-2cm} \\
        \includegraphics[width=3.5in]{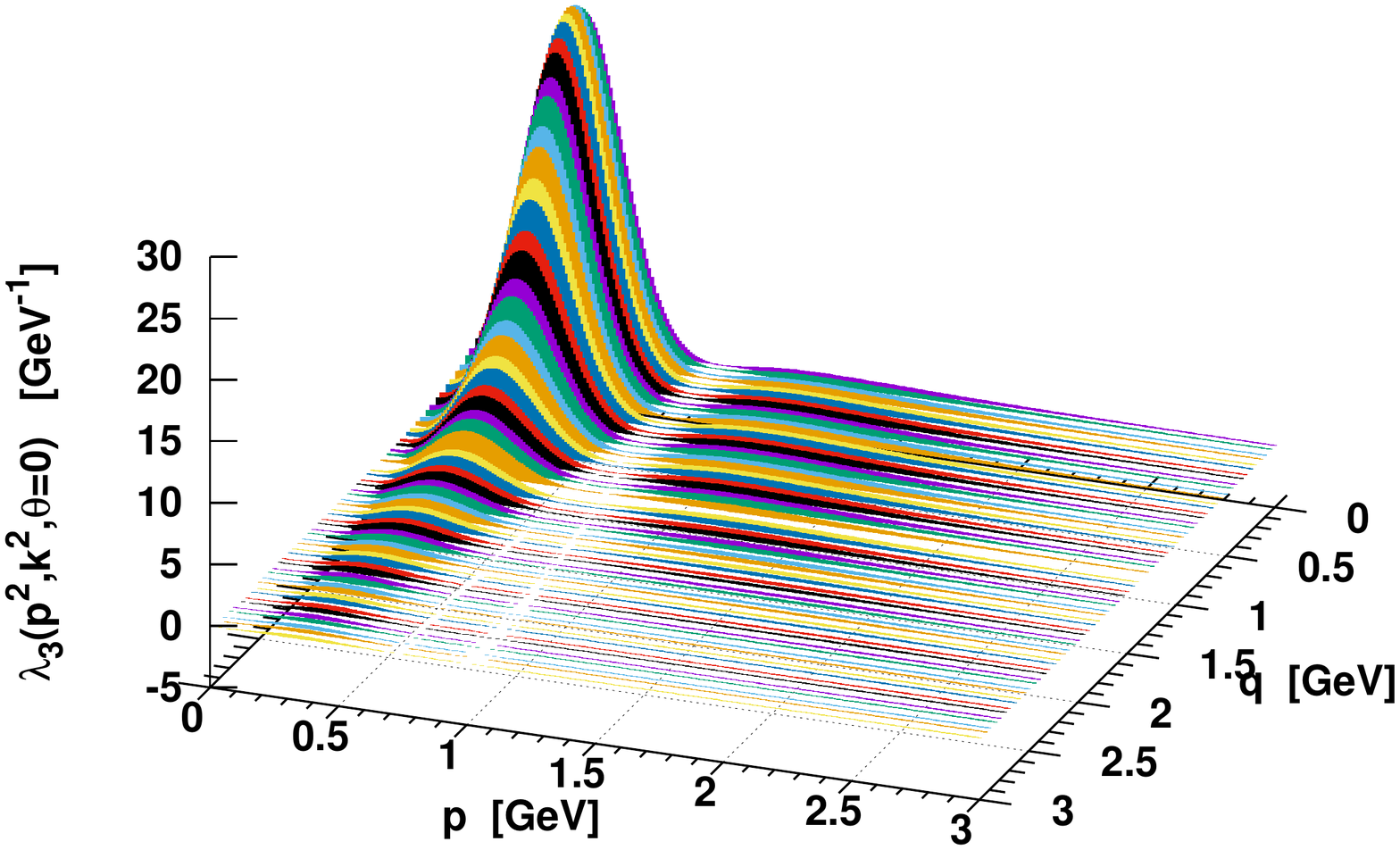} 
\vspace{-2cm} \\
        \includegraphics[width=3.5in]{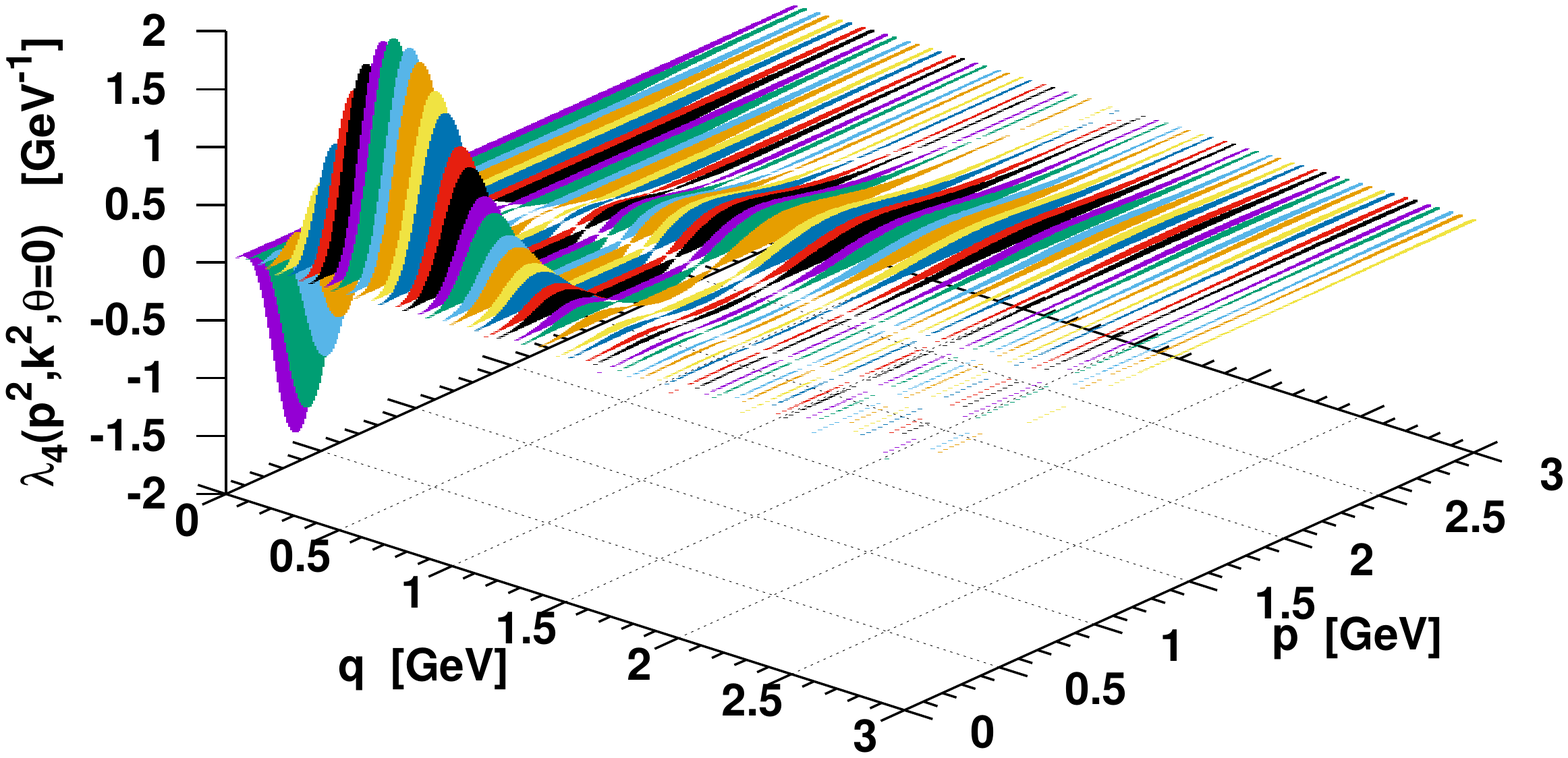}   
\caption{Longitudinal quark-gluon form factor for $\theta = 0$.}    
\label{Fig:3d_all_ang0}
\end{figure}

In the previous section we have computed the quark-ghost kernel form factors $X_0(q^2)$, $Y_1(q^2)$, $Y_3(q^2)$ that, together with
the quark, gluon and ghost propagators,  define the full form factors as given in Eqs. (\ref{Eq:AnsatzX0}),  (\ref{Eq:AnsatzX1}) and (\ref{Eq:AnsatzX3}).
Once the full quark-ghost kernel form factors are known, then the longitudinal quark-gluon form factors can be computed using
Eqs. (\ref{Vertex:L1}) -- (\ref{Vertex:L4}), after performing the rotation to the Euclidean space and identifying the $g_i(p^2_1, p^2_2)$ functions to
\begin{eqnarray}
& & g_0(p^2_1, p^2_2) = 1 \qquad\qquad\mbox{ and } \nonumber \\
& &  g_1(p^2_1, p^2_2) = g_2(p^2_1, p^2_2) = D \left(  \frac{p^2_1 + p^2_2}{2} \right) \ .
\end{eqnarray}
For completeness, we write the full expressions for the longitudinal form factors in Euclidean space
\begin{eqnarray}
& & \lambda_1 (-p, \, k = p-q, \, q  ) =  \frac{F(q^2)}{2} \Bigg\{ \nonumber \\
& & \bigg[ A(p^2)  + A( k^2 ) \bigg]  \, X_0(q^2) \nonumber \\
   & &     \hspace{0.5cm} 
              + \, 2 \, \bigg[ B(p^2)  + B( k^2 ) \bigg] \, D\left( \frac{p^2 + k^2}{2} \right) \, Y_1(q^2)  \nonumber \\
              & &
              \hspace{1cm} + \, \bigg[ A(p^2) \bigg( (pq) - 2 p^2 \bigg)  \nonumber \\
              & & \hspace{3.75cm} + A(k^2) \bigg( 3 (pq) - 2 p^2 - q^2 \bigg)   \bigg]  \times \nonumber \\
              & & \hspace{3cm} \times \, D\left( \frac{p^2 + k^2}{2} \right) \, Y_3(q^2)  \Bigg\} \, , \label{Vertex:FullL1}
\end{eqnarray}
\begin{eqnarray}              
& & \lambda_2 (-p, \, k = p-q, \, q )  =  \frac{F(q^2)}{2( 2 (p \cdot q) - q^2 )} \Bigg\{ \nonumber \\
& &  \bigg[ A(k^2) - A(p^2) \bigg]  \, X_0(q^2)  \nonumber \\
& &  \hspace{1cm}
                              \!        + \bigg[  A(k^2) \bigg( q^2 -  (pq) \bigg) - A(p^2) \, (pq)   \bigg]  \times \nonumber \\
& & \hspace{3.75cm}                              
                                            \times D\left( \frac{p^2 + k^2}{2} \right) \, Y_3(q^2) \Bigg\}  \, , \label{Vertex:FullL2}
\end{eqnarray}
\begin{eqnarray}                                            
& & \lambda_3 (-p, \, k = p-q, \, q ) =  \frac{F(q^2)}{2 (p \cdot q) - q^2 } \Bigg\{  \nonumber \\
& & \bigg[ B(k^2) - B(p^2) \bigg]   \,  X_0(q^2) \nonumber \\
   & &  \hspace{1cm} + \bigg[  A(p^2) \, ( p q )  - A( k^2 ) \, \bigg( q^2 -   (p q) \bigg)   \bigg] \times \nonumber \\
   & & 
   \hspace{3.75cm} \times D\left( \frac{p^2 + k^2}{2} \right)  \, Y_1(q^2) \Bigg\}  ,  \label{Vertex:FullL3}
\end{eqnarray}
\begin{eqnarray}   
& & \lambda_4 (-p, \, k = p-q, \, q ) =  \frac{F(q^2)}{2} \,\, D\left( \frac{p^2 + k^2}{2} \right)  \times\nonumber \\
& & \hspace{0.25cm} \times
  \Bigg\{  \bigg[ A(k^2) - A(p^2) \bigg]    \, Y_1(q^2)  \nonumber \\
  & & \hspace{3.5cm}+ \bigg[ B(k^2) - B(p^2) \bigg]  \, Y_3(q^2) \Bigg\}  \, . \label{Vertex:FullL4} 
\end{eqnarray}
Note that by taking into account  structures of the quark-ghost kernel other than $X_0$ the quark-gluon vertex deviates considerably from a Ball-Chiu type  
and it is now a function both of $p$, $q$ and of the angle between the quark and gluon momenta. 
The angular dependence appears associated to the scalar products $(pq)$ and also on the argument of the gluon propagator $D\left( (p^2 + k^2)/2 \right)$.

For the calculation of $\lambda_1$--$\lambda_4$ we will use Sol. II computed using $\alpha_s( \mu ) = 0.22$; see Sec.~\ref{Sec:tune_alpha} for details.
We recall the reader that the calculation performed here considers only the longitudinal form factors and that the \textit{ansatz} for the vertex takes into account
the dependence between the angle of the incoming quark momentum and the incoming gluon momentum.

\begin{figure}[t]
\vspace{-0.3cm}
        \includegraphics[width=3.5in]{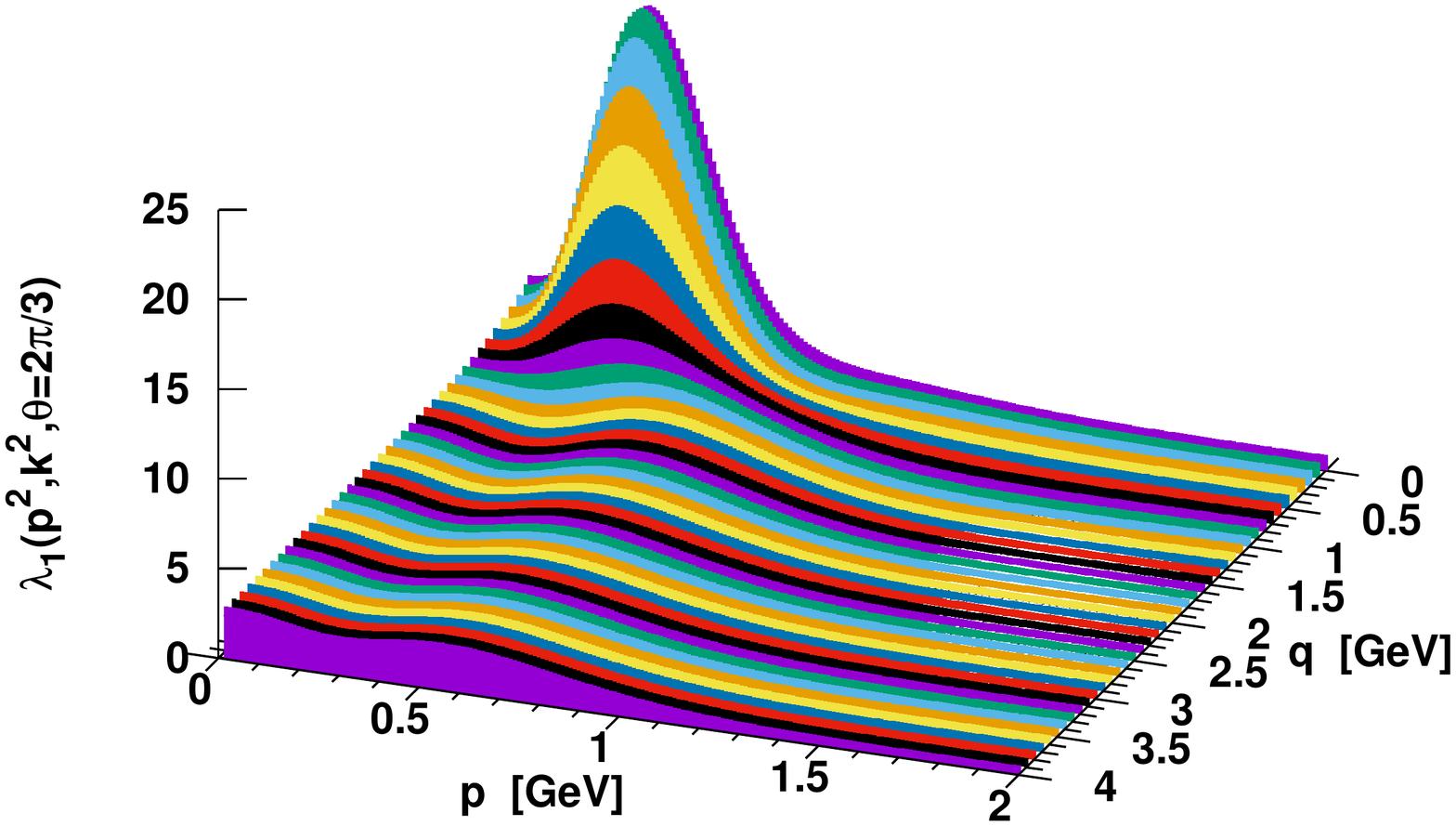} 
\vspace{-2cm} \\
        \includegraphics[width=3.5in]{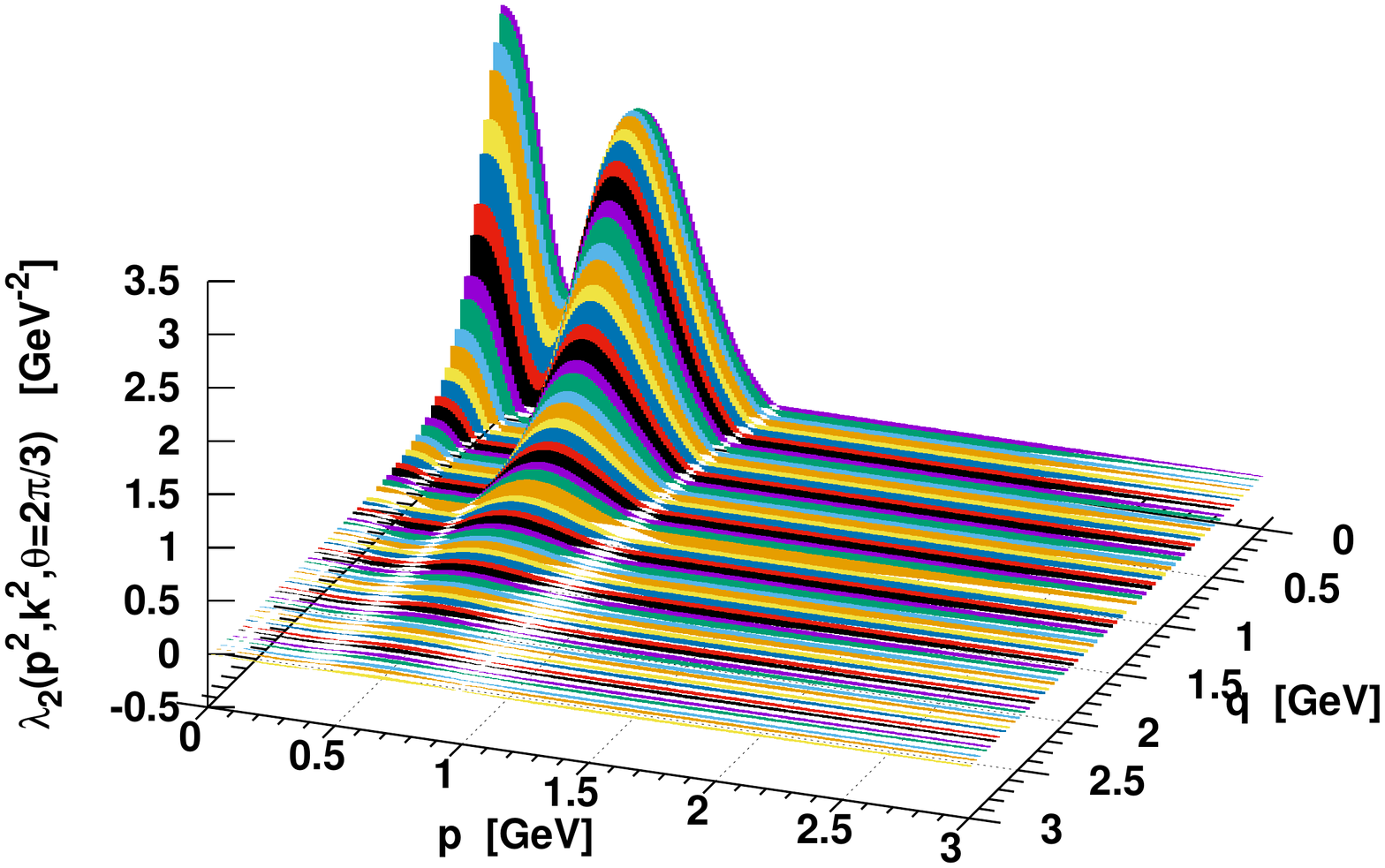} 
\vspace{-2cm} \\
        \includegraphics[width=3.5in]{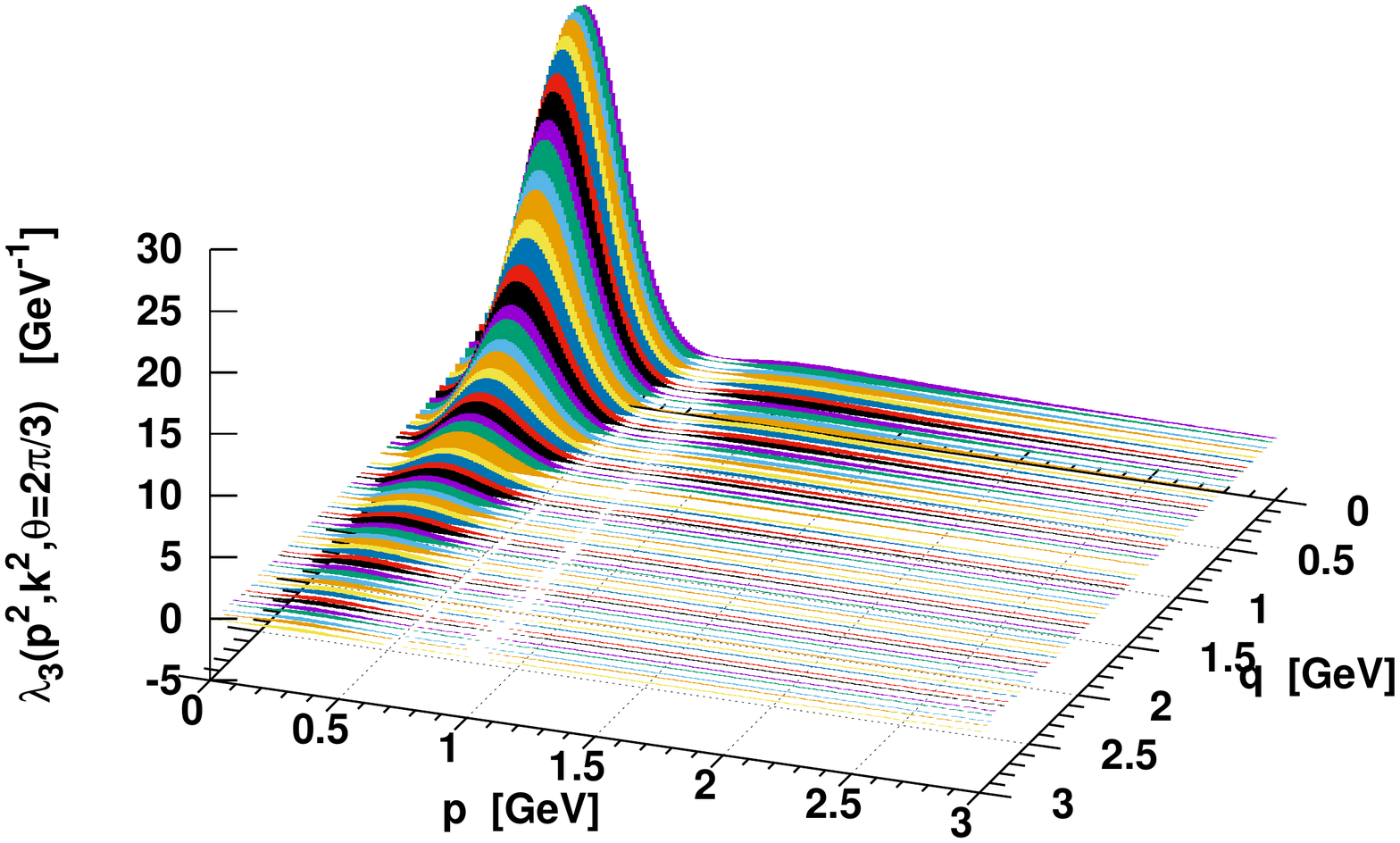} 
\vspace{-2cm} \\
        \includegraphics[width=3.5in]{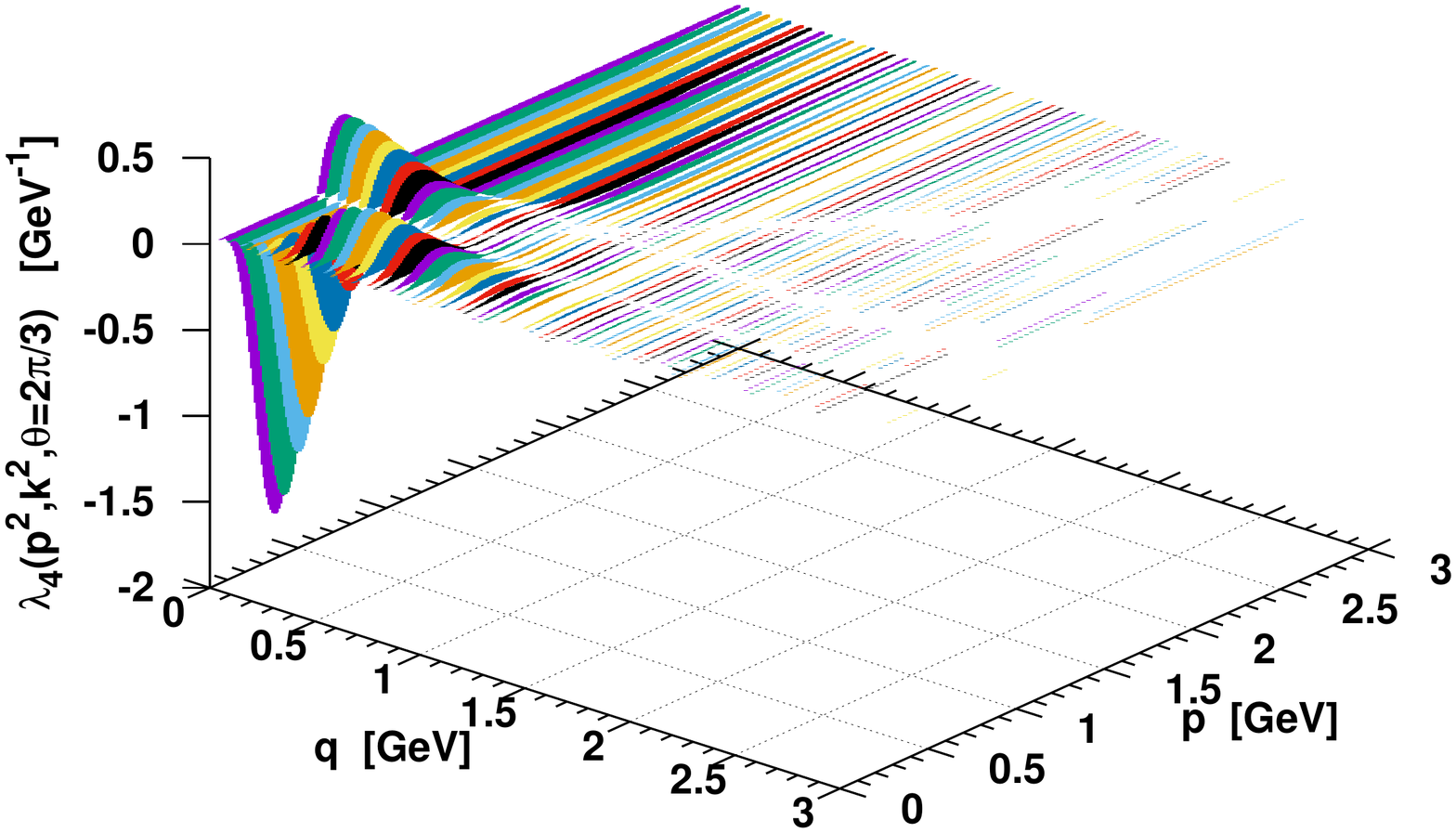} 
\caption{Longitudinal quark-gluon form factor for $\theta = 2 \pi / 3$.}    
\label{Fig:3d_all_ang60}
\end{figure}

The overall picture of the various form factors when the angle between the incoming quark momentum $p$ and the incoming gluon momentum $q$ is $\theta = 0$ can be seen 
on Fig.~\ref{Fig:3d_all_ang0}. On Fig.~\ref{Fig:3d_all_ang60} the $\lambda_1$ to $\lambda_4$ are given for a $\theta = 2 \pi /3$. 
The form factors $\lambda_1$ to $\lambda_4$ are finite for all $p$ and $q$ and approach asymptotically their perturbative values.
Further, for our definition of the operators $L^{(1)}_\mu$ -- $L^{(4)}_\mu$, see Eqs. (\ref{Eq:LongOperators}) for their definition in Minkowski space,
the corresponding form factors are essentially positive defined. The exception being $\lambda_4$ that takes both positive and negative values and whose
maximum absolute value is negative and appears for small $p$ and $q$. The relative magnitude of the $\lambda_i$ suggest that the quark-gluon vertex is
essentially saturated by $\lambda_1$ and $\lambda_3$, with $\lambda_2$ and $\lambda_4$ playing a minor role, i.e. the tensor structures of the longitudinal
part of the vertex seem to be sub-leading; see also the discussion for the soft quark limit, defined by a vanishing quark momentum,
and the symmetric limit below. 

Our result differs significantly from the perturbative estimation of the form factors~\cite{Bermudez:2017bpx}, where all the strength appears 
associated to $\lambda_1$. 
For example, for the kinematical configuration defined by $p^2 = (p-q)^2$ at vanishing $p$ they have $\lambda_1 \approx 1.1$ and $\lambda_2 \approx 0.12$ GeV$^{-2}$ 
and $\lambda_3 \approx 0.18$ GeV$^{-1}$ for a current mass $m_q = 115$ MeV, a renormalisation scale $\mu = 2$ GeV and for $\alpha_s = 0.118$. 
Of course, one should look to the relative values of the various $\lambda$'s and not to their absolute values. 
For the comparison of the contributions from the various form factors one can use the non-perturbative momentum scale of 1 GeV to build dimensionless 
quantities. Then, as seen on Figs.~\ref{Fig:3d_all_ang0} and~\ref{Fig:3d_all_ang60} the scales for $\lambda_1$ and $\lambda_3$ are similar, while 
the maximum of $\lambda_2$ is about 10\% relative to the maxima of $\lambda_1$ and $\lambda_3$ and the maximum for $\lambda_4$ is about half of that 
for $\lambda_2$. 

The comparison of our results with those reported in~\cite{Aguilar:2014lha,Aguilar:2016lbe,Aguilar:2018epe} is difficult to perform but in these 
works $\lambda_1$ clearly dominates. On~\cite{Aguilar:2014lha} $\lambda_2$ reaches at most 16\% of the maximum value of $\lambda_1$, while
$\lambda_3$ seems to have the possibility of taking large values. On~\cite{Aguilar:2016lbe,Aguilar:2018epe}, $\lambda_2$ and $\lambda_3$ take, at most, 
a numerical value that is about 23\% of the maxima of $\lambda_1$, with $\lambda_4$ being essentially negligible. Our solution shows a vertex dominated
by $\lambda_1$ and $\lambda_3$ with these form factors reaching numerical values of the same order of magnitude -- see also Fig.~\ref{Fig:Ls_0_p_eq_q}.

As seen on Figs.~\ref{Fig:3d_all_ang0} and~\ref{Fig:3d_all_ang60} the quark-gluon form factors are significantly enhanced for low values of $p$ and $q$. 
The momentum region where one observes the enhancement of the $\lambda_1$  to $\lambda_4$ happens for $p \lesssim 1 $ GeV and $q \lesssim 1$ GeV,
with its maximum values showing up for $p \approx q \approx \Lambda_{QCD}$ -- see, also, the discussion below on the angular dependence.

The infrared enhancement of $\lambda_1$ to $\lambda_4$ with the gluon momentum is a direct consequence of using the Slavnov-Taylor identity (\ref{STI}) to
rewrite the form factor. Indeed, as can be seen on Eqs. (\ref{Vertex:FullL1}) -- (\ref{Vertex:FullL4}), all the form factors have, as a global factor, the ghost
dressing function $F(q^2)$. The ghost dressing function is enhanced, roughly by a factor of three, 
in the infrared, see Fig.~\ref{fig:gluon_ghost_dress_func},  implying the increase of the $\lambda_i$ as $q = 0$ is approached.

The infrared enhanced of the form factors with the quark incoming momentum is more subtle. It is linked to our \textit{ansatz} 
that relies on the analysis of the soft gluon limit of the Landau gauge lattice data for $\lambda_1$ performed in~\cite{Oliveira2018a}.
Indeed, this work identified a dependence of $\lambda_1$ on the gluon propagator that was incorporated in the \textit{ansatz},
making the quark-ghost kernel form factors $X_1$ and $X_3$  proportional to $D ( (p^2 + (p-q)^2 / 2)$. This term is crucial to have well behaved kernels in
the integral equations, i.e. to ensure that the Dyson-Schwinger equations are finite, and it introduces an additional dependence on the angle between the quark 
and the gluon momenta. The gluon propagator is a decreasing function of its argument and, therefore, for a given $q$ and angle between the 
quark and gluon momenta, the terms proportional to $X_1$ and $X_3$ increase as $p$ decreases. This explains, in part, the observed enhanced of
the quark-gluon form factors together with the increase of the ghost dressing function.

\clearpage
\begin{figure}[t]
\centering
\vspace{-0.3cm}
        \includegraphics[width=3.5in]{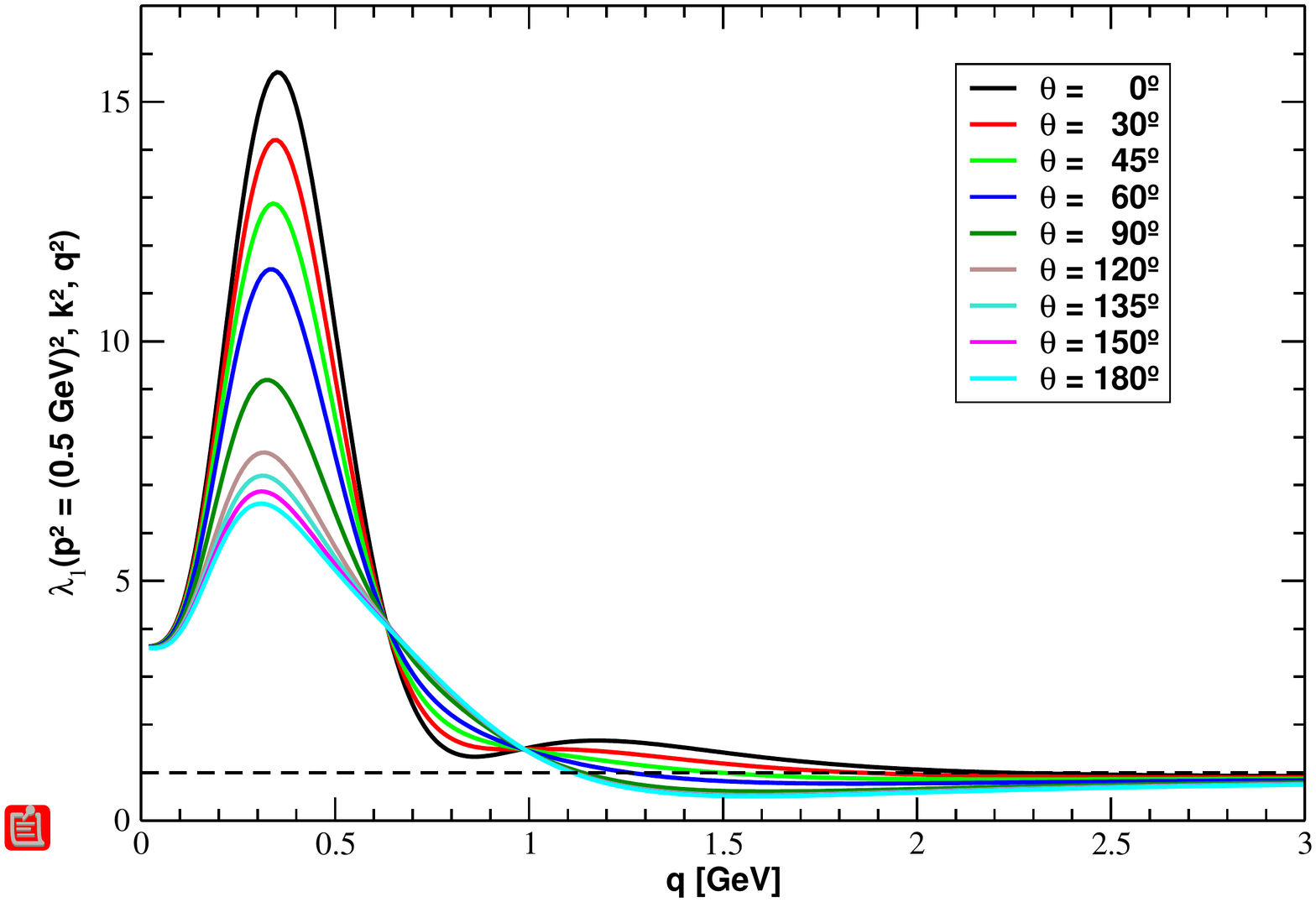} \\
\vspace{-0.5cm}        
        \includegraphics[width=3.5in]{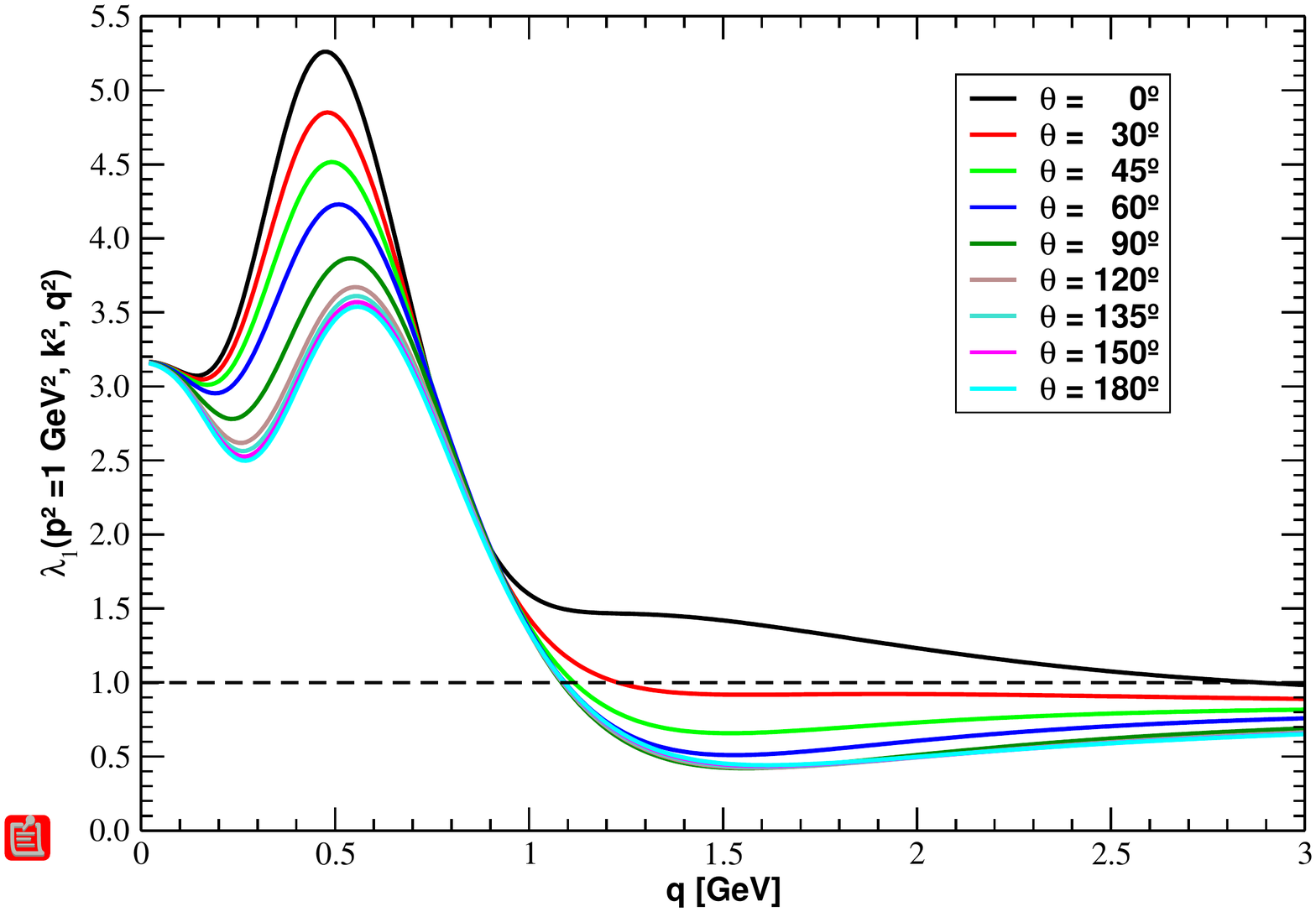}
\caption{$\lambda_1$ for $p = 0.5$ GeV (top) and $p = 1$ GeV (bottom) and various $\theta$.}   
\label{Fig:L1_ang} 
\end{figure}

\begin{figure}[t]
\centering
\vspace{-0.3cm}
        \includegraphics[width=3.5in]{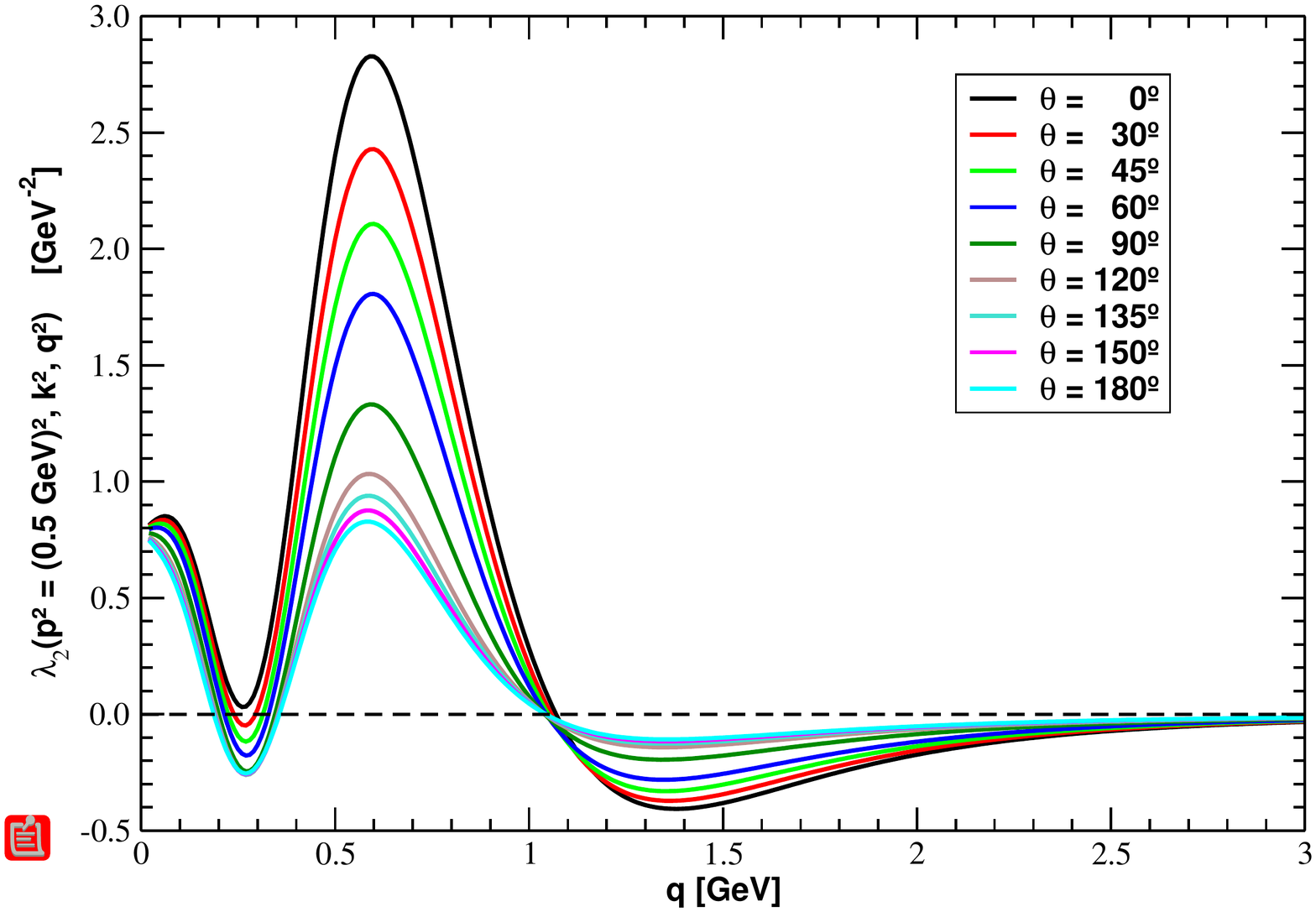} \\
\vspace{-0.5cm}        
        \includegraphics[width=3.5in]{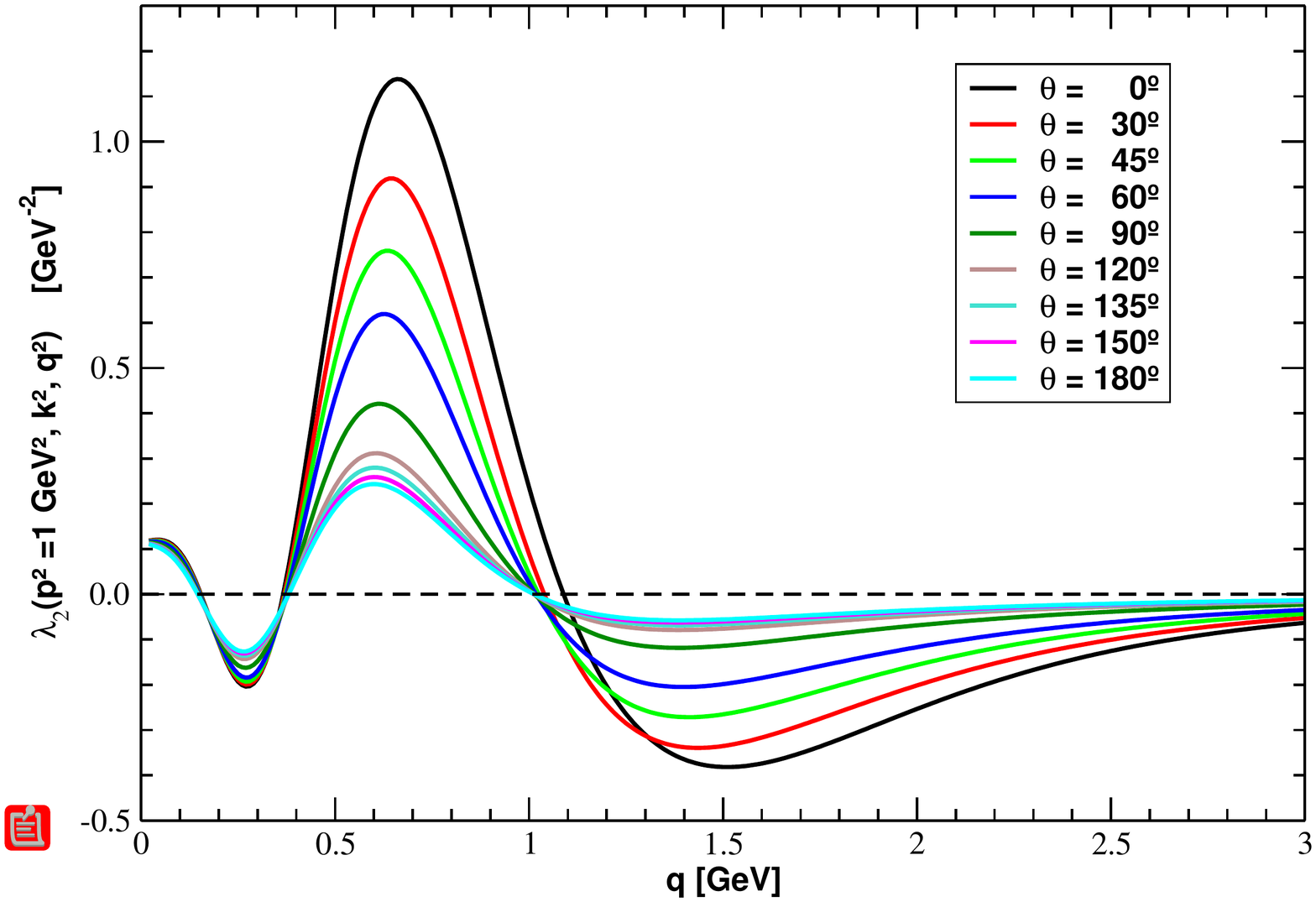}
\caption{$\lambda_2$ for $p = 0.5$ GeV (top) and $p = 1$ GeV (bottom) and various $\theta$.}    
\label{Fig:L2_ang} 
\end{figure}

\begin{figure}[t]
\centering
\vspace{-0.3cm}
        \includegraphics[width=3.5in]{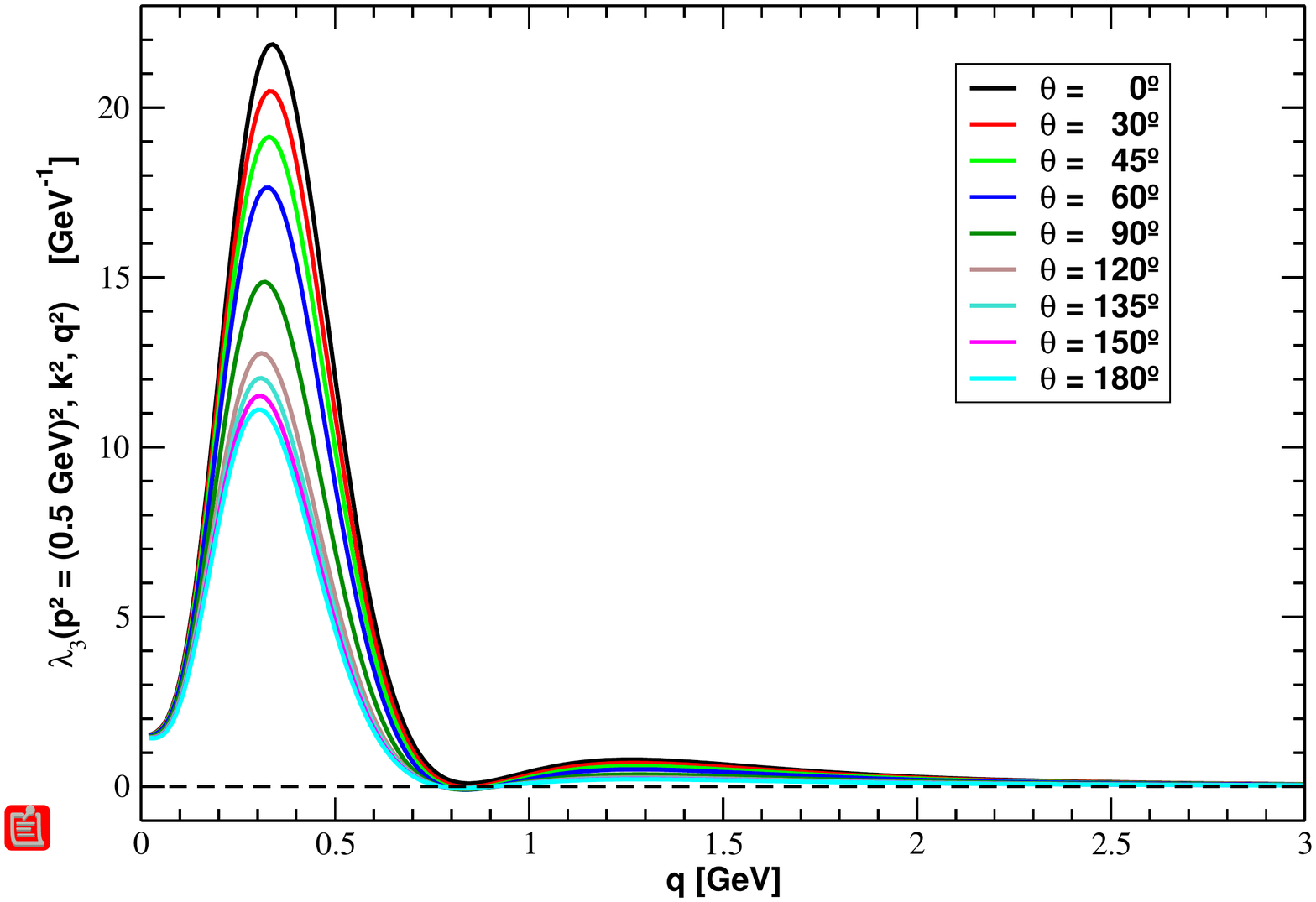} \\
\vspace{-0.5cm}        
        \includegraphics[width=3.5in]{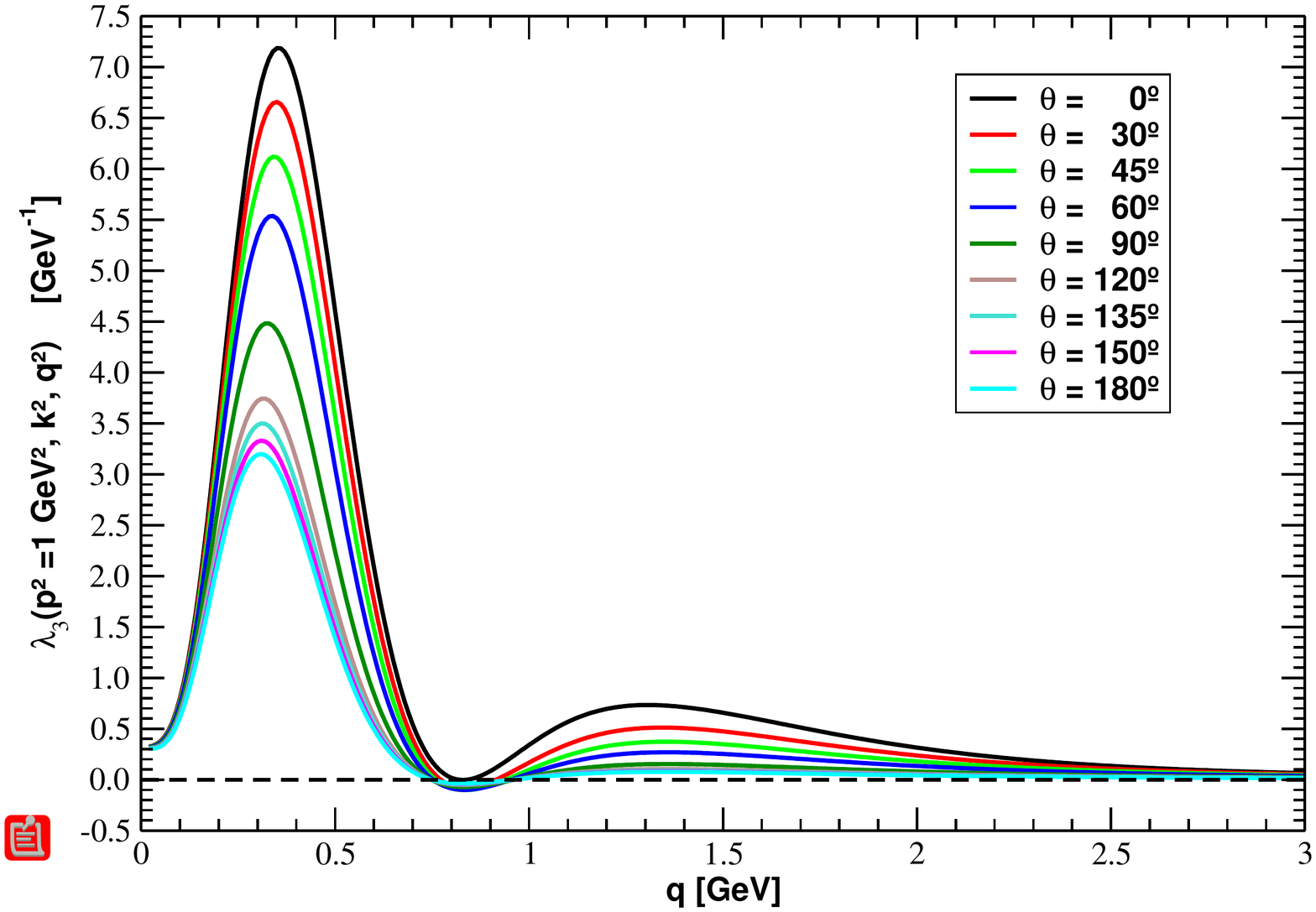}
\caption{$\lambda_3$ for $p = 0.5$ GeV (top) and $p = 1$ GeV (bottom) and various $\theta$.}    
\label{Fig:L3_ang} 
\end{figure}

\begin{figure}[t]
\centering
\vspace{-0.3cm}
        \includegraphics[width=3.5in]{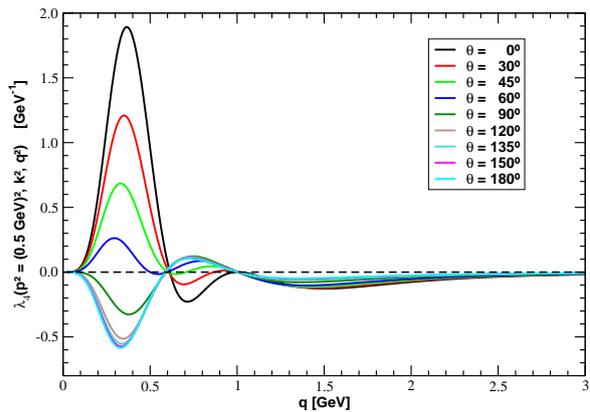} \\
\vspace{-0.5cm}        
        \includegraphics[width=3.5in]{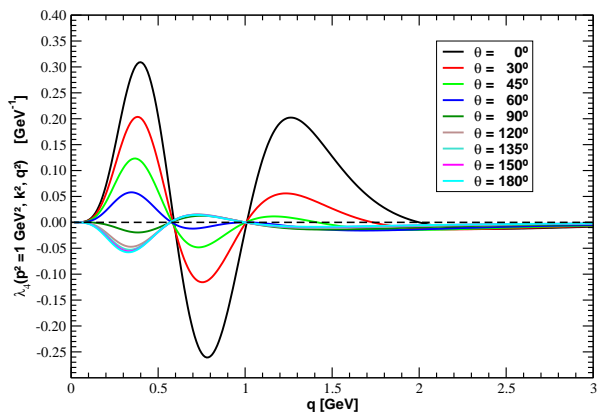}
\caption{$\lambda_4$ for $p = 0.5$ GeV (top) and $p = 1$ GeV (bottom) and various $\theta$.}    
\label{Fig:L4_ang} 
\end{figure}

The dependence of the quark-gluon form factors in the angle between $p$ and $q$ 
can be seen on Figs.~\ref{Fig:L1_ang} -- \ref{Fig:L4_ang}. 
These figures also provide a clear picture of the maxima of the various form factors as functions
of the gluon momenta. For $\lambda_1$ and $\lambda_3$ the maxima are for $q \approx 300$ MeV, while for
$\lambda_2$ the maximum is at $q \approx 600$ MeV. $\lambda_4$ seems to be a more complicated function of $p$, $q$ and 
$\theta$. Indeed, this later form factor shows various maxima of the same order of magnitude for different $p$, $q$ and $\theta$ values.

All the form factors appear to be monotonous decreasing functions of the angle between the incoming quark and incoming gluon momenta $\theta$. If
the pattern of the $q$ dependence of $\lambda_1$, $\lambda_2$ and $\lambda_3$ seems to be independent of $\theta$,
$\lambda_4$ seems to reverse is behaviour relative to the $q-axis$ for $\theta \gtrsim \pi /3$.
Clearly, the maximum values for all the form factors occurs for $\theta = 0$, i.e. the quark-gluon vertex favours the kinematical configurations
with small values of $p$ and $q$ and also of the angle between the quark and gluon 
momentum\footnote{For example, for $p = 0.5$ GeV $\lambda_1$ and $\lambda_3$ 
there is an enhancement of a factor of $\sim 2.7$ and $\sim 2.0$, respectively, between the maxima values for $\theta = 0$ relative to $\theta = \pi$.
For $\lambda_2$ and $\lambda_4$ this enhancement is $\sim 3.5$ and $\sim 3.2$. The corresponding factors for an incoming quark momentum
$p = 1$ GeV are $\sim 1.5$ for $\lambda_1$, $\sim 4.4$ for $\lambda_2$, $\sim 2.3$ for $\lambda_3$ and a suppression by a factor of $\sim 0.6$
for $\lambda_4$. Also, for $\lambda_1$ and $\lambda_3$ the maxima at $\theta = \pi/2$ is about half of the maxima at $\theta = 0$.}.
It follows that the quark-gluon vertex favours small quark and gluon momenta and parallel four-vectors $p$ and $q$.

From the point of view of the momentum dependence, our solution for the quark-gluon vertex is closer to that of the Maris-Tandy model~\cite{Maris99a}
than those computed in~\cite{Aguilar:2014lha,Aguilar:2016lbe,Aguilar:2018epe}, in the sense that we observe a rather strong
enhancement at low momenta. Indeed, compared to these last references, the herein computed form factors are significantly larger.
Recall that the Maris-Tandy model considers a single form factor, that
would be (effective) equivalent to our $\lambda_1$, and ignores the dependence of the vertex on the quark momentum. 
In particular, for this model we also checked 
that the region where our quark-gluon form factors are enhanced occurs essentially within the same range of momenta
as  the corresponding effective form factor of the Maris-Tandy model. Note also that the maxima of the form factors computed in the present work occur for
momenta where the kernels appearing in the original equations take their maximum values -- see Figs.~\ref{fig:NKernels_0},~\ref{fig:NKernelsProp_2}
and~\ref{fig:NKernelsProp_3}.

\begin{figure}[h]
\centering
\vspace{-0.3cm}
        \includegraphics[width=2.9in]{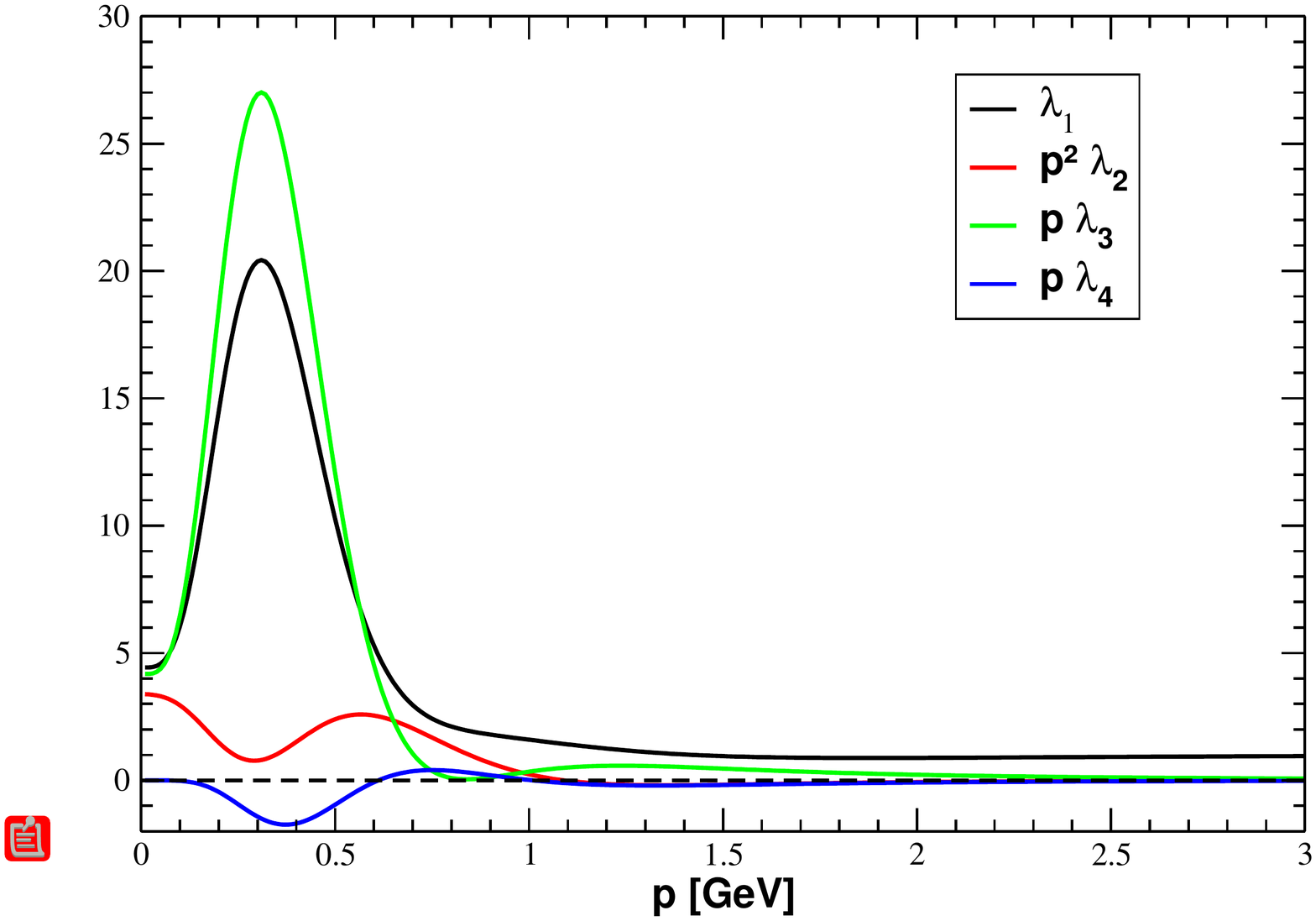} \\
\vspace{-0.5cm}
        \includegraphics[width=2.9in]{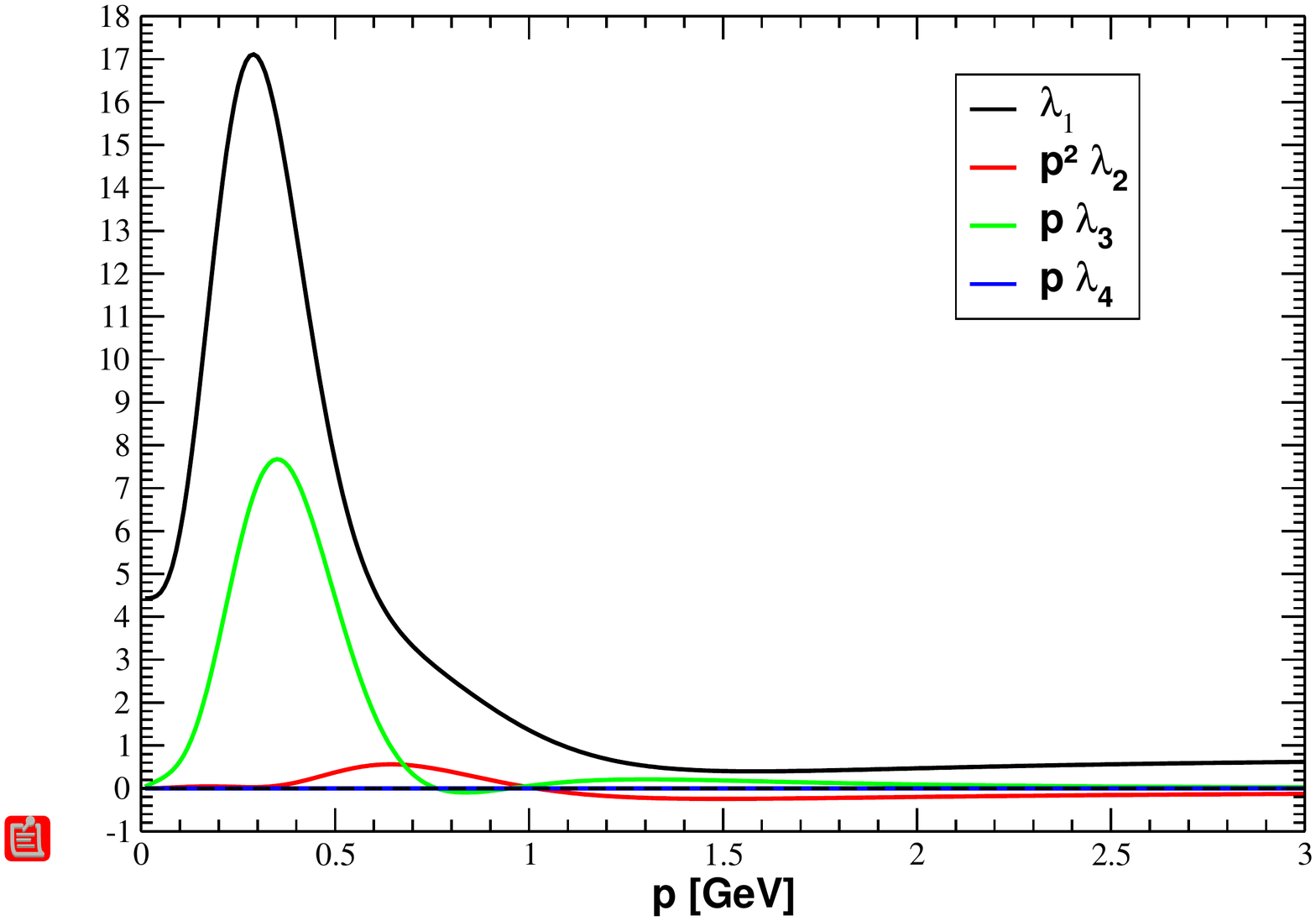}
\caption{The form factors for the soft quark kinematical configuration (top) and for the completely symmetric kinematical configuration (bottom). DDDD}    
\label{Fig:Ls_0_p_eq_q} 
\end{figure}

It is difficult to measure the relative importance of the contribution of the longitudinal form factors $\lambda_1$ -- $\lambda_4$ to the quark-gluon vertex.
However, an idea of their relative importance can be ``measured'' looking at particular kinematical configurations. 
Herein we consider the soft quark limit where the incoming quark momentum vanish and the totally 
symmetric limit where $p^2 = q^2 = k^2$ and $\theta = 2 \pi / 3$. The corresponding form factors multiplied by appropriated powers of momenta to
build dimensionless function can be seen on Fig.~\ref{Fig:Ls_0_p_eq_q} (computed using the $\theta = 2 \pi /3$ data). 
If for the symmetric configuration the dominant form factor seems to be $\lambda_1$, for the soft quark limit that role is played by $p ~  \lambda_3$.
Note that the maximum of the later is about 1.3 times larger than the maximum of the former. Curiously, the maxima of $\lambda_1$ and $p \lambda_3$ occur
at exactly the same momentum scale $p = 310$ MeV. As the figure shows it seems that the quark-gluon vertex is dominated by $\lambda_1$ and $\lambda_3$,
as observed also when studying the solutions associated with the perturbative $X_0$ as discussed at the end of Sec.~\ref{Sec:OneLoopSolution},
with the tensor structures associated to $\lambda_2$ and $\lambda_4$ playing a minor role.

\begin{figure}[t] 
   \centering
      \includegraphics[width=3.5in]{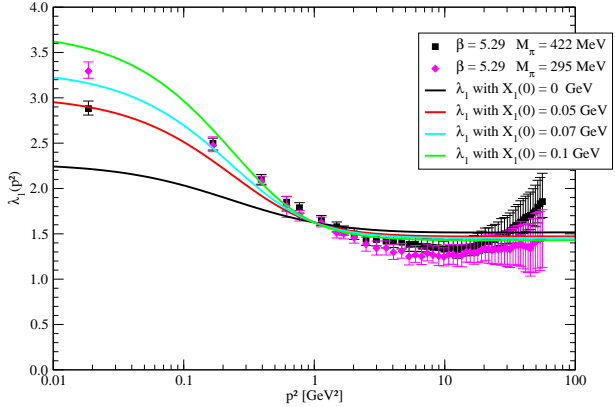}
   \caption{The soft gluon limit prediction for $\lambda_1$. See text for details.}
   \label{fig:L1_softgluon}
\end{figure}

Finally, let us consider the soft gluon limit whose $\lambda_1$ form factor has recently being computed using full QCD lattice simulations~\cite{Oliveira:2016muq}.
The data was investigated in~\cite{Oliveira2018a} revealing an important contribution to $\lambda_1$ linked with the gluon propagator.
Our estimation of $\lambda_1$ in the soft gluon limit can be seen on Fig.~\ref{fig:L1_softgluon} -- see the full curve in black.
This curve was (arbitrarely) normalised to reproduce the lattice data at 1 GeV of the $\beta = 5.29$ and
$M_\pi = 295$ MeV simulation\footnote{The normalisation essentially removes the $F(0)$ that can appear in the expression for $\lambda_1$; see Eq. (\ref{Vertex:FullL1}).}. 
Clearly, our ansatz underestimates $\lambda_1$ in the infrared region. As discussed in~\cite{Oliveira2018a}, in the soft gluon
limit
\begin{equation}
  \lambda_1(p^2) = \frac{F(0)}{Z(p^2)} \bigg\{ 1 + 2 \, M(p^2) \, X_1(p^2) - 2 \, p^2 \, X_3(p^2) \bigg\} \ ,
\end{equation}
where $p$ is the quark incoming momenta, and in our notation
\begin{eqnarray}
& &
  X_1(p^2) \rightarrow D ( p^2 ) \, Y_1(0)
  ~ ~ \mbox{ and } ~ ~
  X_3(p^2) \rightarrow D( p^2 ) \, Y_3(0) \ .
\end{eqnarray}
Our solutions has $Y_1(0) \approx 0$ GeV and $Y_3 (0) \approx 0$ and, therefore, it underestimates $\lambda_1 (p^2)$
in the infrared region.
Note that herein $Y_1$ and $Y_3$ are assumed to be a function only of the gluon momentum and, due to the integration over the gluon momenta $q$ in the Dyson-Schwinger 
equations, these form factor are multiplied by $q^3$ that, possibly, prevent the inversion to resolve correctly $Y_1(q^2)$ and $Y_3(q^2)$ in the deep infrared region. 
If in the calculation of the soft gluon limit one assumes that $Y_1(0)$ deviates from zero by a small quantity, the agreement with the lattice data
is considerably improved both in the infrared and in the ultraviolet. This is represented by the two full curves in colour of Fig.~\ref{fig:L1_softgluon} where $Y_1(0)$ is
set to a small value. The colour curves suggest a $Y_1(0) \sim 0.05$ -- 0.07 GeV. 
Further, the agreement in the ultraviolet region can also be improved if $Y_3(0) $ assumes small and positive values; recall that $Y_3(q^2)$ approaches 
zero from the above when $q^2 \rightarrow 0$ as can be seen on Fig.~\ref{fig:X3_a0p22}.

\section{Summary and Conclusions \label{Sec:summary}}

In this work we investigated the non-perturbative regime of the Landau gauge quark-gluon vertex (QGV), taking into account only its longitudinal components,
and
relying on lattice results for the quark, gluon and ghost propagators, together with continuum exact relations, namely
a Slavnov-Taylor identity and the quark propagator Dyson-Schwinger equation. Furthermore, we incorporate the exact normalisation condition
for the quark-ghost kernel form factor $X_0$ derived in~\cite{Aguilar:2014lha}. 
In addition, we take into account an empirical relation that  links the gluon propagator and 
the soft gluon limit of the form factor $\lambda_1$ checked against full QCD lattice simulations~\cite{Oliveira2018a}.
The full set of the quark-ghost kernel tensor structures are taken into account to build an \textit{ansatz} for the longitudinal quark-gluon vertex 
that is a function of both the incoming quark $p$ and gluon $q$ momenta, and the angle between $p$ and $q$. 

The quark-ghost kernel requires four scalar form factors $X_0$, $X_1$, $X_2$, $X_3$~\cite{Davydychev}. 
For the construction of the quark-ghost kernel a perfect symmetry between
incoming and outgoing quark momentum is assumed, which simplified the description of the QGV in terms of $X_0$, $X_1 = X_2$ and $X_3$.
Charge conjugation demands that for the soft gluon limit, defined by $q = 0$,  $\lambda_4 = 0$ and our construction implements such constraint.
Noteworthy to mention that our \textit{ansatz} goes beyond the Ball-Chiu type of vertex~\cite{Ball:1980ay}
and includes it as a particular case, when $X_1 = X_3 = 0$ and $X_0 = 1$.

The Dyson-Schwinger equations are solved for the quark-gluon vertex that are written in terms of the unknown functions $X_0$, $X_1$ and $X_3$.
From the point of view of the quark-ghost kernel form factors, these are linear integral equations. The corresponding mathematical problem is ill defined 
and needs to be regularised in order to obtain a meaningful solution. The original integral equations for the scalar and vector components of the quark gap equation
are transformed into a set of linear system using Gauss-Legendre quadratures to perform the integrations and after doing the angular integration.
In our approach we rely on the Tikhonov linear regularisation that is equivalent to minimize $|| B - \mathcal{N} X ||^2 + \epsilon ||X||^2$. 
The solutions are found numerically after writing the regularised linear system in its normal form. 
The small parameter $\epsilon$ is set by looking at the balance between the associated error on the Dyson-Schwinger equations, i.e. the difference between the l.h.s and the r.h.s.
$|| B - \mathcal{N} X ||^2$, and the norm of the corresponding quark-ghost form factors, i.e. $||X||^2$, for each solution of the regularised linear system.

The resulting quark-gluon vertex form factors $\lambda_1$ -- $\lambda_4$ show a strong enhancement in the infrared region
and deviate significantly from their tree level results for quark and gluon momenta below $\sim 2$ GeV. 
At high momentum the form factors approach their perturbative values. 
In what concerns the gluon momentum, the observed infrared enhancement for the QGV form factors can be traced back to the multiplicative
contribution of the ghost dressing function introduced through the Slavnov-Taylor identity. Recall that the gluon dressing function peaks at $q = 0$ and,
therefore, favours that the incoming and outgoing quark momentum to be parallel.
On the other hand, the infrared enhancement associated to the quark momentum is linked to the gluon dependence that was observed on the analysis
of the soft gluon limit of the QGV and, clearly, favours small quark momentum $p \sim 0$ and also $p$ parallel to $q$; see Eqs. (\ref{Eq:AnsatzX1}),
(\ref{Eq:AnsatzX3}) and (\ref{Vertex:FullL1}) -- (\ref{Vertex:FullL4}) and, in particular, the argument appearing on the gluon propagator term.
The maxima of the computed form factors are essentially at the maxima of $X_0$, $X_1$ and $X_3$ and they appear for momenta $p, \, q \sim \Lambda_{QCD}$,
which again seems to set the appropriate non-perturbative momentum scale. Recall that  the momentum scale comes from the use of lattice
data for the propagators.
Further, we find that the quark-gluon vertex is dominated by the form factors associated to the tree level vertex $\gamma_\mu$ and to
 $2 \, p_\mu + q_\mu$, with the higher rank tensor structures giving small contributions. Overall, our findings are in qualitative agreement with previous
works both with phenomenological approaches, as in the case of the effective vertex introduced in~\cite{Maris99a}, and those based on first principles \textit{ab initio}
continuum methods, see e.g.~\cite{Aguilar:2018epe} and references therein.

The high momentum behaviour of the quark-gluon vertex form factors reproduces their perturbative values. However, the matching between the computed
form factors and their perturbative tail is not yet implement. In addition, we verified that for the soft gluon limit,  $\lambda_1$ is not able to reproduce 
quantitatively the lattice data from full QCD simulations, apart the qualitative momentum behaviour. This can be traced back to the poor resolution of the
kernel in the deep infrared region, due to the $q^3$ factor coming from the momentum integration. As we have verified,
a small tuning of $X_1$ and $X_3$ at $q = 0$ is enough to reproduce the soft gluon limit lattice data within the present framework.
This two challenging problems, together with inclusion of the transverse part of the vertex,
call for an improvement of the approach devised herein and are to be tackled in a future work. 
Despite of that, we expect that the present results can help understanding the non-perturbative dynamics of quarks and gluons in the infrared region and that
can motivate further applications to the study of hadron phenomenology based on quantum field theoretical approaches as those using
Bethe-Salpeter and/or Faddeev equations -- see e.g.~\cite{Eichmann:2016yit} and references therein.
  
\section*{Acknwlodgements}

This work was partly supported by the
Funda\c{c}\~ao de Amparo \`a Pesquisa do Estado de S\~ao Paulo [FAPESP Grant No. 17/05660-0], Conselho Nacional de Desenvolvimento Cient\'{\i}fico e 
Tecnol\'ogico [CNPq Grant 308486/2015-3] and Coordena\c{c}\~ao de Aperfei\c{c}oamento de Pessoal de N\'{\i}vel Superior (CAPES) of Brazil. 
This work is a part of the project INCT-FNA Proc. No. 464898/2014-5. 
O. Oliveira acknowledge support from CAPES process 88887.156738/2017-00 and FAPESP Grant Number 2017/01142-4. 
JPCBM acknowledges CNPq, contracts 401322/2014-9 and 308025/2015-6. This work was part of the project FAPESP Tem\'atico, No. 2017/05660-0.
 
\begin{appendix}
\section{4D Spherical Coordinates and integration over momentum \label{Sec:integracao}}

In 4D the spherical coordinates are related to the cartesian coordinates as follows
\begin{eqnarray}
 x_1 & = & r \, \cos \phi_1 \nonumber \\
 x_2 & = & r \, \sin \phi_1 \,  \cos \phi_2 \nonumber \\
 x_3 & = & r \, \sin \phi_1 \, \sin \phi_2 \, \cos \phi_3 \nonumber \\
 x_4 & = & r \, \sin \phi_1 \, \sin \phi_2 \, \sin \phi_3 
\end{eqnarray}
where
\begin{equation}
   r \in [0 , \, + \infty [ \, , \qquad \phi_1, \, \phi_2 \in [0 , \, \pi] \qquad\mbox{ and }\qquad \phi_3 \in [ 0 , \, 2 \pi [ \, .
\end{equation} 
The 4D volume element reads
\begin{equation}
  dV = r^3 dr \, \bigg( \sin^2 \phi_1 \, d \phi_1 \bigg) \, \bigg( \sin \phi_2 \, d \phi_2 \bigg) \, \bigg( d \phi_3 \bigg) \ .
\end{equation} 
Setting the outgoing quark momenta $p = (p, \, 0, \, 0 , \, 0 )$ it follows that
\begin{equation}
   p \cdot q = p \, q \, \cos \phi_1
\end{equation}
and the angular integration in the Dyson-Schwinger equations can be written as
\begin{eqnarray}
  \int^\pi_0 \sin^2 \phi_1 \, d \phi_1 ~ \int^\pi_0 \sin \phi_2 \, d \phi_2 ~ \int^{2 \pi}_0 d \phi_3 & = & 4 \pi ~ \int^1_{-1} \sin \phi_1 \, d \bigg( \cos \phi_1 \bigg)
  \nonumber \\ 
  & = & 4 \pi ~ \int^1_{-1} \sqrt{ 1 - x^2} \, dx
\end{eqnarray}
where, in the last identity, we set $ x = \cos \phi_1$. It follows that momentum integration in the Dyson-Schwinger equations reads
\begin{eqnarray}
   \int \frac{d^4q}{(2 \, \pi )^4} & = & \frac{4 \, \pi}{(2 \, \pi )^4} \int^\Lambda_0 \, dq ~ q^3 ~\int^1_{-1} \sin \phi_1 \, d \bigg( \cos \phi_1 \bigg) \nonumber \\
      & = &
   \frac{1}{4 \, \pi^3} \int^\Lambda_0 \, dq ~ q^3 ~\int^1_{-1} \, \sqrt{ 1 - x^2} \, \, d x ~  \cdots
\end{eqnarray}
where $\Lambda$ stands for the cutoff introduced to regulate the theory.

\section{Comparing Propagator Fits with Previous Works \label{Sec:prop_fits}}

\begin{figure}[t] 
   \centering
   \includegraphics[width=3in]{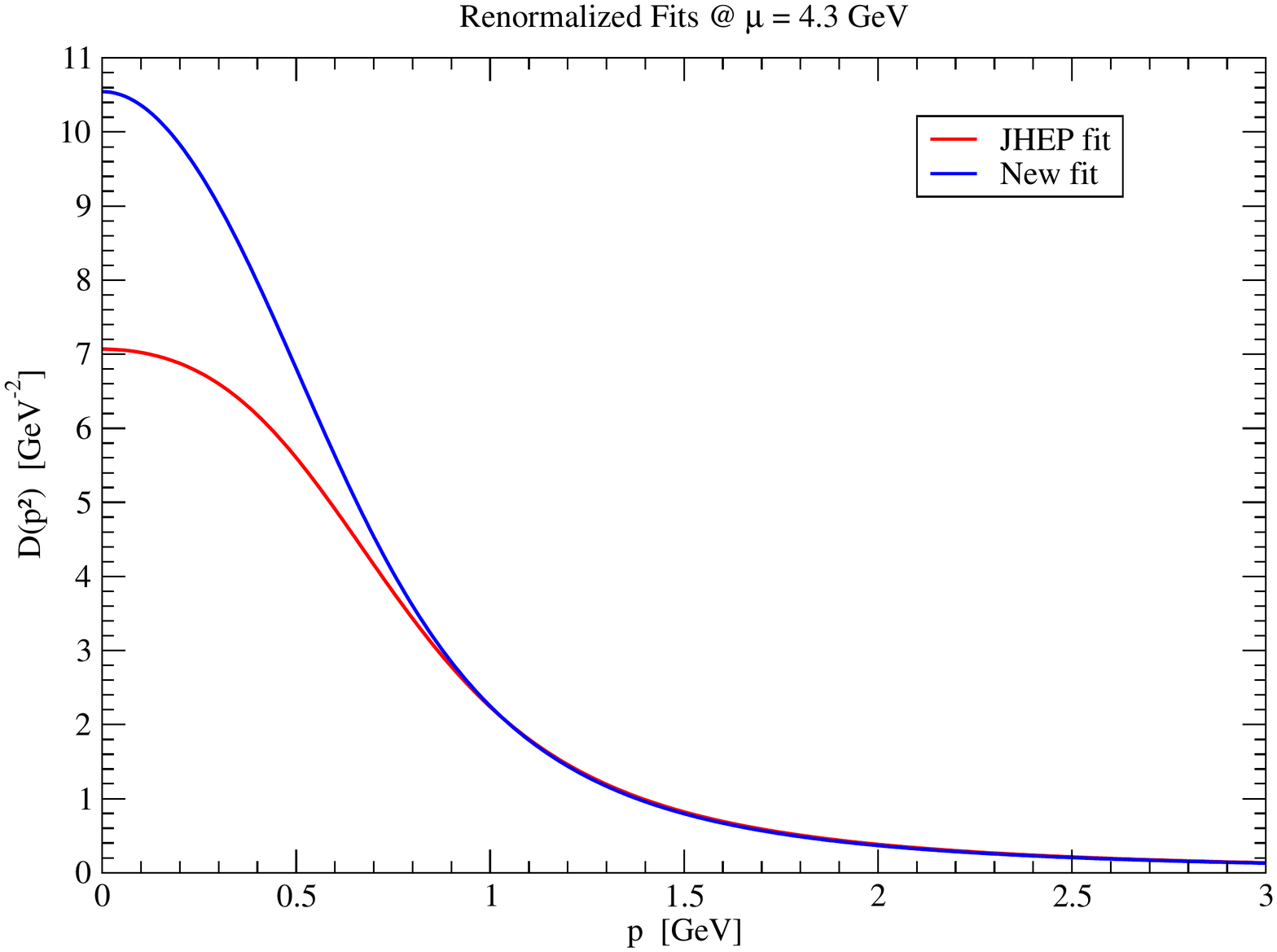} \\
   \includegraphics[width=3in]{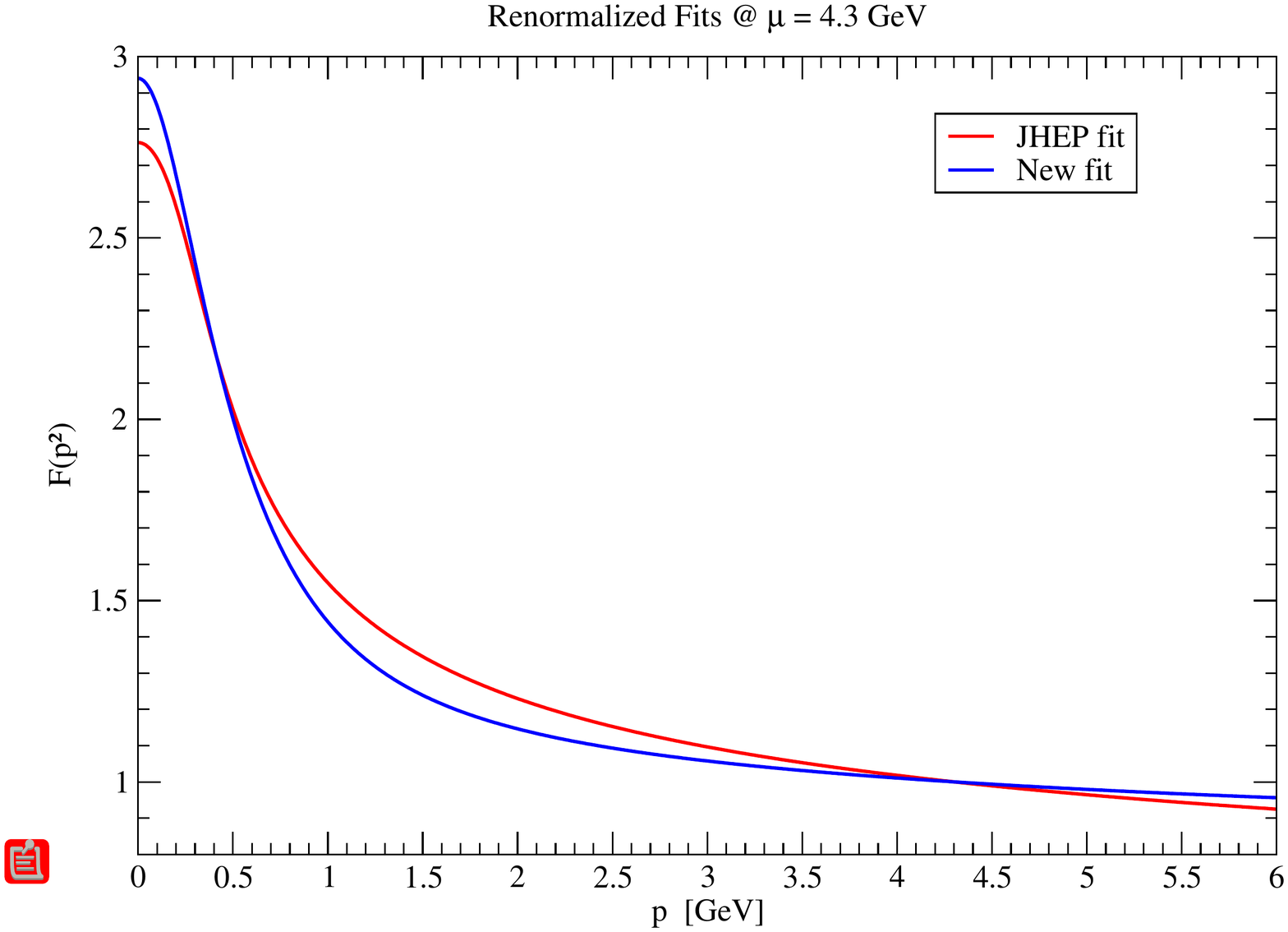} \\
   \includegraphics[width=3in]{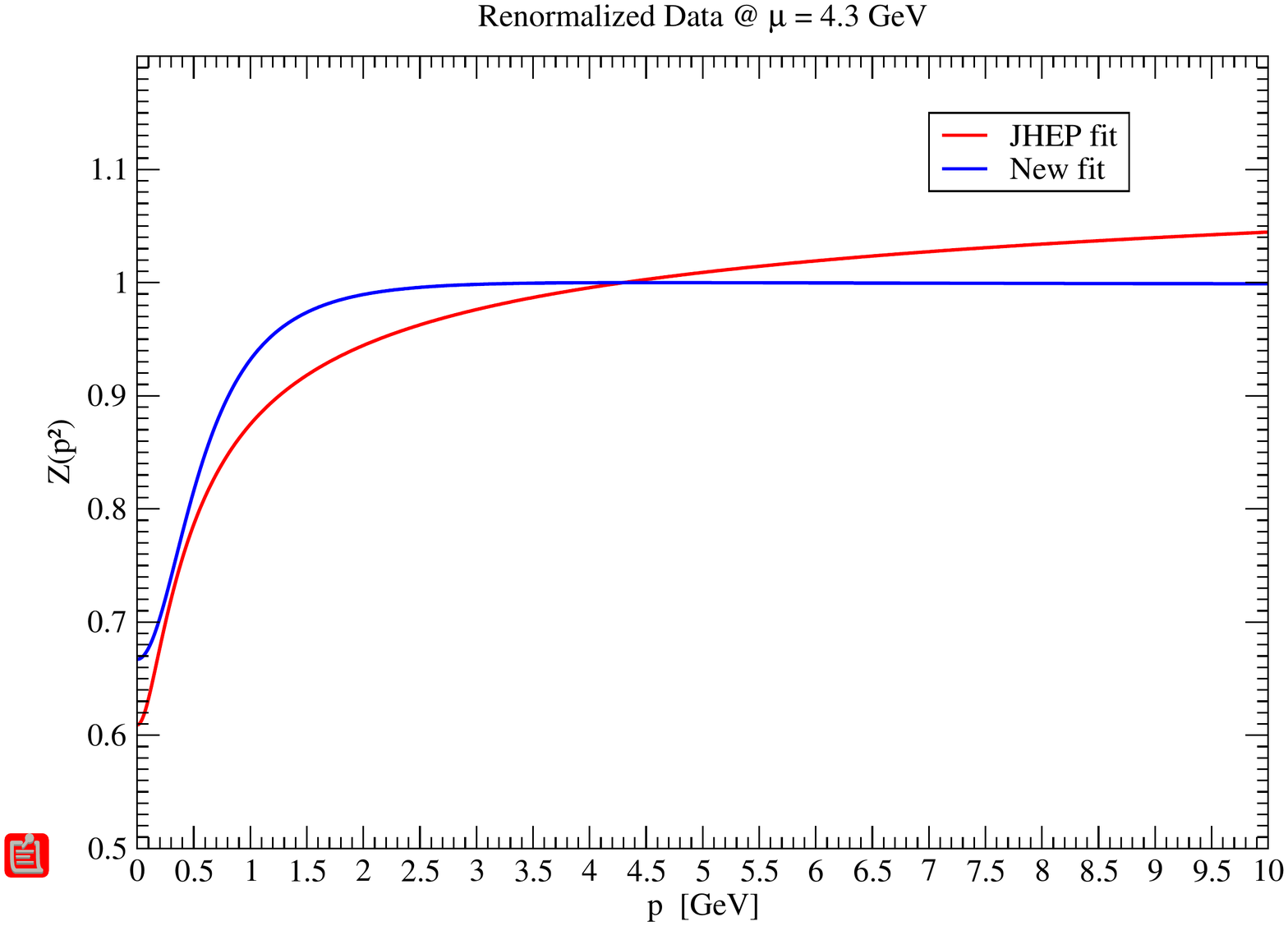} \\
   \includegraphics[width=3in]{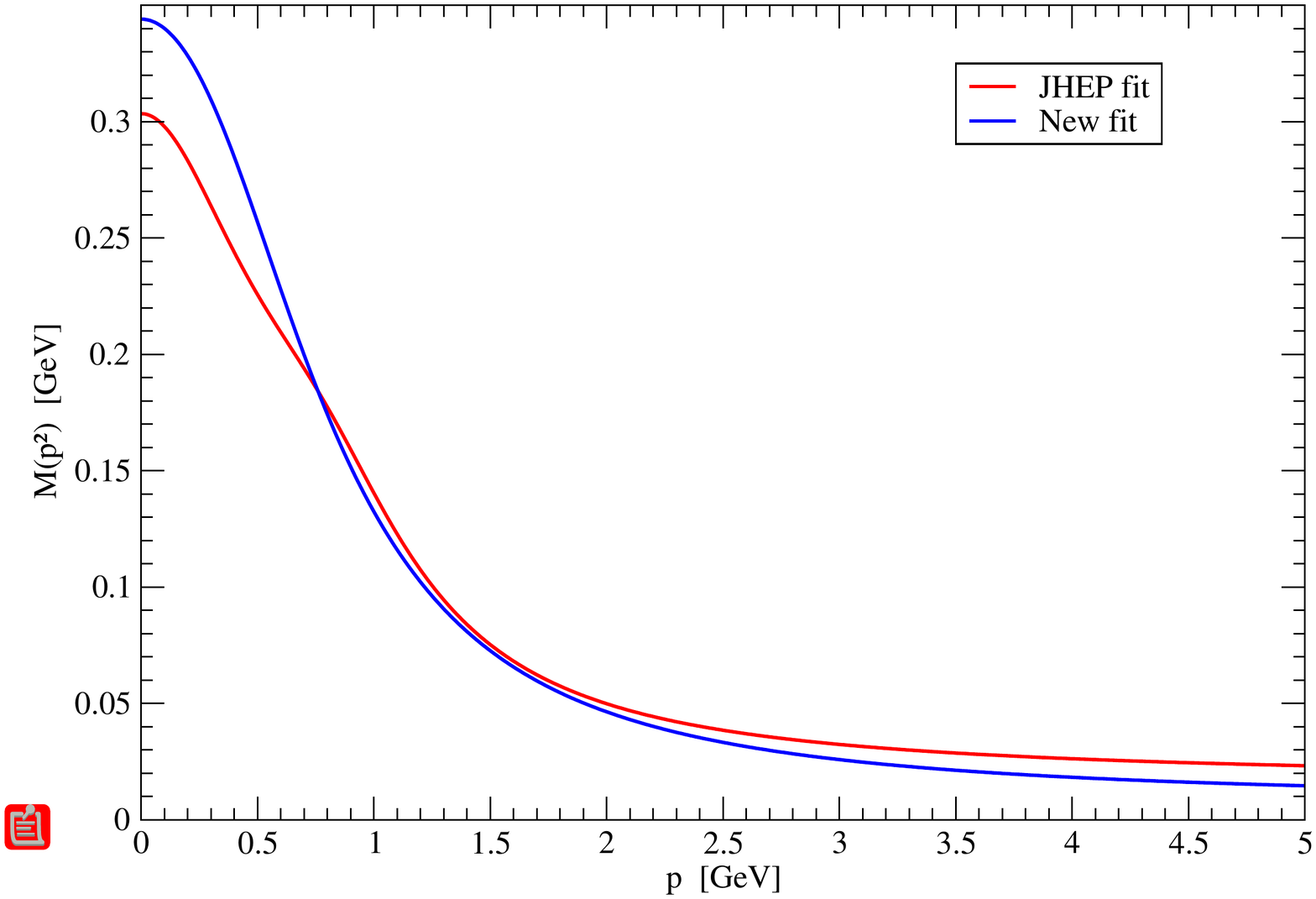} 
   \caption{Fits to the renormalized propagators at $\mu = 4.3$ GeV: (top-left) pure Yang-Mills gluon propagator; (top-right) pure-Yang-Mills ghost dressing
                 function; (bottom-left) quark wave function; (bottom-right) running quark mass.}
   \label{fig:comparefitsprops}
\end{figure}

For completeness and in order to allow for a better comparison between the of the current work with those reported in~\cite{Rojas:2013tza},
we provide the fits used in both works with the gluon propagator, the ghost propagator and the quark wave function curves renormalised
at $\mu = 4.3$ GeV within the MOM scheme. On Fig.~\ref{fig:comparefitsprops} the curves referred to as JHEP are those of~\cite{Rojas:2013tza},
while those designated as NEW are the curves mentioned in Secs.~\ref{Sec:gluonghost} and~\ref{Sec:quarkpropfits}.
As the figure shows, there are differences between the two sets of curves, not only at the infrared region but also on the running at high momentum.

\end{appendix}


\end{document}